\pdfoutput=1
\documentclass[12pt]{article}

\usepackage[utf8]{inputenc}
\usepackage[margin = 1.0in]{geometry}
\usepackage{amsmath}
\numberwithin{equation}{section}
\numberwithin{figure}{section}
\numberwithin{table}{section}
\usepackage[dvipsnames]{xcolor}
\usepackage{tikz}
\usetikzlibrary{arrows.meta, decorations.markings}
\usepackage{standalone}
\usepackage{amsthm}
\usepackage{amssymb}
\usepackage{physics}
\usepackage{textcomp}
\usepackage{mathrsfs}
\usepackage{gensymb}
\usepackage{bbm}
\usepackage{graphicx}
\usepackage[most]{tcolorbox}
\usepackage{mathtools}
\usepackage{bm}
\usepackage{float}
\usepackage{cases}
\usepackage[affil-it]{authblk}
\usepackage{hyperref}
\hypersetup{
    colorlinks,
    citecolor=blue,
    filecolor=black,
    linkcolor=Blue,
    urlcolor=blue,
    linktoc=all
}
\usepackage{caption}
\usepackage{subcaption}
\usepackage{diagbox}
\usepackage{cellspace}
\usepackage{enumitem}
\setlist{topsep=0pt, itemsep=0pt}
\usepackage{makecell}
\usepackage[none]{hyphenat}
\usepackage[scr=stixtwofancy]{mathalpha}
\newcommand{\myemm}{\mathscr{m}}
\newcommand{\myell}{\mathscr{l}}

\usepackage{cite}

\newcommand{\I}{\mathbbm{1}}
\newcommand{\R}{\mathbb{R}}
\newcommand{\Z}{\mathbb{Z}}

\newcommand{\C}{\mathbb{C}}
\newcommand{\F}{\mathbb{F}}
\newcommand{\T}{\mathbb{T}}
\newcommand{\End}{\mathrm{End}}
\newcommand{\Rep}{\mathrm{Rep}}
\newcommand{\Stab}{\mathrm{Stab}}
\newcommand{\Hom}{\mathrm{Hom}}

\newcommand{\Ind}{\mathrm{Ind}}
\renewcommand{\Res}{\mathrm{Res}}

\newcommand{\SL}{\mathrm{SL}}

\newcommand{\DW}{\mathrm{DW}}
\newcommand{\CS}{\mathrm{CS}}

\newcommand{\mil}{\overline{\mu}}

\newcommand{\longto}{\longrightarrow}
\newcommand{\ad}{\mathrm{ad}}
\newcommand{\Ad}{\mathrm{Ad}}
\newcommand{\lie}[1]{\mathfrak{#1}}
\newcommand{\Lie}[1]{\mathsf{#1}}
\newcommand{\orb}[1]{\mathcal{O}_{#1}}
\newcommand{\Heis}{\mathrm{Heis}}
\newcommand{\Fil}{\mathrm{Fil}}
\newcommand{\PD}{\mathrm{PD}}
\DeclareRobustCommand{\wilsonchain}{%
  \tikz[baseline=-0.25em, x=1em, y=1em]{%
    \draw[line width=1.1pt] (0,0) circle (0.28) (1.06,0) circle (0.28)
                            (2.12,0) circle (0.28);
    \draw[line width=0.4pt] (0.28,0) -- (0.78,0) (1.34,0) -- (1.84,0);
  }}

\definecolor{borOrange}{RGB}{235,85,60}
\definecolor{borBlue}{RGB}{40,80,225}
\definecolor{borTeal}{RGB}{40,210,110}

\newtheorem{theorem}{Theorem}[section]
\newtheorem{proposition}{Proposition}[section]
\newtheorem{conjecture}{Conjecture}[section]

\newenvironment{noframetbox}{\begin{tcolorbox}[enhanced, breakable, tile, parbox=false]}{\end{tcolorbox}}

\newcommand\mydots{\makebox[1em][c]{\text{\hspace{-0.17em}$\cdot$\hfil$\cdot$\hfil$\cdot$}}\hspace{-0.17em}}

\title{\textbf{Link invariants in nilpotent group \\ gauge theories} }
\author[1, 2]{Thomas Nicosanti}
\author[3]{Pavel Putrov}
\author[1, 2, 3]{Johann Quenta-Raygada}
\affil[1]{SISSA, Via Bonomea 265, Trieste 34136, Italy}
\affil[2]{INFN -- Sezione di Trieste, Via Valerio 2, Trieste 34127, Italy}
\affil[3]{ICTP, Strada Costiera 11, Trieste 34151, Italy}
\date{}

\begin{document}

\maketitle

\begin{abstract}
We consider 3-dimensional topological gauge theories with nilpotent gauge groups, both finite, that is Dijkgraaf-Witten theories, and continuous, that is Chern-Simons theories. Abelian topological quantum field theories are exactly those which do not detect the Borromean rings, and nilpotent gauge groups provide the natural next steps beyond the abelian case. We show that the untwisted Dijkgraaf-Witten theory with gauge group $G$ detects Brunnian links with up to $n+1$ components, but no Brunnian links with more components, if and only if $G$ is $n$-step nilpotent. Moreover, the expectation values of such links are determined by their Milnor invariants. We also study the properties of Chern-Simons theories with nilpotent Lie groups and draw parallels with those of the Dijkgraaf-Witten theories. Finally, we conjecture, using the Lazard correspondence, a relation between expectation values of line operators in the two types of theories, and test it in examples.
\end{abstract}

\tableofcontents

\section{Introduction and summary} \label{sec:1intro}

Topological quantum field theories (TQFTs) in three dimensions describe the infrared dynamics of gapped phases of matter and, at the same time, provide invariants of knots, links and 3-manifolds. The two basic families of examples are gauge theories. Chern-Simons (CS) theories with compact Lie gauge groups \cite{Witten1988} produce the Jones polynomial and its generalizations, while Dijkgraaf-Witten (DW) theories with finite gauge groups \cite{Dijkgraaf:1989pz,Freed:1991bn} produce invariants counting flat connections. The gauge groups considered in the literature are usually either abelian, or compact semisimple Lie groups, or finite groups treated in full generality. In this paper we consider a class of gauge groups which, in a precise sense, lies in between the abelian and the general case: \textit{nilpotent} groups, both finite and Lie. A group $G$ is $n$-step nilpotent if the \textit{lower central series} $G_{i+1}=[G_i,G]\subseteq G_i$, $G_1=G$, terminates after exactly $n$ steps, that is $G_{n+1}=1$ and $G_n\neq 1$. Abelian groups are 1-step nilpotent. In general, the nilpotency class measures how many nested commutators of group elements one has to take before the result becomes trivial. As we will see, this is precisely the kind of information probed by line operators in the gauge theory supported on Brunnian links.

Let us make this more precise. Consider a TQFT $\mathcal{T}$ and an oriented link $\mathcal{L}\subset S^3$ the components of which are colored by simple line operators of $\mathcal{T}$. We say that $\mathcal{T}$ \textit{detects} $\mathcal{L}$ if for some choice of the colors the expectation value of $\mathcal{L}$ differs from the expectation value of the unlink with the same number of components and the same colors.\footnote{Line operators in a 3d TQFT are in general framed, and the expectation value depends on the framing through the topological spins. Throughout the paper we compare links and the corresponding unlinks with the same framings (fixed to zero self-linking numbers by default).} An abelian TQFT, that is a TQFT in which all simple line operators have quantum dimension one, only sees the linking matrix: the expectation value of a link depends on the pairwise linking numbers and the framings, and on nothing else. In particular, an abelian TQFT does not detect any link with vanishing linking numbers. A link is \textit{Brunnian} if it becomes the unlink after removing any one of its components.\footnote{In this respect Brunnian links are, in a sense, analogous to GHZ states, which become unentangled after tracing out any one of the parties \cite{aravind1997borromean,Kauffman:2002xud,Sugita:2007fca,Quinta:2018scm}. This is only an analogy, however. The states which a TQFT associates to the boundary of the complement of a Brunnian link are in general not of GHZ type; for instance, for the Borromean rings in $SU(2)$ CS theory the state is of W type \cite{Salton:2016qpp}, see also \cite{Balasubramanian:2016sro}.} Pairwise linking numbers of a Brunnian link with at least three components vanish, and the simplest example is the Borromean rings. It is a piece of folklore that the converse holds as well, namely that a TQFT is abelian if and only if it does not detect the Borromean rings \cite{Wang:2014wka,Wang:2016qhf,Putrov:2016qdo,Iqbal:2023wvm}. In Appendix \ref{app:Aabelian} we provide a proof that a unitary TQFT is abelian if and only if it does not detect any 3-component Brunnian link.

The question we address is what happens beyond the first step: which TQFTs detect Brunnian links with up to $n+1$ components but no Brunnian links with more components. For discrete gauge theories we give a complete answer. Denote by $\DW_G$ the untwisted DW theory with finite gauge group $G$. We show in Theorem \ref{th:nilpDWBrunnians} that

\begin{equation} \label{eq:intro-DW-nilpotent-Brunnian}
    \boxed{\;
    \begin{array}{c}
        \text{$\DW_G$ detects Brunnian links} \\
        \text{with up to $n+1$ components}
    \end{array}
    \;\Longleftrightarrow\;
    \text{$G$ is $n$-step nilpotent.}
    \;}
\end{equation}
Moreover, for a given number of components it is sufficient to consider a single Brunnian link, the Milnor link $\mathcal{M}_k$ with $k$ components shown in Figure \ref{fig:milnorlink}. It generalizes the role played by the Borromean rings in the abelian case. When $G$ is $n$-step nilpotent, the expectation value of an $(n+1)$-component Brunnian link is determined by its Milnor invariants of length $n+1$ (Theorem \ref{th:DWBrunnianMilnor}). Using this result we show that any link-homotopically non-trivial Brunnian link is detected by the DW theory of some finite nilpotent group.

The mechanism behind these statements is the following. The expectation value of a link in $\DW_G$ is a sum over homomorphisms from the fundamental group $\pi_1$ of the link complement to $G$, in which the electric charges of the lines are evaluated on the images of the longitudes. For a Brunnian link with $n+1$ components the longitudes lie deep in the lower central series of $\pi_1$. Namely, modulo $(\pi_1)_{n+1}$ each longitude is a product of iterated commutators of the meridians of the other components, with exponents given by the Milnor invariants of the link (Theorem \ref{th:Brlongitude}). This is a consequence of Milnor's theory of $\mil$-invariants \cite{Milnor2015} and of the Magnus expansion \cite{Magnus1937}, which we review in Section \ref{sec:3brunnian}. If $G$ is $n$-step nilpotent, the image of such a longitude is central, and by Schur's lemma the electric charge evaluates on it to a phase governed by the Milnor invariants. If the nilpotency class of $G$ is smaller than $n$, the image is trivial and the link is not detected.

We restrict ourselves to untwisted DW theories. Turning on the twist in general can change the detection properties. In particular, a DW theory with an abelian gauge group and a non-trivial twist can be equivalent, as a TQFT, to an untwisted theory with a non-abelian nilpotent gauge group, see e.g.\ \cite{deWildPropitius:1996gt,Kapustin:2014zva,Yoshida:2015cia} in the physics literature and \cite{naidu2008lagrangian,goff2007gauge,li2025gauge} in the mathematics literature. For twisted theories the nilpotency class of the gauge group is therefore not the relevant quantity, and we leave their systematic treatment for future work.

Finite nilpotent groups have appeared before in the physics of topological phases, in situations where nilpotency is the relevant property rather than an accident. In the study of quantum computation with the anyons of DW theories, Mochon showed that groups which are solvable but not nilpotent, the smallest being $S_3$, allow universal quantum computation by braiding, fusion and measurement \cite{Mochon:2004hoi} (for non-solvable groups see \cite{Mochon:2003zz}), while for nilpotent groups the braiding of fluxes cannot produce a controlled-NOT gate, and universality remains an open question. More recently, nilpotency was found to control the preparation of the corresponding topological orders on quantum processors: the quantum double of a group of nilpotency class 2, such as $D_4$ or $Q_8$, can be prepared with a single round of measurements \cite{Tantivasadakarn:2022hgp}. This is what made $D_4$ the first non-abelian topological order realized experimentally \cite{Iqbal:2023wvm}. See also \cite{Gunn:2026rpm} for the role of the hierarchy of irreducible representations of nilpotent groups in a similar setting, and \cite{Warman:2025hov,Hsin:2025zgn} for applications of non-abelian quantum doubles to fault-tolerant logical gates. In \cite{Hsin:2025zgn} the relevant theory is the $\Z_2^3$ gauge theory with a type-III twist, which, as discussed above, is equivalent to the untwisted DW theory of a non-abelian group of order 8, although nilpotency is not used there explicitly. In \cite{Warman:2025hov} the quantum doubles of the dihedral groups $D_{4N}$ of order $8N$ are used to realize phase gates in constant depth, and the case $8N=2^n$, in which the group is a 2-group and therefore nilpotent (in \cite{Warman:2025hov} this is phrased in terms of a normal series with all successive quotients isomorphic to $\Z_2$), turns out to be special: the gate then lies in the $n$-th level of the Clifford hierarchy and admits a realization in terms of qubits only. In all these cases the relevant feature is that a nilpotent group is obtained from an abelian group by iterated central extensions, so that the non-abelian properties appear ``one commutator at a time''. The detection of Brunnian links discussed above measures precisely how many such steps are present. Mathematically, nilpotency of $G$ is equivalent to nilpotency of the fusion category $\Rep(G)$ in the sense of \cite{gelaki2008nilpotent}, see also \cite{drinfeld2007group} for the corresponding properties of the modular categories $\Rep(D(G))$.

The second half of the paper concerns CS theories with a nilpotent Lie gauge group $\Lie{G}$, with Lie algebra $\lie{g}$. CS theories with non-semisimple\footnote{Throughout the paper ``non-semisimple'' refers to the Lie algebra of the gauge group. This should not be confused with the ``non-semisimple TQFTs'' of \cite{Costantino:2012it}, where the adjective refers to the category of representations of a quantum group, that is to the category of line operators.} gauge groups and ad-invariant metrics were considered in general in \cite{Tseytlin:1995yw,Grassi:1999nb}. The case of a 2-step nilpotent gauge algebra was studied in \cite{Ferrari:2002zq}, where it was observed that the perturbative expansion contains only finitely many Feynman diagrams and that the expectation values of suitably modified Wilson loops produce link invariants of Milnor type. Nilpotent CS theories have also appeared in disguise, as continuum descriptions of twisted DW theories with abelian gauge groups \cite{Wang:2016qhf,Putrov:2016qdo,He:2016xpi} (cf. also \cite{Bergman:2026lnz} for a recent appearance in the context of symmetry TQFTs). However, some of the statements made there require refinements. In this paper we point out that such a CS theory cannot be equivalent as a TQFT to a finite group gauge theory, and that the relation between the two holds rather at the level of expectation values of line operators, up to normalization factors. Several features distinguish nilpotent CS theories from the familiar compact case. A connected nilpotent Lie group of nilpotency class at least two is non-compact. The Killing form of $\lie{g}$ vanishes identically, so the existence of an ad-invariant metric, which is needed to write the CS action, is a non-trivial condition on $\lie{g}$ \cite{ovando2017liealgebrasadinvariantmetrics}. We take the point of view that the metric is part of the data defining the theory. For simply connected $\Lie{G}$ all bundles are trivial and the level is not quantized, so it can be absorbed in the normalization of the metric. At the same time, automorphisms of $\lie{g}$ can rescale the metric and thereby identify theories with different normalizations. Moreover, finite-dimensional irreducible representations of a nilpotent Lie algebra are one-dimensional, so Wilson lines in such representations do not see the non-abelian structure at all. The interesting line operators are labelled by infinite-dimensional unitary irreducible representations. For nilpotent groups the infinite-dimensional ones are classified by coadjoint orbits via Kirillov's orbit method \cite{Kirillov1962,corwin1990representations}, which we review in Section \ref{sec:2nilp}. In general they form a continuous family. The trace of such a representation is a distribution on $\Lie{G}$ rather than a function, and we propose to consider the corresponding ``distribution-valued'' Wilson lines. All this already indicates that a nilpotent CS theory cannot coincide with a non-abelian finite group gauge theory: the latter has a finite number of simple line operators, and their quantum dimensions are positive integers, some of which are greater than 1. In the CS theory the only Wilson lines with a finite quantum dimension are the one-dimensional ones (unless one applies some hypothetical regularization scheme to the infinite-dimensional ones).

The perturbative expansion of expectation values of Wilson lines in CS theory produces finite-type invariants of the link \cite{bar1995vassiliev,Alvarez:1996ir}. For an $(n+1)$-component Brunnian link all finite-type invariants of degree less than $2n$ vanish, and those of degree $2n$ are quadratic forms in the Milnor invariants of length $n+1$ \cite{habiro2008finite} (cf. the appearance of Milnor linking numbers in the expansion of the Jones polynomial observed in \cite{rozansky1994reshetikhin}). We therefore expect the continuous counterparts of nilpotent DW theories to have the same detection properties, with the nilpotency class of $\lie{g}$ controlling which Brunnian links are detected. We discuss the perturbative expansion from this point of view in Section \ref{sec:5CSnilp}.

Finally, we relate the two classes of theories. In the abelian case the relation is well known: the $\Z_p$ DW theory is equivalent to the $U(1)\times U(1)$ CS theory with off-diagonal level matrix, that is to BF theory at level $p$. The corresponding Lie algebra $\R\oplus\R^*$ with its natural pairing is the cotangent Lie algebra of the abelian Lie algebra $\R$. This has the following generalization. The Lazard correspondence (see e.g.\ \cite{Khukhro1998}) identifies finite $p$-groups of nilpotency class $c<p$ with nilpotent Lie rings\footnote{Informally, a Lie ring is a Lie algebra over $\Z$ instead of a field: an abelian group equipped with a $\Z$-bilinear alternating bracket satisfying the Jacobi identity. The precise definition is given in Section \ref{subsec:lazardfiniteorbits}.} of the same class by means of the Baker-Campbell-Hausdorff formula. The additive group of the Lie ring is a $p$-group. Moreover, if the finite group $G$ has exponent $p$, the corresponding Lie ring $\log(G)$ is a Lie algebra over $\F_p$. When it lifts to a real nilpotent Lie algebra $\lie{g}$, the cotangent Lie algebra $T^*\lie{g}=\lie{g}\oplus\lie{g}^*$ carries a canonical ad-invariant metric and has the same nilpotency class as $\lie{g}$. For an arbitrary Lie algebra $\lie{g}$, the CS theory with gauge algebra $T^*\lie{g}$ is, at least at the classical level, the BF theory with gauge algebra $\lie{g}$ \cite{Blau:1989bq}. In most of the previous work on BF theory $\lie{g}$ was assumed to be semisimple, the prime example being $\lie{g}=\mathfrak{so}(2,1)$, for which one obtains the CS formulation of 3d gravity \cite{Witten:1988hc}. Semisimplicity is important for some of the results in the literature, for instance that the perturbative expansion of BF theory produces the Alexander-Conway polynomial of links \cite{Cattaneo:1994pk,Naot:2003hi}. We conjecture that expectation values of line operators both in $\DW_G$ and in the CS theory with Lie algebra $T^*\lie{g}$ have a similar form and are governed by the same polynomial in the electric and magnetic charges (classifying the lines simultaneously in both theories) together with some auxiliary variables. We provide evidence for this in two examples: the Heisenberg group $\Heis(\F_p)$ versus $T^*\lie{h}_3$, and the group $\Fil_4(\F_p)$ of order $p^4$ and nilpotency class 3 versus $T^*\lie{f}_4$. The conjecture is formulated and tested in Section \ref{sec:6corresp}.

Throughout the paper, unless explicitly mentioned otherwise, Lie algebras are defined over $\R$, Lie groups are connected, DW theories are untwisted, and links are considered in $S^3$.

The rest of the paper is organized as follows. In Section \ref{sec:2nilp} we review nilpotent groups and nilpotent Lie algebras, the orbit method and trace distributions of infinite-dimensional unitary representations, and introduce the two examples used later, $\Heis(\F_p)$ and $\Fil_4(\F_p)$. In Section \ref{sec:3brunnian} we review Brunnian links, Milnor invariants and the Magnus expansion, and derive the canonical form of the longitudes of a Brunnian link modulo the lower central series. In Section \ref{sec:DWnilp} we review DW theories and their line operators and prove the statements about Brunnian links in $\DW_G$ described above. In Section \ref{sec:5CSnilp} we consider CS theories with nilpotent gauge groups: the quantization of the level, the role of the metric, distribution-valued Wilson lines and the perturbative expansion. In Section \ref{sec:6corresp} we formulate the conjectured correspondence between the two types of theories and test it in the examples. Appendix \ref{app:Aabelian} contains the proof that a unitary TQFT is abelian if and only if it does not detect any 3-component Brunnian link. In Appendix \ref{app:BT*h3modular} we present explicit calculations of the expectation values of certain links in CS theories. Appendix \ref{app:Cdictionary} contains the details of the algebra isomorphism between the Drinfeld double $D(G)$ and the group algebra of $\exp(T^*\log(G))$, which is used to formulate the conjectural relation between DW and CS theories in Section \ref{sec:6corresp}.

\section{Nilpotent groups} \label{sec:2nilp}

Let us remind the reader what it means for a group to be nilpotent. Let $G_1 \equiv G$ and define $G_i \equiv [G_{i-1}, G]$ recursively for $i \geq 2$.
A group $G$ is \textit{nilpotent} if its lower central series
\begin{equation}
    G = G_1 \unrhd  G_2 \unrhd G_3 \unrhd \cdots
\end{equation}
is such that $G_{n+1} = 1$ for some $n$. Here $A \unrhd B$ means that $B$ is a normal subgroup of $A$ (in fact each $G_i$ is normal in $G$). We say $G$ is $n$-step nilpotent (or nilpotent of class $n$) if moreover $G_n \neq 1$. A group is 0-step nilpotent iff it is the trivial group, and is 1-step nilpotent iff it is abelian. Since $[G_i, G] = G_{i+1}$, the quotient $G_i/G_{i+1}$ is central in $G/G_{i+1}$, so that
\begin{equation}
    1 \longto G_i/G_{i+1} \longto G/G_{i+1} \longto G/G_i \longto 1
\end{equation}
is a central extension for every $i$. An $n$-step nilpotent group is thus obtained from the abelian group $G/G_2$ by $n-1$ successive central extensions by abelian groups, and conversely a central extension of an $(n-1)$-step nilpotent group is nilpotent of class at most $n$.

\subsection{Nilpotent Lie groups and Lie algebras}

A Lie group $\Lie{G}$ is nilpotent if its Lie algebra $\mathfrak{g}$ is nilpotent. More explicitly, let $\mathfrak{g}_1 \equiv \mathfrak{g}$ and define $\mathfrak{g}_i \equiv [\mathfrak{g}, \mathfrak{g}_{i-1}]$ recursively for $i \geq 2$. The Lie algebra $\mathfrak{g}$ is nilpotent if its lower central series
\begin{equation}
    \mathfrak{g} = \mathfrak{g}_1 \supseteq \mathfrak{g}_2 \supseteq \mathfrak{g}_3 \supseteq \dots
\end{equation}
is such that there exists an integer $n$ for which $\mathfrak{g}_{n+1} = 0$. We say $\mathfrak{g}$ is $n$-step nilpotent if moreover $\mathfrak{g}_n \neq 0$. This definition of nilpotency coincides with the group-theoretic definition if $\Lie{G}$ is connected --- we will assume this is the case unless specified otherwise. Moreover, we will always consider our Lie algebras defined over $\R$.

Nilpotent Lie groups and algebras are quite different from their more familiar semisimple counterparts. First of all, the exponential map is a global diffeomorphism from $\mathfrak{g}$ to the simply connected Lie group $\Lie{G} = \exp(\lie{g})$, and one can show that any connected Lie group with the same Lie algebra must be a quotient of $\Lie{G}$ by a discrete central subgroup. This implies that, if $\lie{g}$ has nilpotency class $\geq 2$, any connected Lie group with said Lie algebra must be necessarily non-compact: indeed, such a group surjects onto $\Lie{G}/Z(\Lie{G})$, which is again simply connected and nilpotent, hence diffeomorphic to $\lie{g}/\lie{z}(\lie{g}) \neq 0$ via the exponential map.

Another difference, which will become important in Section \ref{sec:5CSnilp}, is that for semisimple Lie algebras there always exists a non-degenerate ad-invariant symmetric bilinear form, also known as an \textit{ad-invariant metric} --- the Killing form. For nilpotent Lie algebras however the Killing form vanishes identically, since every $\ad_X$ is a nilpotent endomorphism, and in general these algebras may not admit any ad-invariant metrics at all.\footnote{Although there is not a complete classification of Lie algebras admitting ad-invariant metrics, several examples are known in the literature, see e.g.\ the review by Ovando \cite{ovando2017liealgebrasadinvariantmetrics}.} We will refer to a Lie algebra equipped with an ad-invariant metric as a \textit{quadratic} Lie algebra.

More importantly, the representation theory of nilpotent Lie algebras is quite in contrast with that of semisimple ones. In the latter case the classification of finite-dimensional irreducible representations is based on the theory of highest weights, enabled by the fact that, for semisimple algebras, all elements of a Cartan subalgebra are diagonalizable in the adjoint representation. For nilpotent Lie algebras this last condition is of course spoiled, hence a different approach is required to classify their representations.

\subsubsection{Representation theory: the Kirillov orbit method} \label{subsec:Kirillovthy}

The representation theory of nilpotent Lie algebras is well-understood in the mathematics literature, although less familiar for physicists. Indeed, the framework describing the irreducible unitary representations of a nilpotent Lie group $\Lie{G}$ goes under the umbrella of \textit{Kirillov theory}, also known as the orbit  method. In this section we aim to summarize the essential aspects of Kirillov theory, which should serve as a toolkit for physicists interested in the study of nilpotent group gauge theories. The interested reader may find a more rigorous treatment in the excellent textbook by Corwin and Greenleaf \cite{corwin1990representations}.

Before delving into the general theory, let us remark that all finite dimensional complex irreps of a nilpotent Lie algebra are necessarily 1-dimensional, a consequence of Lie's theorem \cite{serre2013complex}. The truly interesting representations (which will be fundamental to our discussions in Section \ref{sec:5CSnilp}) are the \textit{infinite-dimensional} ones. Notably, Kirillov theory classifies \textit{all} unitary irreps of the group, regardless of their dimension.

Let us assume throughout that $\Lie{G}$ is simply connected unless specified otherwise. The basis of Kirillov theory lies in the following theorem \cite{Kirillov1962}:
\begin{theorem}[Kirillov] \label{thm:Kirillov}
    There is a 1-to-1 correspondence between the set of unitary irreps of $\Lie{G}$ and the set $\mathfrak{g}^*/\Ad^*(\Lie{G})$ of coadjoint orbits in the dual $\mathfrak{g}^*$ of its Lie algebra.
\end{theorem}
We remind the reader that the coadjoint representation $\Ad^*:\Lie{G} \to \text{End}(\mathfrak{g}^*)$ is defined as
\begin{equation}
    \qty(\Ad^*(g) \cdot \varphi) (X) = \varphi(\Ad(g^{-1}) \cdot X)
\end{equation}
for any $X \in \mathfrak{g}$ and $g \in \Lie{G}$, where $\Ad$ denotes the usual adjoint representation of $\Lie{G}$. It will also be convenient to use the corresponding right action, $\varphi \cdot g \equiv \Ad^*(g^{-1}) \cdot \varphi$, i.e.\ $(\varphi \cdot g)(X) = \varphi(\Ad(g) \cdot X)$. The orbits of $\mathfrak{g}^*$ under this action, known as the \textit{coadjoint orbits}, classify the unitary irreps of $\Lie{G}$ as follows. For $\varphi \in \mathfrak{g}^*$, let $\mathfrak{m}$ be a \textit{polarizing subalgebra} for $\varphi$, that is a subalgebra of $\lie{g}$ which is a maximal isotropic subspace for the bilinear form $B_\varphi(X,Y) = \varphi([X,Y])$, i.e.\ $B_\varphi$ restricts to zero on $\lie{m}$ and $\lie{m}$ has the largest dimension compatible with this property.\footnote{Maximality is understood with respect to dimension, not inclusion. A subalgebra such that $B_\varphi$ vanishes on it, and such that it is not contained in any larger subalgebra with this property, need not be polarizing, and inducing from it may give a reducible representation.} Then the map
\begin{equation}
    \chi_{\varphi, \lie{m}}(e^Y) = e^{i \varphi(Y)}, \quad Y \in \lie{m},
\end{equation}
is a 1-dimensional representation of $M = \exp(\lie{m})$. In this way, one obtains a representation $\pi_{\varphi, \lie{m}} = \Ind_M^{\Lie{G}}(\chi_{\varphi, \lie{m}})$ of $\Lie{G}$ by induction\footnote{The reader might be familiar with the method of induced representations in the world of finite groups. There is an analogous construction for (locally compact) topological groups, where an induced representation may be built from a representation of a closed subgroup. We refer to \cite{corwin1990representations} for further details --- for our purposes, the explicit description of $\pi_{\varphi, \lie{m}}$ given later on will suffice.}. The following statements essentially prove the theorem above:
\begin{enumerate}
    \item[(1)] $\pi_{\varphi, \lie{m}}$ is irreducible for whatever choice of polarizing subalgebra $\lie{m}$. Moreover,  if $\lie{m}$ and $\lie{m}'$ are two different polarizing subalgebras for $\varphi$, the representations $\pi_{\varphi, \lie{m}}$ and $\pi_{\varphi, \lie{m}'}$ are equivalent. We denote them by $\pi_\varphi$ if we only care about their equivalence class.
    \item[(2)] Any unitary irrep of $\Lie{G}$ is equivalent to $\pi_{\varphi}$ for some $\varphi \in \lie{g}^*$.
    \item[(3)] If $\varphi, \varphi' \in \lie{g}^*$, then $\pi_\varphi \cong \pi_{\varphi'}$ iff $\varphi$ and $\varphi'$ belong to the same coadjoint orbit, which we denote by $\orb{\varphi}$.
\end{enumerate}

It will be useful to discuss some properties of polarizing subalgebras in more detail. The stabilizer $\Stab_\varphi$  of $\varphi$ under the coadjoint action has as Lie algebra the radical of $B_\varphi$,
\begin{equation}
    \lie{r}_\varphi = \{ Y \in \lie{g} \mid B_\varphi(X,Y) = 0, \forall X \in \lie{g} \}.
\end{equation}
It is moreover true that $\Stab_\varphi = \exp \lie{r}_\varphi$. The bilinear form $B_\varphi$ defines a non-degenerate symplectic form on $\lie{g}/\lie{r}_\varphi$ --- this implies that $\lie{r}_\varphi$ is of even codimension. Moreover, it is a fact that polarizing subalgebras for $\varphi$ exist and they all have codimension $\frac{1}{2}\dim{\lie{g}/\lie{r}_\varphi}$. Finally, one can check that the coadjoint orbit $\orb{\varphi}$ is a submanifold of $\lie{g}^*$ of dimension $\dim(\lie{g}/\lie{r}_\varphi)$. From this we conclude that
\begin{equation}
    \dim\orb{\varphi} = 2(n-m),
\end{equation}
where $n \equiv \dim \lie{g}$ and $m \equiv \dim \lie{m}$. In particular, coadjoint orbits are always even-dimensional manifolds.

\subsubsection*{Weak Malcev bases}

Before anything, let us introduce a particular basis that always exists for nilpotent Lie algebras, which will be useful in later computations. Let $\lie{k}_1 \subseteq \dots \subseteq \lie{k}_r$ be a sequence of subalgebras of $\lie{g}$. We say that a basis $\{X_1, \dots, X_n\}$ of $\lie{g}$ is a \textit{weak Malcev basis passing through $\{\lie{k}_j\}$} if the following two conditions are met:
\begin{enumerate}
    \item[(1)] For $1 \leq m \leq n$, $\lie{l}_m \equiv \mathrm{span}_{\R}\{X_1, \dots, X_m\}$ is a subalgebra of $\lie{g}$.
    \item[(2)] For $1 \leq j \leq r$, $\lie{l}_{\dim \lie{k}_j} = \lie{k}_j$.
\end{enumerate}

\subsubsection*{Standard realization of $\pi_{\varphi, \lie{m}}$}

The simplest way to model the representation $\pi = \pi_{\varphi,\lie{m}}$ is via unitary operators acting on the Hilbert space of $L^2$ functions over the coset space $M\backslash \Lie{G} = \{Mg \mid g \in \Lie{G}\}$, as follows. Let $\{X_i\}$ be a weak Malcev basis of $\lie{g}$ passing through $\lie{m}$, and define the coordinate map
\begin{equation}
    g(s_1, \dots, s_n) = e^{s_1X_1} \cdots e^{s_nX_n}, \qquad s \in \R^n.
\end{equation}
The map $\gamma: \R^{n-m} \to \Lie{G}$ given by $\gamma(t) = g(0, t)$ defines a cross-section for $M \backslash \Lie{G}$, which we use to identify $M \backslash \Lie{G}$ with $\R^{n-m}$, and an arbitrary element $x \in \Lie{G}$ may be split as $h_x \cdot \sigma_x \in M \cdot \mathrm{Im}(\gamma)$. The action of $\pi$ on $L^2(M \backslash \Lie{G})$ is then given by
\begin{equation}
    \qty(\pi_{\varphi, \lie{m}}(g) \cdot f)(\zeta) = \chi_{\varphi, \lie{m}}(h_{\gamma(\zeta)g}) f(\zeta \cdot g),
\end{equation}
where $\zeta \in M \backslash \Lie{G}$ and $\zeta \cdot g$ denotes the right action of $\Lie{G}$ on cosets, $Mx \cdot g = Mxg$. %
Of course, since $\Im(\gamma)$ is isomorphic to $\R^{n-m}$, we can further identify the representation space with $L^2(\R^{n-m})$.

\subsubsection*{Traces of irreducible representations}

If $\pi$ is an infinite-dimensional representation, the correct notion of trace is that of a \textit{trace distribution} acting on suitable test functions. Indeed, if $\phi$ is a Schwartz function on $\Lie{G}$, the operator
\begin{equation}
    \pi_\phi = \int_{\Lie{G}} \phi(g) \pi(g) dg
\end{equation}
is trace class for $\pi$ a unitary irrep. We then define the functional $\Tr_\pi$, which is further shown to be a tempered distribution, by
\begin{equation}
    \Tr_\pi [\phi] = \Tr(\pi_\phi), \qquad \phi \in \mathcal{S}(\Lie{G}).
\end{equation}
In the same spirit of physicists denoting the Dirac delta distribution as a function, we will denote the trace functional as a ``function'' $\Tr_\pi(g)$ over $\Lie{G}$, where it is understood that
\begin{equation}
    \Tr_\pi[\phi] \mathrel{\text{``=''}} \int_{\Lie{G}} \Tr_\pi(g) \phi(g) dg.
\end{equation}
In fact, we will see in explicit examples that the trace is usually expressed in terms of Dirac delta functions over $\Lie{G}$.

Quite remarkably, there is a deep connection between the trace of a unitary irrep $\pi = \pi_{\varphi, \lie{m}}$ and the structure of its corresponding coadjoint orbit. More explicitly, let $\{U_1, \cdots, U_r, X_1, \cdots, X_{2k}\}$ be a weak Malcev basis passing through $\lie{r}_\varphi$ (where $k = n-m$ as seen previously), and let $\gamma: \R^{2k} \to \Stab_\varphi \backslash \Lie{G}$ be the corresponding Malcev coordinate map. %
Then the trace of $ \pi_\phi$ can be written as
\begin{equation} \label{eq:traceintegral}
    \Tr(\pi_\phi) = \frac{1}{(2\pi)^k} \abs{\det B}^{1/2} \int_{\R^{2k}} \hat{\phi}(\varphi \cdot \gamma(s)) d^{2k}s.
\end{equation}
Here $B$ is the matrix with entries $(B)_{ij} = B_\varphi(X_i, X_j)$, and $\hat{\phi}$ denotes the \textit{Euclidean Fourier transform} of $\phi$,
\begin{equation}
    \hat{\phi}(\varphi) = \int_\lie{g} e^{i \varphi(X)} \phi(e^X) dX,
\end{equation}
with $dX$ any Euclidean measure on $\lie{g}$.\footnote{Although a different choice of Euclidean measure on $\lie{g}$ does change the resulting Fourier transform by an overall factor, the latter cancels in $\Tr(\pi_\phi)$ since the associated Haar measure on $\Lie{G}$ changes accordingly.} Note that $2k$ here is exactly the dimension of the coadjoint orbit $\orb{\varphi}$ --- indeed, we can rewrite Eq.\ (\ref{eq:traceintegral}) as an integral over the coadjoint orbit as follows. The map $F: \R^{2k} \to \orb{\varphi}$ given by $F(s) = \varphi \cdot \gamma(s)$ can be used to push forward the Euclidean measure $(2\pi)^{-k}\abs{\det B}^{1/2} d^{2k}s$ to an $\Ad^*\Lie{G}$-invariant measure on $\orb{\varphi}$, which we denote by $d\varphi'$. By definition of the pushforward, it is easy to see that the trace can be rewritten as
\begin{equation}
    \Tr(\pi_\phi) = \int_{\orb{\varphi}} \hat{\phi}(\varphi') d\varphi'.
\end{equation}

\subsection{The Lazard correspondence and the finite orbit method} \label{subsec:lazardfiniteorbits}

Quite remarkably, it turns out that under certain assumptions the Kirillov orbit method can be generalized to the world of finite groups. That is, given a $p$-group $G$ of nilpotency class $c < p$ (remember that $p$-groups are always nilpotent), there is an appropriate notion of a ``Lie algebra'', commonly denoted by $\log(G)$, and the coadjoint orbits of its Pontryagin dual are in 1-to-1 correspondence with the irreducible representations of $G$. The construction of the $\log$ map (and its inverse, the $\exp$ map) leads to a mathematical result known as the \textit{Lazard correspondence}, as we now explain.

Let $p$ be a fixed prime. The Lazard correspondence gives a 1-to-1 map between $p$-groups of nilpotency class $c < p$, and Lie rings of nilpotency class $c < p$ whose additive group is a $p$-group.\footnote{A Lie ring is an abelian group equipped with a bracket operation that is $\Z$-bilinear, alternating and satisfies the Jacobi identity.} The map $\exp$ taking the latter to the former is constructed by means of the Baker-Campbell-Hausdorff (BCH) formula, as follows. If $L$ is a Lie ring as above, we define a group structure on its underlying set via the BCH formula:
\begin{equation}
    \exp(X) \cdot \exp(Y) \equiv \exp(X + Y + \frac{1}{2}[X,Y] + \frac{1}{12} [X, [X,Y]] - \frac{1}{12} [Y,[X,Y]] + \dots),
\end{equation}
for any $X, Y \in L$. The reason why this leads to a well-defined product is the following: if $c < p$, the BCH formula ends after finitely many steps, namely once commutators of weight $c$ are reached. Furthermore, one can check that the coefficients accompanying the commutators are all rational numbers which never contain $p$ in their denominator, hence they give well-defined operations on the additive $p$-group of $L$. The fact that this product yields a $p$-group of nilpotency class $c$ is also a direct consequence of the BCH formula. The resulting group is commonly denoted as $\exp(L)$.

Conversely, one can use the inverse BCH formula (IBCH) to obtain the map $\log$ going in the reverse direction. This formula arises when expressing the exponential $e^{A+B}$ solely in terms of $e^A$ and $e^B$. More explicitly,
\begin{equation}
    e^{A+B} = e^Ae^B [e^A, e^B]^{-\frac{1}{2}}[e^A,[e^A,e^B]]^{\frac{5}{12}}[e^B,[e^A,e^B]]^{\frac{7}{12}} \cdots,
\end{equation}
where here $[g, h] = ghg^{-1}h^{-1}$ denotes the group commutator in $G$, and the fractional powers in this formula are to be understood formally as e.g.\
\begin{equation}
    [e^A, e^B]^{-\frac{1}{2}} = \exp(-\frac{1}{2}\log([e^A, e^B])).
\end{equation}
Note of course that the IBCH formula contains an infinite product of group elements. In a similar way, we can also express the exponential $e^{[A,B]}$ in terms of $e^A$ and $e^B$, in which case one finds
\begin{equation} \label{eq:commIBCH}
    e^{[A,B]} = [e^A,e^B][e^A,[e^A,e^B]]^{-\frac{1}{2}}[e^B,[e^A,e^B]]^{-\frac{1}{2}} \cdots.
\end{equation}

Following the above, given a $p$-group $G$ of nilpotency class $c < p$ we may define a Lie ring structure on its underlying set of elements --- the resulting Lie ring is commonly denoted by $\log(G)$. The abelian group structure is given by the IBCH formula:
\begin{equation}
    \log(g)+\log(h) \equiv gh[g,h]^{-\frac{1}{2}}[g,[g,h]]^{\frac{5}{12}}[h,[g,h]]^{\frac{7}{12}} \cdots,
\end{equation}
and the Lie bracket is defined via Eq.\ (\ref{eq:commIBCH}), namely
\begin{equation}
    [\log(g), \log(h)] \equiv [g,h][g,[g,h]]^{-\frac{1}{2}}[h,[g,h]]^{-\frac{1}{2}} \cdots,
\end{equation}
for any $g,h \in G$. Once again, these operations are well-defined as a consequence of the strategic choice of nilpotency class. Indeed, one can check that the ``roots'' of group elements appearing in the above formulae are well-defined and unique as they are $n$-th roots with $p \nmid n$. That the underlying additive group is a $p$-group and that the nilpotency class of $\log(G)$ is $c$ follows easily from the definitions above.

We would like to mention that the Lazard map satisfies the following important property: conjugation in $G$ corresponds to the adjoint action on $\log(G)$, that is
\begin{equation} \label{eq:lazardconjadjintertwine}
    g e^X g^{-1} = e^{\Ad_g X}.
\end{equation}
This implies that $e^X$ and $e^Y$ commute if and only if $[X,Y] = 0$. In particular, the centralizer of an element $e^X$ in $G$ can be studied in the Lie ring by considering
\begin{equation} \label{eq:lazardcentralizer}
    \log(C_G(e^X)) = \{ Y \in \log(G) \mid [X,Y] = 0\}.
\end{equation}

As mentioned at the start of this section, the Lazard correspondence has a remarkable application in the representation theory of $p$-groups of nilpotency class $c < p$: a generalization of the Kirillov theory described in Section \ref{subsec:Kirillovthy} for finite groups \cite{Kirillov2004, Kazhdan1977}. We refer the interested reader to \cite{BoyarchenkoSabitova2008} for details --- the result is a finite analog of Theorem \ref{thm:Kirillov} and Eq.\ (\ref{eq:traceintegral}), whose statement is as follows.
\begin{theorem} \label{thm:KirillovFinite}
    If $G$ is a $p$-group of nilpotency class $c<p$, there is a 1-to-1 correspondence between the irreducible representations of $G$ and the set $\widehat{\lie{g}}/\Ad^*(G)$ of coadjoint orbits in the Pontryagin dual $\widehat{\lie{g}} \equiv \Hom_{\Z}(\lie{g}, U(1))$ of its Lazard Lie ring $\lie{g} = \log(G)$. Moreover, the character $\pi_\varphi$ of the irreducible representation associated to the coadjoint orbit $\mathcal{O}_\varphi$ of $\varphi \in \widehat{\lie{g}}$ is given by
    \begin{equation} \label{eq:finiteorbitchar}
        \pi_\varphi(e^{X}) = \frac{1}{\abs{\orb{\varphi}}^{1/2}} \sum_{f \in \orb{\varphi}} f(X), \qquad \forall X \in \lie{g}.
    \end{equation}
\end{theorem}
It is worth noting that the construction of the irrep $\pi_\varphi$ is completely analogous to that of the continuous case. That is, given an element $\varphi$ of $\widehat{\lie{g}}$ we define the bicharacter $B_\varphi : \lie{g} \times \lie{g} \to U(1)$ given by $B_\varphi(X,Y) = \varphi([X,Y])$. A Lie subring $\lie{m} \subseteq \lie{g}$ is then a \textit{polarizing subring} for $\varphi$ if it is isotropic, i.e.\ $B_\varphi$ is trivial on $\lie{m}$, and if it is maximal with respect to size among all subrings with this property. Denoting $M = \exp \lie{m}$, the induced representation
\begin{equation}
    \pi_{\varphi, \lie{m}} = \Ind_{M}^G (\chi_{\varphi, \lie{m}}),
\end{equation}
with $\chi_{\varphi,\lie{m}}$ the abelian character sending $e^Y$ to $\varphi(Y)$ (for $Y \in \lie{m}$), is then irreducible and independent of $\lie{m}$ up to equivalence.

As a remark, in the particular case that the $p$-group $G$ has exponent exactly $p$ (meaning all of its non-trivial elements have order $p$) we can upgrade $\log(G)$ to a Lie algebra over $\F_p$. In that case, we can identify the Pontryagin dual $\widehat{\lie{g}}$ as the dual space $\Hom_{\F_p}(\lie{g}, \F_p)$ via a (non-zero) character $\psi$ of $\F_p$ so that
\begin{equation}
    \begin{aligned}
        \Hom_{\F_p}(\lie{g}, \F_p) & \overset{\simeq}{\longto} \widehat{\lie{g}} \\
        \varphi & \longmapsto \psi \circ \varphi,
    \end{aligned}
\end{equation}
and polarizing subrings become polarizing subalgebras over $\F_p$ as well.

\subsection{Examples}

\subsubsection{The finite group \texorpdfstring{$\Heis(\F_p)$}{Heis(F\_p)}} \label{subsec:HeisFp}

The Heisenberg group over $\F_p$ is the matrix group consisting of all $3\times 3$ matrices of the form
\begin{equation}
    \mqty(1 & a & c \\
    0 & 1 & b \\
    0 & 0 & 1 \\) \equiv [a,b,c],
\end{equation}
with $a, b, c \in \F_p$. This is a finite $p$-group of order $p^3$, and is easily checked to be 2-step nilpotent. It will be useful to consider the group presentation
\begin{equation} \label{eq:heisFppresent}
    \Heis(\F_p) = \langle a,b,c \mid a^p = b^p = c^p = 1, [a,b] = c, [a,c] = [b,c] = 1 \rangle,
\end{equation}
which makes its description as a central extension of $\Z_p \times \Z_p$ by $\Z_p$ manifest.

Let $p > 2$. The conjugacy class structure of $\Heis(\F_p)$ is summarized in Table \ref{tab:HeisFpconj}. The centralizer $\Z_p \times \Z_p$ of the non-central conjugacy class $[a,b,*]$ is, more explicitly, the subgroup
\begin{equation}
    C_{\Heis(\F_p)}([a,b,*]) = \qty{[x,y,z] \in \Heis(\F_p) \mid ay-bx = 0} \cong \langle [a,b,*] \rangle \times \langle [0,0,1] \rangle.
\end{equation}
We will make a convenient choice of representatives for these conjugacy classes by considering all $[a,b,ab/2]$ with $(a,b) \neq (0,0)$ --- this will be useful in our calculations in Section \ref{sec:6corresp}.
\begin{table}[h!]
    \centering
    \begin{tabular}{Sc|Sc|Sc}
         Representatives & \makecell{$[0,0,c]$ \\ \footnotesize{$(c \in \F_p)$}}  & \makecell{$[a,b,*]$ \\ \footnotesize{$((a,b) \in \F_p^2 \setminus \{(0,0)\})$}} \\
         \hline
         No.\ of classes & $p$ & $p^2-1$  \\
         \hline
         Cardinality & $1$ & $p$ \\
         \hline
         Centralizer & $\Heis(\F_p)$ & $\Z_p\times \Z_p$
    \end{tabular}
    \caption{Conjugacy classes of $\Heis(\F_p)$, $p > 2$.}
    \label{tab:HeisFpconj}
\end{table}

Since there are $p^2+p-1$ conjugacy classes, there is an equal number of irreducible representations: the values of their associated characters on conjugacy classes are shown in Table \ref{tab:HeisFpchar}.
\begin{table}[h!]
    \centering
    \begin{tabular}{Sc|Sc|Sc}
        & \makecell{$[0,0,c]$ \\ \footnotesize{$(c \in \F_p)$}}  & \makecell{$[a,b,*]$ \\ \footnotesize{$((a,b) \in \F_p^2 \setminus \{(0,0)\})$}} \\
        \hline
        \makecell{$\psi_{x,y}$ \\ \footnotesize{$((x,y) \in \F_p^2)$}} & 1 & $e^{2\pi i \frac{ax+by}{p}}$ \\
        \hline
        \makecell{$\chi_z$ \\ \footnotesize{$(z \in \F_p^\times)$}} & $pe^{2\pi i \frac{cz}{p}}$ & 0
    \end{tabular}
    \caption{Character table of $\Heis(\F_p)$, $p > 2$.}
    \label{tab:HeisFpchar}
\end{table}
It is instructive to understand (and will also be important later on) how these representations arise from the finite Kirillov method, which is possible since $\Heis(\F_p)$ is a $p$-group with nilpotency class $2 < p$. Its Lazard Lie ring, which is moreover an $\F_p$-Lie algebra as $\Heis(\F_p)$ has exponent $p$, is the mod $p$ Heisenberg Lie algebra $\lie{h}_3$ with basis $\qty{X_1, X_2, X_3}$ and commutation relations
\begin{equation}
    [X_1,X_2] = X_3.
\end{equation}
The coadjoint action of an element $\exp(a \cdot X)$ on a functional $\varphi = \alpha \cdot X^*$ is
\begin{equation}
    \alpha \longmapsto \alpha + a \times (0,0, \alpha_3).
\end{equation}
Orbits are then of 2 types: point orbits with representatives $\varphi_{\alpha_1, \alpha_2} = \alpha_1 X_1^* + \alpha_2 X_2^*$, and orbits of size $p^2$ with representatives $\varphi_{\alpha_3} = \alpha_3 X_3^*$, with $\alpha_3 \neq 0$. As one can verify by means of the character formula (\ref{eq:finiteorbitchar}), the associated irreducible representations are, in the notation of Table \ref{tab:HeisFpchar},
\begin{equation}
    \psi_{x,y} = \pi_{\varphi_{x,y}}, \qquad \chi_z = \pi_{\varphi_z}.
\end{equation}

\subsubsection{The finite group \texorpdfstring{$\Fil_4(\F_p)$}{Fil\_4(F\_p)}} \label{subsec:F7}

Among the 10 non-abelian $p$-groups of order $p^4$ (with $p > 3$), there is only one with maximal nilpotency class $c = 3$ and exponent $p$ \cite{burnside2012theory}, which we denote by $\Fil_4(\F_p)$. It has the group presentation
\begin{equation}
    \Fil_4(\F_p) = \langle a,b,c,d \mid a^p = b^p = c^p = d^p = 1, [a,b]=c, [a,c]=d, [b,c] = [a,d] = [b,d] = 1 \rangle,
\end{equation}
from which it can be understood as a central extension of $\Heis(\F_p)$ by $\Z_p$, with $d$ being the additional generator twisting the presentation in (\ref{eq:heisFppresent}). This group has no standard name in the literature: the notation we use is instead motivated by the Lazard correspondence. Indeed, its Lazard Lie ring, which is again a Lie algebra over $\F_p$, is the mod $p$ \textit{filiform Lie algebra} $\lie{f}_4$ of dimension $4$, with basis $\qty{X_1, X_2, X_3, X_4}$ and commutation relations
\begin{equation}
    [X_1,X_2] = X_3, \qquad [X_1,X_3] = X_4.
\end{equation}
To be precise, we have performed a slight change of basis after application of the Lazard map for convenience. The logarithm actually takes
\begin{equation}
    \log(a) = X_1, \quad \log(b) = X_2, \quad \log(c) = X_3 + \frac{1}{2}X_4, \quad \log(d) = X_4.
\end{equation}

Let us study the conjugacy classes of this group. Finding them explicitly is easy to do using the Lie algebra coordinates
\begin{equation}
    [a_1,a_2,a_3,a_4] \equiv e^{a \cdot X}.
\end{equation}
In the Lie algebra, finding conjugacy classes translates to finding the adjoint orbits of $\lie{f}_4$. We summarize the end results in Table \ref{tab:F7conjclasses}.
\begin{table}[h!]
    \centering
    \small
    \begin{tabular}{Sc|Sc|Sc|Sc|Sc}
         Representatives & \makecell{$[0,0,0,d]$ \\ \footnotesize{($d \in \F_p$)}} & \makecell{$[0,0,c,0]$ \\ \footnotesize{($c \in \F_p^\times$)}}  & \makecell{$[0,b,0,d]$ \\ \footnotesize{($b \in \F_p^\times, d \in \F_p$)}}& \makecell{$[a,b,0,0]$ \\ \footnotesize{($a \in \F_p^\times, b \in \F_p$)}}
         \\
         \hline
         No.\ of classes & $p$ & $p-1$ & $p(p-1)$ & $p(p-1)$\\
         \hline
         Cardinality & $1$ & $p$ & $p$ & $p^2$\\
         \hline
         Centralizer & $\Fil_4(\F_p)$ & $\Z_p^3$ & $\Z_p^3$ & $\Z_p^2$
    \end{tabular}
    \caption{Conjugacy classes of $\Fil_4(\F_p)$.}
    \label{tab:F7conjclasses}
\end{table}
Their centralizers are given more explicitly as follows. $\Z_p^3$ in both cases is the subgroup
\begin{equation} \label{eq:FilZp3centralizer}
    C_{\Fil_4(\F_p)}([0,0,c,0]) = C_{\Fil_4(\F_p)}([0,b,0,d]) = \langle e^{X_2} \rangle \times  \langle e^{X_3} \rangle \times \langle e^{X_4} \rangle.
\end{equation}
Meanwhile, $\Z_p^2$ in the last entry is the subgroup
\begin{equation} \label{eq:FilZp2centralizer}
    C_{\Fil_4(\F_p)}([a,b,0,0]) = \langle e^{aX_1+bX_2} \rangle \times \langle e^{X_4} \rangle.
\end{equation}
We note that there is a total of $2p^2-1$ conjugacy classes.

We now proceed to find the irreducible representations of $\Fil_4(\F_p)$, again by means of the finite orbit method. The coadjoint action of $e^{a\cdot X}$ on $\varphi = \alpha \cdot X^*$ reads
\begin{equation}
    \begin{aligned}
        \alpha_1 & \longmapsto \alpha_1 + a_2\alpha_3 + \qty(a_3 -\frac{a_1a_2}{2}) \alpha_4, \\
        \alpha_2 & \longmapsto \alpha_2 - a_1\alpha_3 + \frac{1}{2}a_1^2 \alpha_4, \\
        \alpha_3 & \longmapsto \alpha_3 - a_1\alpha_4, \\
        \alpha_4 & \longmapsto \alpha_4.
    \end{aligned}
\end{equation}
Without going into details, one may find that there are 3 kinds of orbits, with representatives
\begin{equation}
    \begin{aligned}
        \varphi^{(1)}_{\alpha_1,\alpha_2} & = \alpha_1 X_1^* + \alpha_2 X_2^*, & \\
        \varphi^{(2)}_{\alpha_3} &= \alpha_3 X_3^*, & \alpha_3 \neq 0, \\
        \varphi^{(3)}_{\alpha_2, \alpha_4} &= \alpha_2 X_2^* + \alpha_4 X_4^*, & \alpha_4 \neq 0.
    \end{aligned}
\end{equation}
As expected, the total number of representatives equals $2p^2-1$. The values of their associated Kirillov characters are shown in Table \ref{tab:F7chartable}, where we introduced the notation $\zeta = e^{2\pi i/p}$.
\begin{table}[h!]
    \centering
    \begin{tabular}{Sc|Sc|Sc|Sc|Sc}
        & \makecell{$[0,0,0,d]$ \\ \footnotesize{($d \in \F_p$)}} & \makecell{$[0,0,c,0]$ \\ \footnotesize{($c \in \F_p^\times$)}}  & \makecell{$[0,b,0,d]$ \\ \footnotesize{($b \in \F_p^\times, d \in \F_p$)}}& \makecell{$[a,b,0,0]$ \\ \footnotesize{($a \in \F_p^\times, b \in \F_p$)}} \\
        \hline
        \makecell{$\chi^{(1)}_{x,y}$ \\ \footnotesize{($(x,y)\in \F_p^2$)}} & 1 & 1 & $\zeta^{by}$ & $\zeta^{ax+by}$ \\
        \hline
        \makecell{$\chi^{(2)}_z$ \\ \footnotesize{($z \in \F_p^\times$)}} & $p$ & $p\zeta^{cz}$ & 0 & 0 \\
        \hline
        \makecell{$\chi^{(3)}_{y,w}$ \\ \footnotesize{($y \in \F_p, w \in \F_p^\times$)}} & $p \zeta^{dw}$ & 0 & $\zeta^{by+dw} \sum_{s\in\F_p} \zeta^{\frac{1}{2}bws^2}$ & 0
    \end{tabular}
    \caption{Character table of $\Fil_4(\F_p)$, for $p > 3$.}
    \label{tab:F7chartable}
\end{table}

\subsubsection{The Lie algebra \texorpdfstring{$T^*\lie{h}_3$}{T*h\_3}} \label{subsec:T*h3}

The real Heisenberg Lie algebra $\lie{h}_3$ is a 3-dimensional algebra with basis $\{X_1,X_2,X_3\}$ and commutation relations
\begin{equation}
    [X_1,X_2] = X_3.
\end{equation}
Note that the change of basis $X_i \to \lambda^{-1} X_i$ rescales all structure constants by $\lambda$. We will keep $\lambda$ general throughout our computations, as it will be useful for our later purposes. The corresponding cotangent Lie algebra $T^*\lie{h}_3$ (cf.\ Section \ref{sec:6corresp}) is a 6-dimensional 2-step nilpotent Lie algebra. If we denote the basis dual to $X_i$ by $Y_i$, the commutation relations read
\begin{equation} \label{eq:T*h3commrelOG}
    [X_1,X_2]= \lambda X_3, \qquad [Y_3,X_1]= \lambda Y_2, \qquad [Y_3,X_2] = -\lambda Y_1.
\end{equation}
A slightly different but particularly convenient basis is given by $\{E_i, F_i\}$ such that
\begin{equation} \label{eq:T^*h3EFbasis}
    E_{1,2} = X_{1,2}, \qquad F_{1,2} = Y_{1,2}, \qquad E_3 = Y_3, \qquad F_3 = X_3.
\end{equation}
In this basis, the commutation relations (\ref{eq:T*h3commrelOG}) can be written succinctly as
\begin{equation}
    [E_i, E_j] = \lambda \epsilon_{ijk} F_k.
\end{equation}

We now delve into the representation theory of the simply connected Lie group $\Lie{G} = \exp(T^*\lie{h}_3)$, via the Kirillov method. The coadjoint action of an element $\exp(a \cdot E + b \cdot F)$ on a linear functional $\varphi = \alpha \cdot E^* + \beta \cdot F^*$ reads
\begin{equation}
    \alpha \longmapsto \alpha + \lambda a \times \beta, \qquad \beta \longmapsto \beta.
\end{equation}
There are thus two kinds of coadjoint orbits: point orbits whose representatives are of the form $\varphi = \alpha \cdot E^*$, and 2-dimensional orbits with representatives $\varphi = \kappa \beta \cdot E^* + \beta\cdot F^*$, for $\beta \in \R^3 \setminus \{0\}$ and $\kappa \in \R$. We denote the associated irreducible representations of $\Lie{G}$ collectively by $\pi_{(\alpha,\beta)}$, where we understand that $\pi_{(\alpha,0)}$ corresponds to the 0-dimensional orbits, while $\pi_{(\alpha,\beta \neq 0)}$ corresponds to the 2-dimensional ones, keeping in mind the identifications
\begin{equation}
    \pi_{(\alpha,\beta)} \simeq \pi_{(\alpha+ \lambda a \times \beta, \beta)}, \qquad \forall a \in \R^3.
\end{equation}

The characters of the above irreps can be computed following the general procedure discussed in Section \ref{subsec:Kirillovthy}. The character of $\pi_{(\alpha,0)}$ reads
\begin{equation}
    \pi_{(\alpha,0)}(e^{a \cdot E + b \cdot F}) = e^{i (\alpha \cdot a)}.
\end{equation}
These representations are exactly the ones that are trivial on the center, and correspond to the one-dimensional (abelian) irreps of $\Lie{G}$. The irreps $\pi_{(\alpha,\beta \neq 0)}$ are instead infinite-dimensional, and their character distribution reads
\begin{equation}
    \pi_{(\alpha,\beta)}(e^{a\cdot E + b \cdot F}) = \frac{2\pi}{\lambda|\beta|} \delta^{(2)}(a_\perp) e^{i (\alpha\cdot a + \beta \cdot b)},
\end{equation}
with $a_\perp$ the component of $a$ perpendicular to $\beta$. For reference, this delta function can be given a working definition by choosing a direction $\beta_i \neq 0$, so that
\begin{equation}
    \frac{1}{|\beta|}\delta^{(2)}(a_\perp) = |\beta_i| \delta\qty((\beta \times a)_j) \delta\qty((\beta \times a)_k),
\end{equation}
where $j, k$ are the other 2 directions $\neq i$. With this in mind, we will often use the notation $\delta(\beta \times a)$ as an alternative to $\delta^{(2)}(a_\perp)$.

\subsubsection{The Lie algebra \texorpdfstring{$T^*\lie{f}_4$}{T*f\_4}} \label{subsec:T*f4}

The real filiform Lie algebra $\lie{f}_4$ of dimension 4 with basis $\{X_1,X_2,X_3,X_4\}$ is defined by the commutation relations
\begin{equation}
    [X_1,X_2] = X_3, \qquad [X_1,X_3] = X_4,
\end{equation}
which is easily verified to be 3-step nilpotent. Again, the change of basis $X_i \to \lambda^{-1} X_i$ rescales all structure constants by $\lambda$, and we will put this factor explicitly throughout calculations. The cotangent Lie algebra $T^*\lie{f}_4$ is an 8-dimensional Lie algebra, and denoting the dual basis by $Y_i$, its commutation relations read
\begin{equation}
    \begin{aligned}
        &[X_1,X_2]=\lambda X_3, \qquad & [X_1,X_3]= \lambda  X_4, \qquad & [X_1,Y_3]=- \lambda Y_2, \\
        &[X_1,Y_4]=-\lambda  Y_3, \qquad & [X_2,Y_3]=\lambda  Y_1, \qquad & [X_3,Y_4]= \lambda Y_1.
    \end{aligned}
\end{equation}

The representation theory of $\Lie{G} = \exp(T^*\lie{f}_4)$ can be derived through the Kirillov method, although calculations are lengthy. The coadjoint action of $\exp(a \cdot X + b\cdot Y)$ on a linear functional $\varphi = e \cdot X^* + q \cdot Y^*$ is given by
\begin{equation}
    \begin{aligned}
        e_1 & \longmapsto e_1 + \lambda a_2 e_3+\qty(\lambda a_3-\frac{1}{2}\lambda^2a_1 a_2) e_4+\lambda^2 a_2 b_4 q_1+\qty(-\lambda b_3-\frac{1}{2}\lambda^2 a_1 b_4) q_2-\lambda b_4 q_3, \\
        e_2 & \longmapsto e_2 -\lambda a_1e_3 +  \frac{1}{2} \lambda^2 a_1^2 e_4 +\qty(\lambda b_3- \frac{1}{2}\lambda^2 a_1 b_4) q_1, \\
        e_3 & \longmapsto e_3 -\lambda a_1 e_4+\lambda b_4 q_1, \\
        e_4 & \longmapsto e_4, \\
        q_1 & \longmapsto q_1, \\
        q_2 & \longmapsto q_2, \\
        q_3 & \longmapsto q_3 -\lambda a_2 q_1+\lambda a_1 q_2, \\
        q_4 & \longmapsto  q_4 + \qty(-\frac{1}{2}\lambda^2 a_1a_2-\lambda a_3) q_1 + \frac{1}{2}\lambda^2 a_1^2 q_2 +\lambda a_1q_3.
    \end{aligned}
\end{equation}
One can check that there exist essentially 5 types of coadjoint orbits, with representatives
\begin{equation}
    \begin{aligned}
        \varphi^{(1)}_{e_1,e_2,q_4} &= e_1X_1^* + e_2X_2^* + q_4Y_4^*, & \\
        \varphi^{(2)}_{e_3,q_3,\kappa} &= e_3X_3^* + q_3Y_3^* + \kappa q_3 X_2^* + \kappa e_3 Y_4^*, & (e_3,q_3) \neq 0, \\
        \varphi^{(3)}_{e_1,e_4,q_1,q_2} &= e_1X_1^* + e_4X_4^*+q_1Y_1^* + q_2Y_2^*, & q_1 \neq 0, \\
        \varphi^{(4)}_{e_2,q_2,q_4;0,e_3} &= e_2X_2^* + e_3X_3^* + q_2Y_2^* + q_4Y_4^*, & q_2 \neq 0, \\
        \varphi^{(5)}_{e_2,q_2,q_4;e_4,q_3} &= e_2X_2^* + e_4X_4^* + q_2Y_2^* + q_3Y_3^* + q_4Y_4^*, & e_4 \neq 0.
    \end{aligned}
\end{equation}
Note that the type-1 representations correspond to abelian representations, while type-2 representations instead behave similarly to the non-abelian irreps of $\exp(T^*\lie{h}_3)$. The orbits of type-4 and type-5 representations are 2-dimensional, while those of type-3 representations are 4-dimensional.

One can compute the character distributions for these representations using Eq.\ (\ref{eq:traceintegral}), however the orbit integrals cannot be evaluated in a simple way in most cases. For example, the characters for the type-5 representations read
\begin{align} \label{eq:T*f4type5char}
    \chi^{(5)}_{e_2,q_2,q_4; e_4,q_3}(e^{a\cdot X + b \cdot Y}) = \delta(a_1) e^{i (e \cdot a + q \cdot b)} \int_\R ds\ e^{i(Ps^2 + Qs)},
\end{align}
with
\begin{equation}
    P = \frac{\lambda^2}{2} (e_4a_2 + q_2b_4), \qquad Q = \lambda (e_4a_3 - q_2b_3-q_3b_4).
\end{equation}
The integral in Eq.\ (\ref{eq:T*f4type5char}) is a Fresnel integral, which evaluates to a finite result when $P \neq 0$, but instead yields a Dirac delta for $Q$ if $P = 0$. In any case, the expression above suffices for our later purposes.

\section{Brunnian links} \label{sec:3brunnian}

A (non-trivial) $n$-component link $\mathcal{L}$ is \textit{Brunnian} if it becomes trivial after removing any one of its components, meaning the remaining $n-1$ components become all unlinked. In particular, this implies that for Brunnian links with at least 3 components, all pairwise linking numbers vanish.

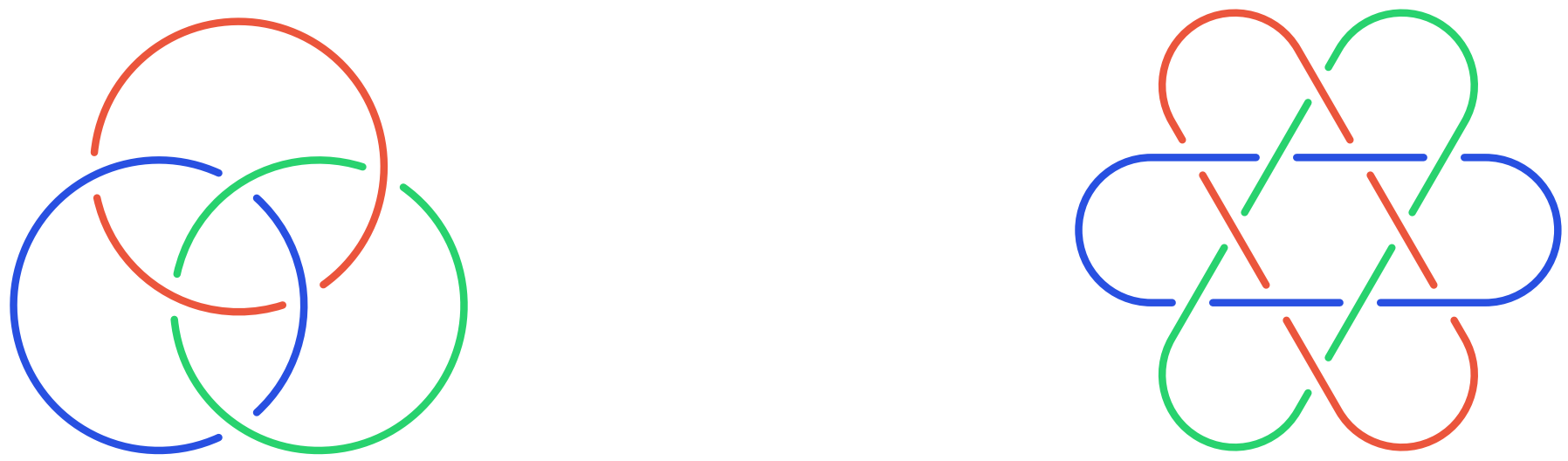
\begin{figure}[h!]
    \centering
    \begin{subfigure}[t]{0.49\textwidth}
        \centering
        \begin{tikzpicture}[line width=1.6pt, line cap=round]
  \draw[borOrange] (90:0.7) ++(192.4:1.1)
    arc[start angle=192.4, end angle=287.6, radius=1.1];
  \draw[borOrange] (90:0.7) ++(305.6:1.1)
    arc[start angle=305.6, end angle=534.4, radius=1.1];
  \draw[borBlue] (210:0.7) ++(65.6:1.1)
    arc[start angle=65.6, end angle=294.4, radius=1.1];
  \draw[borBlue] (210:0.7) ++(312.4:1.1)
    arc[start angle=312.4, end angle=407.6, radius=1.1];
  \draw[borTeal] (330:0.7) ++(72.4:1.1)
    arc[start angle=72.4, end angle=167.6, radius=1.1];
  \draw[borTeal] (330:0.7) ++(185.6:1.1)
    arc[start angle=185.6, end angle=414.4, radius=1.1];
\end{tikzpicture}
        \caption{The Borromean rings.}
        \label{fig:brunn3comp1}
    \end{subfigure}
    \begin{subfigure}[t]{0.49\textwidth}
        \centering
        \begin{tikzpicture}[line width=1.6pt, line cap=round, scale=0.55]

  \newcommand{\gappedstadium}[2]{%
    \begin{scope}[rotate=#2, color=#1]
      \draw (-0.297,1) -- (1.452,1);
      \draw (2.012,1) -- (2.3,1)
        arc[start angle=90, end angle=-90, radius=1] -- (0.857,-1);
      \draw (0.297,-1) -- (-1.452,-1);
      \draw (-2.012,-1) -- (-2.3,-1)
        arc[start angle=270, end angle=90, radius=1] -- (-0.857,1);
    \end{scope}}
  \gappedstadium{borBlue}{0}
  \gappedstadium{borTeal}{60}
  \gappedstadium{borOrange}{120}
\end{tikzpicture}
        \caption{A 3-component Brunnian link, inequivalent to the Borromean rings.}
        \label{fig:brunn3comp2}
    \end{subfigure}
    \caption{Examples of 3-component Brunnian links.}
    \label{fig:brunnian3comp}
\end{figure}

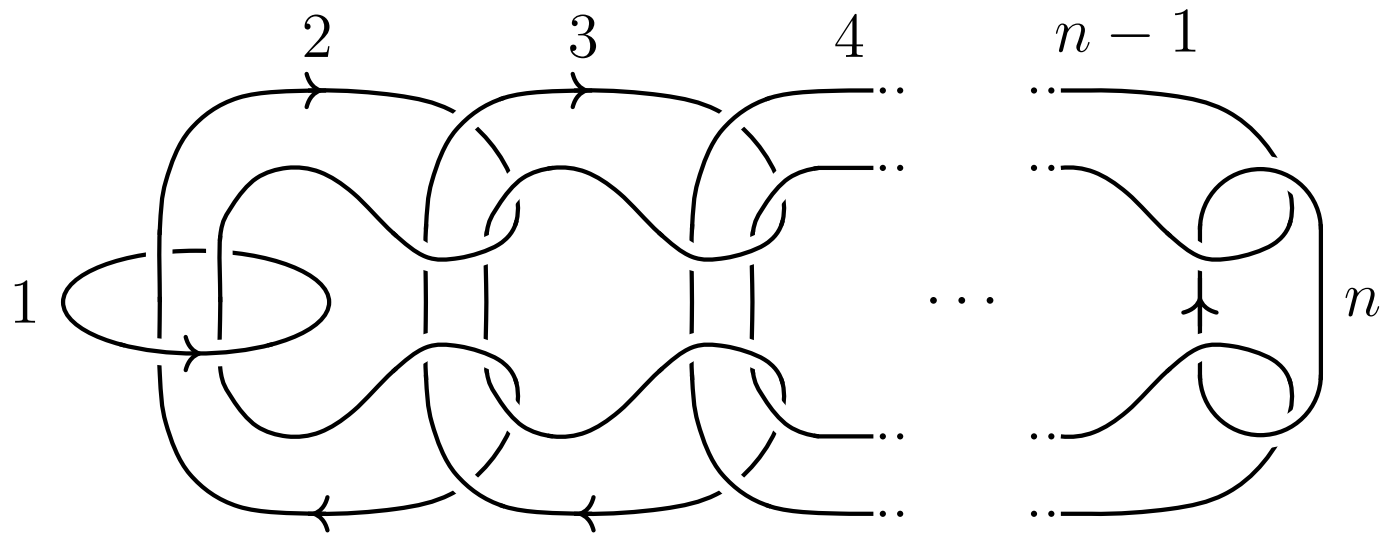
\begin{figure}[h!]
    \centering
    \begin{tikzpicture}[thick,
    baseline={([yshift=-.5ex]current bounding box.center)}, scale=0.4,
    overstrand/.style={preaction={draw, white, line width=5pt,
        shorten <=1pt, shorten >=1pt}},
    midarrow/.style={decoration={markings,
        mark=at position 0.5 with {\arrow[xshift=1pt]{>}}},
      postaction=decorate}]

  \newcommand{\milnorring}[1]{%
    \draw (#1,0)
      .. controls (#1,0.4) and (#1-0.02,0.87) .. (#1,1.2) .. controls (#1+0.02,1.53) and (#1+0.02,1.73) .. (#1+0.1,2)
      .. controls (#1+0.18,2.28) and (#1+0.3,2.62) .. (#1+0.5,2.85) .. controls (#1+0.7,3.08) and (#1+0.95,3.27) .. (#1+1.3,3.38)
      .. controls (#1+1.65,3.49) and (#1+2.17,3.49) .. (#1+2.6,3.5) .. controls (#1+3.03,3.51) and (#1+3.53,3.47) .. (#1+3.9,3.42)
      .. controls (#1+4.27,3.37) and (#1+4.54,3.31) .. (#1+4.8,3.18) .. controls (#1+5.06,3.05) and (#1+5.28,2.86) .. (#1+5.45,2.65)
      .. controls (#1+5.63,2.44) and (#1+5.78,2.17) .. (#1+5.85,1.95) .. controls (#1+5.93,1.73) and (#1+5.94,1.52) .. (#1+5.9,1.35)
      .. controls (#1+5.86,1.18) and (#1+5.76,1.05) .. (#1+5.6,0.95) .. controls (#1+5.44,0.85) and (#1+5.15,0.76) .. (#1+4.95,0.73)
      .. controls (#1+4.75,0.7) and (#1+4.58,0.68) .. (#1+4.4,0.75) .. controls (#1+4.22,0.82) and (#1+4.05,0.97) .. (#1+3.85,1.15)
      .. controls (#1+3.65,1.33) and (#1+3.4,1.63) .. (#1+3.2,1.8) .. controls (#1+3,1.97) and (#1+2.83,2.08) .. (#1+2.65,2.15)
      .. controls (#1+2.47,2.22) and (#1+2.28,2.24) .. (#1+2.1,2.22) .. controls (#1+1.93,2.2) and (#1+1.75,2.15) .. (#1+1.6,2.05)
      .. controls (#1+1.45,1.95) and (#1+1.3,1.76) .. (#1+1.2,1.6) .. controls (#1+1.1,1.44) and (#1+1.03,1.37) .. (#1+1,1.1)
      .. controls (#1+0.97,0.83) and (#1+1,0.37) .. (#1+1,0) .. controls (#1+1,-0.37) and (#1+0.97,-0.83) .. (#1+1,-1.1)
      .. controls (#1+1.03,-1.37) and (#1+1.1,-1.44) .. (#1+1.2,-1.6) .. controls (#1+1.3,-1.76) and (#1+1.45,-1.95) .. (#1+1.6,-2.05)
      .. controls (#1+1.75,-2.15) and (#1+1.93,-2.2) .. (#1+2.1,-2.22) .. controls (#1+2.28,-2.24) and (#1+2.47,-2.22) .. (#1+2.65,-2.15)
      .. controls (#1+2.83,-2.08) and (#1+3,-1.97) .. (#1+3.2,-1.8) .. controls (#1+3.4,-1.63) and (#1+3.65,-1.33) .. (#1+3.85,-1.15)
      .. controls (#1+4.05,-0.97) and (#1+4.22,-0.82) .. (#1+4.4,-0.75) .. controls (#1+4.58,-0.68) and (#1+4.75,-0.7) .. (#1+4.95,-0.73)
      .. controls (#1+5.15,-0.76) and (#1+5.44,-0.85) .. (#1+5.6,-0.95) .. controls (#1+5.76,-1.05) and (#1+5.86,-1.18) .. (#1+5.9,-1.35)
      .. controls (#1+5.94,-1.52) and (#1+5.93,-1.73) .. (#1+5.85,-1.95) .. controls (#1+5.78,-2.17) and (#1+5.63,-2.44) .. (#1+5.45,-2.65)
      .. controls (#1+5.28,-2.86) and (#1+5.06,-3.05) .. (#1+4.8,-3.18) .. controls (#1+4.54,-3.31) and (#1+4.27,-3.37) .. (#1+3.9,-3.42)
      .. controls (#1+3.53,-3.47) and (#1+3.03,-3.51) .. (#1+2.6,-3.5) .. controls (#1+2.17,-3.49) and (#1+1.65,-3.49) .. (#1+1.3,-3.38)
      .. controls (#1+0.95,-3.27) and (#1+0.7,-3.08) .. (#1+0.5,-2.85) .. controls (#1+0.3,-2.62) and (#1+0.18,-2.28) .. (#1+0.1,-2)
      .. controls (#1+0.02,-1.72) and (#1+0.02,-1.53) .. (#1,-1.2) .. controls (#1-0.02,-0.87) and (#1,-0.4) .. (#1,0) -- cycle;}

  \newcommand{\milnorhooks}[1]{%
    \draw[overstrand] (#1,1.2)
      .. controls (#1+0.02,1.53) and (#1+0.02,1.73) .. (#1+0.1,2) .. controls (#1+0.18,2.28) and (#1+0.3,2.62) .. (#1+0.5,2.85)
      .. controls (#1+0.7,3.08) and (#1+0.95,3.27) .. (#1+1.3,3.38);
    \draw[overstrand] (#1+1.3,-3.38)
      .. controls (#1+0.95,-3.27) and (#1+0.7,-3.08) .. (#1+0.5,-2.85) .. controls (#1+0.3,-2.62) and (#1+0.18,-2.28) .. (#1+0.1,-2)
      .. controls (#1+0.02,-1.72) and (#1+0.02,-1.53) .. (#1,-1.2);
    \draw[overstrand] (#1+2.1,2.22)
      .. controls (#1+1.93,2.2) and (#1+1.75,2.15) .. (#1+1.6,2.05) .. controls (#1+1.45,1.95) and (#1+1.3,1.76) .. (#1+1.2,1.6)
      .. controls (#1+1.1,1.44) and (#1+1.03,1.37) .. (#1+1,1.1);
    \draw[overstrand] (#1+1,-1.1)
      .. controls (#1+1.03,-1.37) and (#1+1.1,-1.44) .. (#1+1.2,-1.6) .. controls (#1+1.3,-1.76) and (#1+1.45,-1.95) .. (#1+1.6,-2.05)
      .. controls (#1+1.75,-2.15) and (#1+1.93,-2.2) .. (#1+2.1,-2.22);
    \draw[overstrand] (#1+1.5,1.35)
      .. controls (#1+1.46,1.18) and (#1+1.36,1.05) .. (#1+1.2,0.95) .. controls (#1+1.04,0.85) and (#1+0.75,0.76) .. (#1+0.55,0.73)
      .. controls (#1+0.35,0.7) and (#1+0.18,0.68) .. (#1,0.75) .. controls (#1-0.18,0.82) and (#1-0.35,0.97) .. (#1-0.55,1.15);
    \draw[overstrand] (#1-0.55,-1.15)
      .. controls (#1-0.35,-0.97) and (#1-0.18,-0.82) .. (#1,-0.75) .. controls (#1+0.18,-0.68) and (#1+0.35,-0.7) .. (#1+0.55,-0.73)
      .. controls (#1+0.75,-0.76) and (#1+1.04,-0.85) .. (#1+1.2,-0.95) .. controls (#1+1.36,-1.05) and (#1+1.46,-1.18) .. (#1+1.5,-1.35);}

  \draw (0.6,0) ellipse [x radius=2.2, y radius=0.85];
  \milnorring{0}
  \milnorring{4.4}
  \draw (11.8,-3.5) -- (11.4,-3.5)
    .. controls (10.97,-3.49) and (10.45,-3.49) .. (10.1,-3.38) .. controls (9.75,-3.27) and (9.5,-3.08) .. (9.3,-2.85)
    .. controls (9.1,-2.62) and (8.98,-2.28) .. (8.9,-2) .. controls (8.82,-1.72) and (8.82,-1.53) .. (8.8,-1.2)
    .. controls (8.78,-0.87) and (8.8,-0.4) .. (8.8,0) .. controls (8.8,0.4) and (8.78,0.87) .. (8.8,1.2)
    .. controls (8.82,1.53) and (8.82,1.73) .. (8.9,2) .. controls (8.98,2.28) and (9.1,2.62) .. (9.3,2.85)
    .. controls (9.5,3.08) and (9.75,3.27) .. (10.1,3.38) .. controls (10.45,3.49) and (10.97,3.49) .. (11.4,3.5) -- (11.8,3.5);
  \draw (11.8,2.22) -- (10.9,2.22)
    .. controls (10.73,2.2) and (10.55,2.15) .. (10.4,2.05) .. controls (10.25,1.95) and (10.1,1.76) .. (10,1.6)
    .. controls (9.9,1.44) and (9.83,1.37) .. (9.8,1.1) .. controls (9.77,0.83) and (9.8,0.37) .. (9.8,0)
    .. controls (9.8,-0.37) and (9.77,-0.83) .. (9.8,-1.1) .. controls (9.83,-1.37) and (9.9,-1.44) .. (10,-1.6)
    .. controls (10.1,-1.76) and (10.25,-1.95) .. (10.4,-2.05) .. controls (10.55,-2.15) and (10.73,-2.2) .. (10.9,-2.22) -- (11.8,-2.22);
  \draw (14.9,3.5) -- (15.4,3.5)
    .. controls (15.83,3.51) and (16.33,3.47) .. (16.7,3.42) .. controls (17.07,3.37) and (17.34,3.31) .. (17.6,3.18)
    .. controls (17.86,3.05) and (18.08,2.86) .. (18.25,2.65) .. controls (18.43,2.44) and (18.58,2.17) .. (18.65,1.95)
    .. controls (18.73,1.73) and (18.74,1.52) .. (18.7,1.35) .. controls (18.66,1.18) and (18.56,1.05) .. (18.4,0.95)
    .. controls (18.24,0.85) and (17.95,0.76) .. (17.75,0.73) .. controls (17.55,0.7) and (17.38,0.68) .. (17.2,0.75)
    .. controls (17.02,0.82) and (16.85,0.97) .. (16.65,1.15) .. controls (16.45,1.33) and (16.2,1.63) .. (16,1.8)
    .. controls (15.8,1.97) and (15.63,2.08) .. (15.45,2.15) .. controls (15.27,2.22) and (15.08,2.24) .. (14.9,2.22);
  \draw (14.9,-2.22)
    .. controls (15.08,-2.24) and (15.27,-2.22) .. (15.45,-2.15) .. controls (15.63,-2.08) and (15.8,-1.97) .. (16,-1.8)
    .. controls (16.2,-1.63) and (16.45,-1.33) .. (16.65,-1.15) .. controls (16.85,-0.97) and (17.02,-0.82) .. (17.2,-0.75)
    .. controls (17.38,-0.68) and (17.55,-0.7) .. (17.75,-0.73) .. controls (17.95,-0.76) and (18.24,-0.85) .. (18.4,-0.95)
    .. controls (18.56,-1.05) and (18.66,-1.18) .. (18.7,-1.35) .. controls (18.74,-1.52) and (18.73,-1.73) .. (18.65,-1.95)
    .. controls (18.58,-2.17) and (18.43,-2.44) .. (18.25,-2.65) .. controls (18.08,-2.86) and (17.86,-3.05) .. (17.6,-3.18)
    .. controls (17.34,-3.31) and (17.07,-3.37) .. (16.7,-3.42) .. controls (16.33,-3.47) and (15.83,-3.51) .. (15.4,-3.5) -- (14.9,-3.5);
  \draw (17.2,-1.2) -- (17.2,1.2) arc[start angle=180, end angle=0, radius=1]
    -- (19.2,-1.2) arc[start angle=0, end angle=-180, radius=1] -- cycle;

  \draw[overstrand] (0,0) .. controls (0,0.4) and (-0.02,0.87) .. (0,1.2);
  \draw[overstrand] (1,1.1) .. controls (0.97,0.83) and (1,0.37) .. (1,0);
  \draw[overstrand, midarrow] (0.6,0) ++(235:2.2 and 0.85)
    arc[start angle=235, end angle=305, x radius=2.2, y radius=0.85];
  \milnorhooks{4.4}
  \milnorhooks{8.8}
  \draw[overstrand] (17.2,1.2) arc[start angle=180, end angle=0, radius=1];
  \draw[overstrand] (17.2,-1.2) arc[start angle=180, end angle=360, radius=1];
  \draw[overstrand] (18.7,1.35)
    .. controls (18.66,1.18) and (18.56,1.05) .. (18.4,0.95) .. controls (18.24,0.85) and (17.95,0.76) .. (17.75,0.73)
    .. controls (17.55,0.7) and (17.38,0.68) .. (17.2,0.75) .. controls (17.02,0.82) and (16.85,0.97) .. (16.65,1.15);
  \draw[overstrand] (16.65,-1.15)
    .. controls (16.85,-0.97) and (17.02,-0.82) .. (17.2,-0.75) .. controls (17.38,-0.68) and (17.55,-0.7) .. (17.75,-0.73)
    .. controls (17.95,-0.76) and (18.24,-0.85) .. (18.4,-0.95) .. controls (18.56,-1.05) and (18.66,-1.18) .. (18.7,-1.35);

  \foreach \x in {0, 4.4} {
    \draw[midarrow] (\x+2.45,3.5) -- (\x+2.75,3.5);
    \draw[midarrow] (\x+2.75,-3.5) -- (\x+2.45,-3.5);
  }
  \draw[midarrow] (17.2,-0.4) -- (17.2,0.4);

  \foreach \y in {3.5, 2.22, -2.22, -3.5} {
    \node at (12.1,\y) {$. .$};
    \node at (14.6,\y) {$. .$};
  }
  \node at (13.35,0) {$\cdots$};

  \node[left] at (-1.6,0) {$1$};
  \node[above=2pt] at (2.6,3.5) {$2$};
  \node[above=2pt] at (7,3.5) {$3$};
  \node[above=2pt] at (11.4,3.5) {$4$};
  \node[above=2pt] at (16,3.5) {$n-1$};
  \node[right] at (19.2,0) {$n$};
\end{tikzpicture}
    \caption{The $n$-component Milnor link $\mathcal{M}_n$.}
    \label{fig:milnorlink}
\end{figure}

The most famous example of a Brunnian link is the \textit{Borromean rings}, depicted in Figure \ref{fig:brunn3comp1}, while Figure \ref{fig:brunn3comp2} shows a 3-component Brunnian link that is inequivalent to the Borromean rings. It is worth noting that for a fixed $n \geq 3$, there is an infinite number of inequivalent Brunnian links with that number of components.
A family of examples which will be important throughout this work are the \textit{Milnor links} $\mathcal{M}_n$, constructed iteratively as in Figure \ref{fig:milnorlink}.

The fundamental group of the link complement $S^3\backslash \mathcal{L}$, which we denote by $\pi_1$ throughout, has interesting properties if $\mathcal{L}$ is Brunnian. First of all, remember that an explicit presentation of $\pi_1$ (known as the \textit{Wirtinger presentation}) can be constructed from a link projection of $\mathcal{L}$ as follows \cite{rolfsen2003knots}. To every arc in the projection, we associate a generator $a_{i,j}$. The index $i$ denotes the component it belongs to, while $j$ denotes an arc for said component. Then, to each crossing we associate a relation whose form depends on the sign of the crossing. We depict this procedure for the Borromean rings in Figure \ref{fig:borrwirtinger}.
\begin{figure}[h!]
    \centering
    \begin{tikzpicture}[scale=1.5, line cap=round,
    comp/.style={line width=1.3pt,
      preaction={draw, white, line width=4pt, line cap=butt}},
    gen/.style={draw=OrangeRed, line width=1pt, ->}]

  \newcommand{\wgen}[3]{%
    \draw[gen] ([shift={(#1:0.7)}]#2:1.3) -- ([shift={(#1:0.7)}]#2:0.9);
    \node[text=OrangeRed] at ([shift={(#1:0.7)}]#2:1.5) {#3};}
  \newcommand{\orient}[2]{%
    \path[line width=1.2pt, decoration={markings,
        mark=at position 0.5 with {\arrow[xshift=2pt]{>}}}, decorate]
      ([shift={(#1:0.7)}]{#2-10}:1.1) arc[start angle=#2-10, end angle=#2+10, radius=1.1];}

  \wgen{90}{90}{$x_1$}
  \wgen{90}{215}{$x_2$}
  \wgen{210}{210}{$y_1$}
  \wgen{210}{335}{$y_2$}
  \wgen{330}{330}{$z_1$}
  \wgen{330}{95}{$z_2$}

  \draw[comp] (90:0.7) ++(190.4:1.1) arc[start angle=190.4, end angle=289.6, radius=1.1];
  \draw[comp] (90:0.7) ++(303.6:1.1) arc[start angle=303.6, end angle=536.4, radius=1.1];
  \draw[comp] (210:0.7) ++(63.6:1.1) arc[start angle=63.6, end angle=296.4, radius=1.1];
  \draw[comp] (210:0.7) ++(310.4:1.1) arc[start angle=310.4, end angle=409.6, radius=1.1];
  \draw[comp] (330:0.7) ++(70.4:1.1) arc[start angle=70.4, end angle=169.6, radius=1.1];
  \draw[comp] (330:0.7) ++(183.6:1.1) arc[start angle=183.6, end angle=416.4, radius=1.1];

  \orient{90}{20}  \orient{90}{268}
  \orient{210}{140} \orient{210}{28}
  \orient{330}{260} \orient{330}{148}

  \node at (100:2)  {$1$};
  \node at (190:2) {$2$};
  \node at (350:2) {$3$};

  \def\eqx{2.8}
  \def\eqsep{1.4}
  \node[anchor=west] at (\eqx,0.3) {%
    \renewcommand{\arraystretch}{\eqsep}%
    $\left\{\begin{array}{r@{}l}
      x_2 &{}= y_1^{-1} x_1 y_1 \\
      y_1 &{}= z_2 y_2 z_2^{-1} \\
      z_2 &{}= x_1^{-1} z_1 x_1 \\
      z_1 &{}= x_2 z_2 x_2^{-1} \\
      x_1 &{}= y_2 x_2 y_2^{-1} \\
      y_2 &{}= z_1^{-1} y_1 z_1
    \end{array}\right.$};
\end{tikzpicture}
    \caption{Wirtinger presentation of $\pi_1(S^3 \backslash \mathcal{L})$, for $\mathcal{L}$ the Borromean rings. The group presentation can be simplified to $\pi_1 = \langle x,y,z \mid [[x^{-1},z],y] = [[y^{-1},x],z] = [[z^{-1},y],x] = 1 \rangle$, where $x \equiv x_1$, $y \equiv y_1$, and $z \equiv z_1$.}
    \label{fig:borrwirtinger}
\end{figure}
A meridian $\myemm_i \in \pi_1$ for the $i$-th component can be chosen to be whichever of the $a_{i,j}$ (in fact, they are all related to each other via conjugation, as per the Wirtinger relations). Each component has also a canonically defined longitude $\myell_i$, fixed by the requirement that it links trivially with the component itself. In $\pi_1$, they can be expressed as words in the meridians, and will turn out to have special properties as discussed later on. For example, if for the Borromean rings we set $\myemm_1 \equiv x_1, \myemm_2 \equiv y_1,$ and $\myemm_3 \equiv z_1$, the zero-linking longitude along the first component reads $\myell_1 = [\myemm_2, \myemm_3^{-1}]$.

\subsection{Milnor invariants} \label{ssec:Milnorinv}

In \cite{Milnor2015}, Milnor defined certain numerical invariants of links which are now known as the \textit{Milnor invariants} (or Milnor numbers), as follows. For each component, fix a word in the meridians $\myemm_j$ representing the longitude $\myell_i$. The \textit{Magnus expansion} of $\mathscr{l}_i$ is obtained by formally replacing
\begin{equation}
    \myemm_i \longmapsto 1+x_i, \qquad \myemm_i^{-1} \longmapsto 1-x_i + x_i^2 -x_i^3 + \dots,
\end{equation}
where the $x_i$ are $n$ non-commuting variables. Thus, $\myell_i$ will be mapped to a formal power series with general form
\begin{equation}
    \myell_i \longmapsto 1 + \sum_n \sum_{j_1, \dots, j_n} \mu( j_1,\dots,j_n, i) x_{j_1} \cdots x_{j_n}.
\end{equation}
Now, let $\Delta(i_1, \dots, i_n)$ denote the greatest common divisor of $\mu(j_1, \dots, j_s)$ ($2 \leq s < n$), with $(j_1, \dots, j_s)$ ranging over all sequences obtained by deleting at least one index out of $(i_1, \dots, i_n)$, and permuting the remaining ones cyclically. The \textit{Milnor invariant} $\overline{\mu}(i_1, \dots, i_n)$ is defined as the value of $\mu(i_1, \dots, i_n)$ modulo $\Delta(i_1, \dots, i_n)$, and these are isotopy invariants of the link $\mathcal{L}$.

For a fixed length $n$, the $n$-tuple Milnor numbers are not all independent. They have two basic symmetry relations \cite{Milnor2015}:
\begin{itemize}
    \item[(1)] Cyclicity: $\mil(i_1 i_2 \cdots i_n) = \mil(i_2 \cdots i_n i_1)$.
    \item[(2)] For any two sequences $I_r = (i_1, \dots, i_r)$ and $J_s = (j_1, \dots, j_s)$,
    \begin{equation}
        \sum_{(h_1, \dots, h_{r+s})} \bar{\mu}(h_1, \dots, h_{r+s}, k) = 0 \ \text{mod gcd$(\Delta(h_1, \dots, h_{r+s}, k))$},
    \end{equation}
    where the sum ranges over all riffle shuffles of $I_r$ and $J_s$\footnote{A riffle shuffle of $I_r$, $J_s$ is a sequence of length $r+s$ obtained by interleaving the elements of the two sequences. There are $\frac{(r+s)!}{r!s!}$ such sequences in total.}.
\end{itemize}

We now list a few important properties of fundamental groups of (not necessarily Brunnian) link complements, following \cite{cha_transfinite_2023}:
\begin{theorem} \label{th:Brpi1prop}
    Let $\mathcal{L}$ be an $n$-component link, and denote by $F$ the free group on $n$ generators. The following are equivalent:
    \begin{itemize}[left=2\parindent .. 2.5\parindent]
        \item[$(1)_k:$] There is an isomorphism $\pi_1/(\pi_1)_{k+1} \cong F/F_{k+1}$.
        \item[$(2)_k:$] The zero-linking longitudes $\myell_i \in (\pi_1)_k$.
        \item[$(3)_k:$] The length-$(k+1)$ Milnor invariants of $\mathcal{L}$ are well defined as integers.
    \end{itemize}
    Moreover, if $(1)_{k-1}-(3)_{k-1}$ hold, then $(1)_k-(3)_k$ are equivalent to
    \begin{itemize}[left=2\parindent .. 2.5\parindent]
        \item[$(4)_k:$] All length-$k$ Milnor invariants of $\mathcal{L}$ vanish.
    \end{itemize}
\end{theorem}
If $\mathcal{L}$ is Brunnian, all Milnor numbers of length $k < n$ vanish, and therefore $(1)_k-(3)_k$ hold for all $k \leq n-1$. In particular, statement $(2)_{n-1}$ implies that the longitudes lie in $(\pi_1)_{n-1}$, and statement $(3)_{n-1}$ implies that the $n$-tuple Milnor numbers are well defined as integers. Hence, the Magnus expansion of e.g.\ $\myell_1$ reads
\begin{equation} \label{eq:BrunnMagnus}
    \myell_1 \longmapsto 1 + \sum_{\sigma \in S_{n-1}} \mil(\sigma(2) \cdots \sigma(n) 1) x_{\sigma(2)} \cdots x_{\sigma(n)} + \mathcal{O}(x^n),
\end{equation}
and similarly for all other longitudes $\myell_i$. Statement $(1)_{n-1}$ will be useful in the proof of Theorem \ref{th:nilpDWBrunnians}.

The reader may find a more geometric interpretation of these invariants illuminating. Note first that the double Milnor numbers $\mu(ij)$ are exactly the pairwise linking numbers $\ell k(i,j)$. Hence, triple Milnor numbers are well defined as integers for a 3-component Brunnian link. In that case one can understand the latter as giving a good measure of ``genuine" triple linking ---  a higher-component generalization of the linking numbers. Analogously, $n$-tuple Milnor numbers give a good measure of ``genuine" $n$-tuple linking whenever all $(k<n)$-tuple linkings vanish.

Having said this, we will now show that longitudes of Brunnian links can be written in a canonical way when taken modulo $(\pi_1)_n$.
\begin{noframetbox}
    \begin{theorem} \label{th:Brlongitude}
        For an $n$-component Brunnian link,
        \begin{equation}
            \myell_1 = \underbrace{\prod_{\sigma \in S_{n-2}} [\mydots [[\myemm_2,\myemm_{\sigma(3)}],  \dots], \myemm_{\sigma(n)}]^{\mil(2\sigma(3)\cdots\sigma(n)1)}}_{ \equiv \tilde{\myell}_1} \text{ mod } (\pi_1)_{n},
        \end{equation}
        and similarly for all other longitudes $\myell_i$. The ordering of elements in the product is irrelevant, since any two different orderings are equivalent mod $(\pi_1)_n$.
    \end{theorem}
\end{noframetbox}
The proof goes as follows. Let $F$ denote the free group in the meridians $\myemm_i$. A theorem by Magnus \cite{Magnus1937} tells us that the subgroup of elements of $F$ whose Magnus expansion is of the form $1 + \mathcal{O}(x^k)$ is equal to $F_k$. In particular $F_k/F_{k+1}$ is generated (as an abelian group) by the commutators
\begin{equation}
    [ \mydots [[\myemm_{i_1},\myemm_{i_2}],  \dots], \myemm_{i_k}],
\end{equation}
whose Magnus expansion reads
\begin{equation}
    1 + [ \mydots [[x_{i_1},x_{i_2}],  \dots], x_{i_k}] + \mathcal{O}(x^{k+1}).
\end{equation}
The commutator here is the ring-theoretic commutator $[x_i, x_j] = x_ix_j - x_jx_i$. Thus, up to terms of order larger than $k$, the most general Magnus expansion for an element in $F_k$ will be of the form $1 + P_k(x)$, with $P_k(x)$ a homogeneous degree-$k$ polynomial in the $x_i$ arising as a linear combination of commutators.

Now let $w_1 \in F$ be a word in the meridians representing the longitude $\myell_1$. As discussed in Section \ref{ssec:Milnorinv}, $\myell_1 \in (\pi_1)_{n-1}$ and therefore its Magnus expansion (\ref{eq:BrunnMagnus}) has first non-trivial term a polynomial $P_{n-1}(x)$ as explained above, which in this case does not contain $x_1$. A basis for the space of all such polynomials is given by the commutators \cite[Sec.\ 5.6.2]{reutenauer1993free}
\begin{equation}
    [\mydots [[x_2,x_{\sigma(3)}],  \dots], x_{\sigma(n)}], \qquad \sigma \in S_{n-2}.
\end{equation}
By matching coefficients of the terms $x_{2}x_{\sigma(3)}\cdots x_{\sigma(n)}$ in Eq.\ (\ref{eq:BrunnMagnus}), we see that the same Magnus expansion arises from $\tilde{\myell}_1$.\footnote{The reader might notice that this matching implies certain (a priori non-obvious) equalities between Milnor numbers. Indeed, they are equivalent to the symmetry relations discussed in Section \ref{ssec:Milnorinv}.} Thus, $\myell_1\tilde{\myell}_1^{-1}$ has Magnus expansion $1 + \mathcal{O}(x^n)$. This implies that $\myell_1 \tilde{\myell}_1^{-1} \in F_n$, and hence $\myell_1 \tilde{\myell}^{-1}_1 \in (\pi_1)_n$. \qed

From the proof above we also learn the following fact: the number of independent $n$-tuple Milnor numbers for an $n$-component Brunnian link is exactly $(n-2)!$. Indeed, by cyclicity every length-$n$ Milnor number can be rewritten as $\mu(i_1 \dots i_{n-1} 1)$, and the latter are all fixed by the polynomial $P_{n-1}(x)$ appearing in the Magnus expansion of $\myell_1$, which we know has $(n-2)!$ independent parameters.

\section{Dijkgraaf-Witten theories with nilpotent gauge groups} \label{sec:DWnilp}

Discrete gauge theories in 3 dimensions were introduced by Dijkgraaf and Witten and are defined as follows \cite{Dijkgraaf:1989pz}. Let $G$ be a finite group, and $M$ be an oriented closed 3-manifold. Fix also a cohomology class $\omega \in H^3(G,U(1))$ --- we will commonly refer to it as the \textit{twist} or the \textit{topological action} of the theory. The Dijkgraaf-Witten theory twisted by $\omega$ is defined through its partition function as
\begin{equation}
    Z_{\DW}(M) = \frac{1}{\abs{G}} \sum_{\varphi \in \Hom(\pi_1(M),G)} \langle \varphi^*\omega, [M] \rangle,
\end{equation}
where we also use $\varphi$ to denote the classifying map $M \to BG$ of the corresponding flat connection, and $[M]$ denotes the fundamental class in $H_3(M, \Z)$. The above expression is manifestly a topological invariant of $M$. Indeed, Dijkgraaf-Witten theories are examples of bona fide 3-dimensional TQFTs \cite{Freed:1991bn}.

If the group $G$ is non-trivial, the theory comes equipped with non-trivial topological line operators. All the information about the spectrum of lines, together with information on properties such as mutual fusion or braiding, is encoded in a mathematical structure known as a \textit{modular tensor category} (MTC). In particular, the underlying MTC of line operators in a Dijkgraaf-Witten theory is given by the category $\Rep(D^\omega(G))$ of (finite-dimensional complex) representations of the twisted Drinfeld double of $G$ \cite{Roche:1990hs}. Rather than a complete account, we will only present the details that are necessary for our purposes in what follows. In particular, we will limit ourselves to the untwisted case.

The simple objects of the category $\Rep(D(G))$, which physically correspond to the simple line operators of the theory (i.e.\ those which cannot be decomposed as direct sums of other line operators) are given as follows. Let $R$ be a set of representatives for the conjugacy classes of $G$, and let $C_G(g)$ be the centralizer of the element $g$ in $G$. Simple lines are then labeled by pairs $(a, \chi)$ for $a \in R$ and $\chi$ an irreducible character of $C_G(a)$. Note that if we change conjugacy class representatives, say $a \mapsto gag^{-1}$ for some $g \in G$, we should identify the labels $(a, \chi)$ and $(gag^{-1}, \chi^g)$, where $\chi^g(h) = \chi(g^{-1} h g)$ is now an irreducible character of $C_G(gag^{-1})$. A special role is played by the object $1 \equiv (e, \chi_{\mathbf{1}})$, which corresponds to the identity line operator.

Physically, the labels $(a, \chi)$ represent the ``magnetic" and ``electric" charges of the line operator, in the following sense. Inserting a line with magnetic charge $a$ in the path integral enforces holonomies of small loops around it to be in the conjugacy class $[a]$. Likewise, inserting a line with electric charge $\chi$ corresponds to inserting the value of its holonomy in the $\chi$ representation. To be more explicit, let us write down the general expression for the expectation value of a configuration of lines in $S^3$. Consider a link $\mathcal{L}$ consisting of lines $(g_i, \chi_i)$ for $i = 1, \dots, N$. Denote as before the fundamental group of the link complement of $\mathcal{L}$ by $\pi_1$. Then, the (normalized) expectation value of the link $\mathcal{L}$ reads
\begin{equation}
    \langle \mathcal{L} \rangle = \frac{Z(S^3; \mathcal{L})}{Z(S^3)} = \sum_{\substack{\varphi \in \Hom(\pi_1, G), \\ \varphi(\myemm_i) \in [g_i]}} \prod_{i=1}^N \chi^{k_i}_i(\varphi(\myell_i)),
\end{equation}
where $\myemm_i, \myell_i \in \pi_1$ are a choice of meridians and zero-linking longitudes for $\mathcal{L}$ as discussed in Section \ref{sec:3brunnian}, $[g_i]$ denotes the conjugacy class of $g_i$, and $k_i \in G$ are such that $\varphi(\myemm_i) = k_ig_ik_i^{-1}$. We also used the fact that for a discrete $G$-gauge theory, $Z(S^3) = 1/\abs{G}$.

Another piece of information included in the category of line operators is the \textit{modular data} of the theory. This data consist of two square matrices $S$ and $T$, with entries parametrized by the simple objects in $\Rep(D(G))$. They are defined as \cite{Kitaev:2005hzj}
\begin{equation} \label{eq:definitionST}
    S_{ij} = \frac{1}{\mathcal{D}} \expval
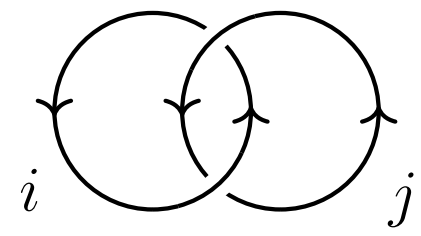
{\begin{tikzpicture}[thick,
    baseline={([yshift=-.5ex]current bounding box.center)}, scale=0.65,
    twoarrows/.style={decoration={markings,
        mark=at position 0.25 with {\arrow[xshift=1pt]{>}},
        mark=at position 0.75 with {\arrow[xshift=1pt]{>}}},
      postaction=decorate}]
  \draw[twoarrows] (0,-1) arc[start angle=-90, end angle=270, radius=1];
  \draw[twoarrows] (1.3,-1) arc[start angle=-90, end angle=270, radius=1];
  \draw[preaction={draw, white, line width=5pt}]
    (1.3,0) ++(105:1) arc[start angle=105, end angle=155, radius=1];
  \draw[preaction={draw, white, line width=5pt}]
    (0,0) ++(-25:1) arc[start angle=-25, end angle=-75, radius=1];
  \node at (-1.25,-0.8) {$i$};
  \node at (2.55,-0.9) {$j$};
\end{tikzpicture}}, \qquad T_{ij} = \delta_{ij} \frac{1}{d_i} \expval
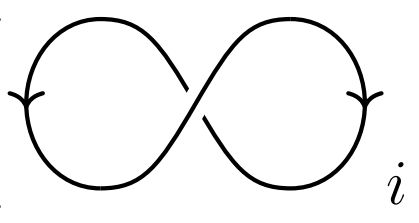
{\begin{tikzpicture}[thick,
    baseline={([yshift=-.5ex]current bounding box.center)}, scale=0.7,
    twoarrows/.style={decoration={markings,
        mark=at position 0.25 with {\arrow[xshift=1pt]{>}},
        mark=at position 0.75 with {\arrow[xshift=1pt]{>}}},
      postaction=decorate}]

  \draw[twoarrows] (0,0)
    .. controls (0.35,0.6) and (0.5,0.8) .. (0.9,0.8) %
    to[out=0, in=90]     (1.6,0) %
    to[out=-90, in=0]    (0.9,-0.8) %
    .. controls (0.5,-0.8) and (0.35,-0.6) .. (0,0) %
    .. controls (-0.35,0.6) and (-0.5,0.8) .. (-0.9,0.8) %
    to[out=180, in=90]   (-1.6,0) %
    to[out=-90, in=180]  (-0.9,-0.8) %
    .. controls (-0.5,-0.8) and (-0.35,-0.6) .. (0,0); %
  \draw[preaction={draw, white, line width=5pt}]
    (-0.9,-0.8) .. controls (-0.5,-0.8) and (-0.35,-0.6) .. (0,0)
                .. controls (0.35,0.6) and (0.5,0.8) .. (0.9,0.8);
  \node at (1.9,-0.75) {$i$};
\end{tikzpicture}}.
\end{equation}
where $d_i$ stands for the \textit{quantum dimension} of the $i$-th line, defined as the expectation value of $i$ when placed along a circle (with 0-framing), and $\mathcal{D} = \sqrt{\sum_\ell d_\ell^2}$ is the \textit{total quantum dimension} of the TQFT, which coincides with $1/Z(S^3)$. Note that quantum dimensions can also be read directly from the $S$-matrix,
\begin{equation} \label{eq:qdim}
    d_i = \frac{S_{i,1}}{S_{1,1}}.
\end{equation}
The above matrices are ``modular" in the sense that, together with the charge conjugation matrix $C$, they define a representation of the modular group $SL(2,\Z)$:
\begin{equation}
    S^2 = (ST)^3 = C, \quad C^2 = \I.
\end{equation}
The matrix $C$ is a permutation matrix that exchanges the line $(a, \chi)$ by its charge conjugate $(a^{-1}, \chi^*)$.\footnote{As mentioned previously $a^{-1}$ might not be an element in our initial set of representatives, so we really mean $(ga^{-1}g^{-1}, \chi^{g*})$ for some $g$ such that $ga^{-1}g^{-1} \in R$.} Following the above definition, one can check that the entries of the $S$ and $T$ matrices read \cite{Coste:2000tq}
\begin{align}
    S_{(a,\chi),(b,\psi)} &= \frac{1}{\abs{C_G(a)}\abs{C_G(b)}} \sum_{g \in G(a,b)} \chi(gbg^{-1})^*\psi(g^{-1}ag)^*, \label{eq:DWSmatrix} \\
    T_{(a,\chi),(b,\psi)} &= \delta_{a,b}\delta_{\chi,\psi} \frac{\chi(a)}{\chi(e)}, \label{eq:DWTmatrix}
\end{align}
where we defined $G(a,b) = \qty{g \in G \mid agbg^{-1}=gbg^{-1}a}$. From Eq.\ (\ref{eq:qdim}) it also follows that quantum dimensions are given by $d_{(g, \chi)} = \abs{[g]} \dim\chi$.

\subsection{Some theorems for Brunnian links}

We now proceed to study expectation values of Brunnian links in nilpotent DW theories. The main result we will prove in this section is the following:
\begin{noframetbox}
    \begin{theorem} \label{th:nilpDWBrunnians}
        Let $G$ be a finite group, and let $\DW_G$ denote the untwisted $G$ Dijkgraaf-Witten theory. The following statements are equivalent:
        \begin{itemize}
            \item[(1)] $\DW_G$ detects all Milnor links with $k \leq n+1$ components, but not those with $k > n+1$.
            \item[(2)] $\DW_G$ detects some Brunnian links with $k$ components for each $k \leq n+1$, but none with $k> n+1$.
            \item[(3)] $G$ is $n$-step nilpotent.
        \end{itemize}
    \end{theorem}
\end{noframetbox}
Let us first prove the equivalence between $(1)$ and $(3)$. Consider a configuration of $N$ lines forming the Milnor link $\mathcal{M}_N$ with the $N$-th line being purely electric, $L_N = (1,\chi)$, and the rest purely magnetic, of the form $L_i = (g_i, \chi_{\mathbf{1}})$. Note that since one of the magnetic charges is trivial, $\pi_1$ effectively becomes the fundamental group of the complement of only $N-1$ components, which due to the Brunnian property are unlinked. We can then replace $\pi_1$ by the free group $F$ in the $N-1$ meridians around $L_1, \dots, L_{N-1}$. Moreover, one can check that the zero-linking longitude along the $N$-th component reads $[\mydots [[h_1,h_2],  \dots], h_{N-1}]$. The expectation value then reduces to
\begin{equation} \label{eq:Milnorlinkexpval}
    \langle \mathcal{M}_N \rangle = \sum_{\substack{h_i \in [g_i] \\ (i = 1, \dots, N-1)}} \chi([ \mydots [[h_1,h_2],  \dots], h_{N-1}]) = \sum_{g \in G} \nu_{\{g_i\}}(g) \chi(g),
\end{equation}
where we defined the function
\begin{equation}
    \nu_{\{g_i\}}(g) = |\{(h_1, \dots, h_{N-1}) \in [g_1] \times \cdots \times [g_{N-1}] \mid [ \mydots [[h_1,h_2],  \dots], h_{N-1}] = g \}|,
\end{equation}
which is moreover a class function. Now suppose this link is not detected by the $G$-gauge theory, meaning no choice of line operators $L_i$ makes the expectation value non-trivial. This implies
\begin{equation}
    \expval{\mathcal{M}_N} = \expval{\bigcirc^{\sqcup N}} = |[g_1]|\cdots |[g_{N-1}]| \chi(1).
\end{equation}
Together with Eq.\ (\ref{eq:Milnorlinkexpval}), this leads to the equality
\begin{equation}
    \sum_{g \in G}\qty(\nu(g)-|[g_1]|\cdots|[g_{N-1}]|\delta_{g,1})\chi(g) = 0,
\end{equation}
valid for any $\chi$. Since the term in parenthesis is a class function, and this sum is proportional to its inner product (in the space of class functions) with $\chi$, we conclude that $\nu_{\{g_i\}}(g) = |[g_1]|\cdots|[g_{N-1}]| \delta_{g,1}$. Moreover, this holds for whatever choice of conjugacy classes $[g_i]$, so evaluating these functions at $g=1$ allows us to conclude that all commutators of the form $[ \mydots [[h_1,h_2],  \dots], h_{N-1}]$ are trivial, for any $h_i \in G$. Hence $G_{N-1} = \{1\}$.

If the theory does detect the $(N-1)$-component Milnor link, then $G_{N-2} \neq \{1\}$, otherwise the expectation value trivializes. We conclude that $G$ is $(N-2)$-step nilpotent, and the desired equivalence follows by setting $N = n+2$.

We now show that $(3)$ implies $(2)$. First, note that $(1)$ implies that the $\DW_G$ theory can detect at least one Brunnian link with $n+1$ components, namely the Milnor link $\mathcal{M}_{n+1}$. Therefore, it suffices to show that it cannot detect Brunnian links $\mathcal{L}$ with $k > n+1$ components.

Consider now a map $\varphi \in \Hom(\pi_1, G)$. Since $G$ is $n$-step nilpotent, $\varphi((\pi_1)_{k}) \subseteq G_{k} = \{1\}$, so $\varphi$ factors uniquely through $\pi_1/(\pi_1)_k \cong F/F_k$, where $F$ is the free group in $k$ generators (cf.\ Theorem \ref{th:Brpi1prop}). Conversely, any map from $F/F_k$ to $G$ defines a unique map from $F$ to $G$. We conclude that $\Hom(\pi_1,G) \cong \Hom(F,G)$. Note, moreover, that the latter is exactly the space of gauge field configurations for the $k$-component unlink. Finally, we also know that $\myell_i \in (\pi_1)_{k-1}$, and thus $\varphi(\myell_i) = 1$. With all this information, we conclude
\begin{equation}
    \expval{\mathcal{L}} = \sum_{\substack{h_i \in [g_i] \\ (i = 1, \dots, k)}} \prod_{i=1}^k \chi_i(1) = \expval{\bigcirc^{\sqcup k}}.
\end{equation}

Finally, we finish by proving the implication $(2) \implies (3)$. Since the theory does not detect Brunnian links with more than $n+1$ components, in particular it does not detect the Milnor link $\mathcal{M}_{n+2}$. Moreover, it does detect the Milnor link $\mathcal{M}_{n+1}$ --- otherwise $G$ would be (at most) $(n-1)$-step nilpotent, which by the above implies it does not detect Brunnian links with $n+1$ components, in contradiction with the starting hypothesis. This implies that $G$ is exactly $n$-step nilpotent. \qed

Theorem \ref{th:nilpDWBrunnians} above deserves some further comments. The implication $(1) \Rightarrow (2)$ is telling us that, in order to understand if a DW theory detects $n$-component Brunnian links, it suffices to know if it detects the $n$-component Milnor link. This should be understood as a generalization of Theorem \ref{th:abelianBrunnian} proved in Appendix \ref{app:Aabelian} --- restricted to the world of Dijkgraaf-Witten TQFTs --- in which probing the Borromean rings expectation value was enough to conclude that a TQFT is abelian.

Consider now an $(n+1)$-component Brunnian link $\mathcal{L}$ in an untwisted $G$ Dijkgraaf-Witten theory. We know by the above that its expectation value can be non-trivial, but so far we do not know what topological information of the link is encoded in this quantity. The following theorem answers this question:
\begin{noframetbox}
    \begin{theorem} \label{th:DWBrunnianMilnor}
        If $G$ is an $n$-step nilpotent group, the expectation value in $\DW_G$ of an $(n+1)$-component Brunnian link is completely determined by its length-$(n+1)$ Milnor invariants.
    \end{theorem}
\end{noframetbox}
This is an easy conclusion of Theorem \ref{th:Brlongitude}, which we present schematically for the sake of brevity. See that, under $\varphi \in \Hom(\pi_1, G)$, we have $\varphi(\myell_1) = \varphi(\tilde{\myell}_1)$ since $G_{n+1} = \{1\}$. Moreover, each commutator inside $\tilde{\myell}_1$ is mapped by $\varphi$ into $G_n \subseteq Z(G)$. Thus, if the first line operator has electric charge $\chi_1$,
\begin{equation}
    \chi_1(\varphi(\tilde{\myell}_1)) = \dim\chi_1 \prod_{\sigma \in S_{n-1}} \qty(\zeta_{\{h_i\}; \sigma})^{\mu(2\sigma(3)\cdots\sigma(n+1)1)},
\end{equation}
where we used Schur's lemma to write the commutator $[\mydots[[h_2,h_{\sigma(3)}],  \dots], h_{\sigma(n+1)}]$ as a phase $\zeta_{\{h_i\}; \sigma}$ times the identity in the $\chi_1$ representation, with $\varphi(\myemm_i) = h_i$. Note also that $\Hom(\pi_1, G) \cong \Hom(F, G)$, with $F$ the free group in $n+1$ generators, as discussed in the proof of Theorem $\ref{th:nilpDWBrunnians}$. Hence, the expectation value of $\mathcal{L}$ simplifies to
\begin{equation} \label{eq:DWbrunnianexpvalMilnor}
    \expval{\mathcal{L}} = \sum_{\substack{h_i \in [g_i] \\ i = 1,\dots, n+1}} \qty(\dim\chi_1 \prod_{\sigma \in S_{n-1}} \qty(\zeta_{\{h_i\}; \sigma})^{\mu(2\sigma(3)\cdots\sigma(n+1)1)} ) \times \dots,
\end{equation}
where each line operator will contribute with a similar factor, omitted above. \qed

The reader might wonder if, given an arbitrary Brunnian link, there exists a $\DW_G$ theory that detects it. The last theorem of this section gives a positive answer to this question up to link homotopy, that is up to isotopies and deformations that allow self-crossings of individual components.
\begin{noframetbox}
    \begin{theorem}
        Let $\mathcal{L}$ be an $n$-component Brunnian link that is not link-homotopic to the trivial link. Then Dijkgraaf-Witten theory for $G = F_{r,c,p}$, the free group in $r$ generators of nilpotency class $c$ and exponent $p$, detects $\mathcal{L}$ for appropriate choices of $r, c$ and $p$.
    \end{theorem}
\end{noframetbox}
The group $F_{r,c,p}$ above is defined as the following quotient of the free group $F$ in $r$ generators:\footnote{In the mathematical literature, this group is also known as the relatively free group in the variety of groups of exponent $p$ and class $c$, see e.g.\ \cite{Ito2006}.}
\begin{equation}
    F_{r,c,p} = \frac{F}{F_{c+1}F^p}.
\end{equation}

The proof of the theorem goes as follows. Let $r = c = n-1$ in the group above, and denote the $n-1$ generators of $F$ as $x_2, \dots, x_n$. Consider now a configuration of lines of the form $L_1 = (1, \chi)$ and $L_i = (x_i, \mathbf{1})$ for $i=2, \dots, n$. The expectation value of $\mathcal{L}$ reads
\begin{equation} \label{eq:FrcpExpval}
    \expval{\mathcal{L}} = \sum_{\substack{h_i \in [x_i] \\ i = 2, \dots, n}} \dim \chi \prod_{\sigma \in S_{n-2}} \qty(\zeta_{\{h_i\}; \sigma})^{\mu(2\sigma(3)\cdots\sigma(n)1)},
\end{equation}
and will be non-trivial as long as any one of the terms in the sum is a non-trivial phase, say in particular the term where all $h_i = x_i$. This is because a non-trivial phase for a single term in the sum forces $\mathrm{Re}\expval{\mathcal{L}} < \dim \chi \cdot|[x_2]|\ldots |[x_n]|$, with the right-hand side being the expectation value of the unlink.

As mentioned in the proof of Theorem \ref{th:Brlongitude}, the $(n-2)!$ commutators $[\mydots[[x_2,x_{\sigma(3)}],  \dots], x_{\sigma(n)}]$ are all independent from each other as elements of $F_{c}/F_{c+1}$ --- in particular, they generate a subgroup of the free abelian group generated by all commutators of weight $c$. If we moreover quotient by $F^p$, it is a known result \cite[Th.\ 2.6.3]{Mattarei92} that no extra relations are introduced between these generators as long as $p > c$, in the sense that $(F_{r,c,p})_c$ is now the free elementary abelian $p$-group generated by the commutators of weight $c$.

The final step is to construct an irreducible representation of $F_{r,c,p}$ in such a way that the product of phase factors in (\ref{eq:FrcpExpval}) is non-trivial. We construct it via induction, as follows. Remember that Milnor numbers classify Brunnian links up to link homotopy \cite{milnor_link_1954}. Since our Brunnian link is not link-homotopically trivial, there must be a non-zero Milnor number of length $n$, which without loss of generality we take to be $\mu(23\cdots n1)$. Consider the character $\psi$ of the subgroup $H = \Z_p^{(n-2)!}$ of $(F_{r,c,p})_c$ assigning $e^{2\pi i/p}$ to $[\mydots[[x_2,x_3],  \dots], x_n]$, and $1$ to all other generators. Now let $\chi = \Ind_H^G\psi=\bigoplus_i n_i\chi_i$ be the induced representation on the whole group, with $\chi_i$ being its distinct irreducible components entering with the multiplicities $n_i\in \mathbb{Z}_{>0}$. Let $\langle\,\cdot\,,\cdot\,\rangle_{G}$ be the standard inner product on the space of characters of a finite group $G$. Then by Schur's orthogonality and Frobenius reciprocity we have
\begin{equation}
   n_i= \langle  \Ind_H^G\psi,\chi_i\rangle_{G}=\langle  \psi,\Res^G_H\chi_i\rangle_{H}.
\end{equation}
Moreover, since $H$ is central, by Schur's lemma the restriction of an irreducible representation of $G$ to it should be a multiple of a one-dimensional representation of $H$, so that $\Res^G_H\chi_i=(\dim\chi_i)\,\psi_i$ for some character $\psi_i$ of $H$, and $n_i = \dim\chi_i\, \langle \psi, \psi_i \rangle_H$. Hence every component with $n_i > 0$ satisfies $\Res^G_H\chi_i=n_i\psi$. Taking any such $\chi_i$ as the electric charge of $L_1$, the phases in (\ref{eq:FrcpExpval}) are those of $\psi$, and the term with all $h_i = x_i$ equals $\dim\chi_i \cdot e^{2\pi i\, \mu(23\cdots n1)/p}$. Choosing $p > c$ coprime to $\mu(23\cdots n1)$, which holds for all sufficiently large $p$, we obtain an expectation value for $\expval{\mathcal{L}}$ different from that of the unlink, as promised.

\qed

\section{Chern-Simons theories with nilpotent gauge groups} \label{sec:5CSnilp}

Chern-Simons theory is a 3-dimensional topological gauge theory defined as follows \cite{Witten1988}. Let $\Lie{G}$ be a Lie group whose Lie algebra $\lie{g}$ admits an ad-invariant metric $\langle \cdot, \cdot \rangle$, and $M$ an oriented closed 3-manifold. Let also $\mathcal{A}$ be a connection on a principal $\Lie{G}$-bundle over $M$. If this bundle is trivial, we can write the Chern-Simons action as
\begin{equation} \label{eq:CSaction}
    S_{\CS}[\mathcal{A}] = \frac{k}{4\pi}\int_M \langle \mathcal{A} \wedge {}, d\mathcal{A} + \frac{1}{3}[\mathcal{A},\mathcal{A}] \rangle.
\end{equation}
The parameter $k$ is known as the \textit{level}, and if chosen appropriately, one can obtain a well-defined quantum field theory by path-integrating over the moduli space $\mathscr{A}/\mathscr{G}$ of all $\Lie{G}$-connections on $M$, weighted by the Chern-Simons action:
\begin{equation}
    Z_{\CS}(M) = \int_{\mathscr{A}/\mathscr{G}} \mathcal{D}\mathcal{A} \ e^{i S_{\CS}[\mathcal{A}]}.
\end{equation}
Note that we do not require the gauge group $\Lie{G}$ to be compact, in order to accommodate for nilpotent Lie groups. This introduces some subtleties that are not present in the more familiar case of semisimple and compact gauge groups, as we will now discuss.

The first feature of nilpotent CS theories is that the level is not always necessarily quantized. First, if $\Lie{G}$ is a simply-connected nilpotent group, it is diffeomorphic to $\R^n$ and therefore all $\Lie{G}$-bundles over $M$ are trivial. We can then rewrite the CS action as an integral over a 4-manifold $X$ whose boundary is $M$:
\begin{equation}
    S_{\CS}[A] = \frac{k}{4\pi} \int_X \langle \mathcal{F} \wedge {}, \mathcal{F}\rangle.
\end{equation}
This quantity is independent of the extension chosen because $\langle \mathcal{F} \wedge {}, \mathcal{F} \rangle$ is a 4-form with quantized periods, but $H^4(B\Lie{G}, \Z) = 0$ since $\Lie{G}$ is contractible. We conclude that the level $k$ is not quantized.

As a remark, we stress that we are taking the point of view where the ad-invariant metric on $\lie{g}$ is part of the data defining the CS theory. However, we are still allowed to make field redefinitions which, in general, may rescale the metric. To be more precise, an isomorphism $\phi$ between quadratic Lie algebras $(\lie{g}_1, \langle \cdot,\cdot \rangle_1)$ and $(\lie{g}_2, \langle \cdot,\cdot \rangle_2)$ is an isomorphism $\phi: \lie{g}_1 \xrightarrow{\sim} \lie{g}_2$ satisfying
\begin{equation}
    \langle X , Y \rangle_1 = \langle \phi(X), \phi(Y) \rangle_2, \qquad \forall X, Y \in \lie{g}_1.
\end{equation}
For simple $\lie{g}$, where the Killing form $\kappa$ is the unique ad-invariant metric up to an overall normalization, every automorphism of $\lie{g}$ leaves the metric invariant. Thus $(\lie{g}, \kappa)$ and $(\lie{g}, \lambda \kappa)$ are only isomorphic if $\lambda = 1$. In contrast, for nilpotent $\lie{g}$ one may find automorphisms that rescale the metric, as we will see in our explicit examples.

If $\Lie{G}$ is connected but not simply-connected, one may find a quantization condition coming from its compact directions. Remember that such $\Lie{G}$ are always a quotient of a simply-connected group by a discrete central subgroup --- therefore, the maximal compact subgroup of $\Lie{G}$ is isomorphic to $U(1)^l$, with $l$ the number of compact directions. The relevant cohomology group is therefore $H^4(BU(1)^l, \Z)$, but the particular characteristic classes involved in the quantization is completely determined by the metric. That is, if we choose a basis $\{Z_a\}_{a=1,\dots,l}$ of the subalgebra of $\lie{g}$ spanning the compact directions, then
\begin{equation}
    \langle F \wedge {},  F \rangle \supset K^{ab} F_a \wedge F_b,
\end{equation}
where $K^{ab} = \langle Z_a, Z_b \rangle$ acts as an abelian Chern-Simons level matrix. In the examples that concern us the metric happens to vanish when restricted to the compact directions, hence no quantization condition will be found.

\subsection{Distributional line operators} \label{subsec:distriblines}

Another important aspect of nilpotent CS theories is that the non-compactness of the gauge group gives rise to more exotic families of line operators. For compact Lie groups, the Peter-Weyl theorem (see e.g.\ \cite{Knapp1996}) implies that unitary irreducible representations are all finite-dimensional, and these can be labeled by a discrete set following the theory of highest weights. In contrast, for non-compact $\Lie{G}$ there generally exist a continuum of infinite-dimensional unitary irreps. In particular, for such representations the naive definition of Wilson line operators as traces of holonomies is unavailable. One option is to discard said representations altogether --- however, the theory loses most of its interesting features as finite-dimensional unitary irreps of nilpotent Lie groups are all abelian. Our proposal is to instead consider ``distribution-valued" line operators given by the trace functionals associated to such representations, which are well-defined objects as reviewed in Section \ref{sec:2nilp}.

The definition of these distributional line operators is straightforward once one has an explicit formula for the characters of the infinite-dimensional irreps of the gauge group. If we want to evaluate the trace of a path-ordered exponential --- following the usual definition of a Wilson line --- we can use the \textit{Magnus expansion} to rewrite
\begin{equation}
    \mathcal{P}e^{\oint_\gamma A} \equiv e^{\Omega_1 + \Omega_2 + \dots},
\end{equation}
where the $\Omega_i \in \lie{g}$ are determined by matching with the usual Dyson expansion of the path-ordered exponential. Parametrizing the support of the Wilson loop as $\gamma: [0,1] \to M$, the first few terms read as follows:
\begin{align}
    \Omega_1 &= \int_0^1 ds_1 \mathcal{A}(s_1), \\
    \Omega_2 &= -\frac{1}{2} \int_0^1 ds_1 \int_0^{s_1} ds_2 [\mathcal{A}(s_1), \mathcal{A}(s_2)], \\
    \Omega_3 &= \frac{1}{6} \int_0^1 ds_1 \int_0^{s_1} ds_2 \int_0^{s_2} ds_3 \qty([\mathcal{A}(s_1),[\mathcal{A}(s_2),\mathcal{A}(s_3)]] + [[\mathcal{A}(s_1),\mathcal{A}(s_2)],\mathcal{A}(s_3)]).
\end{align}
Here by $\mathcal{A}(s)$ we mean $\equiv \mathcal{A}_\mu(x(s)) \frac{dx^\mu}{ds}$, while commutators act only on Lie algebra indices. Going further in the Magnus expansion involves higher commutators, so the series eventually terminates for nilpotent Lie algebras. The line operators are then obtained by plugging this finite result into the irreducible characters.

The following general remark will also be useful later on. Equation \ref{eq:traceintegral} tells us that the character of $\pi_\varphi$ is proportional to
\begin{equation}
    \int_{\R^{2k}} e^{i (\varphi \cdot \gamma(s)) (X)} d^{2k}s,
\end{equation}
where we are considering this as a distribution on the Lie algebra $\lie{g}$ rather than on $\Lie{G}$, with $X$ the free variable. First of all, note that the coadjoint action is always of the form
\begin{equation}
    \varphi \cdot \gamma(s) = \varphi + \mathcal{O}(s),
\end{equation}
so there is always a global factor of $e^{i \varphi(X)}$ that we can pull out from the integral. The other terms will generically yield distributions on $\lie{g}$ after integrating over $s$ (unless the orbit is zero-dimensional, in which case the representation corresponds to an abelian irrep). If we want to evaluate this distribution on $\Omega_1 + \Omega_2 + \dots$, as we would to find the distributional line operators of the theory, the resulting operator always has a dependence of the form
\begin{equation} \label{eq:generallineopfactor}
    e^{i \varphi(\Omega_1 + \Omega_2 + \cdots)},
\end{equation}
together with constraints on the components of the $\Omega_i$ --- which involve integrals of polynomials in the gauge fields --- implemented by distributions.

Lastly, let us point out that there exist certain subtleties in considering this kind of distributional line operators. Basic quantities like the quantum dimensions are divergent, and it is not clear if a regularization procedure to render them finite exists. A similar problem occurs for the fusion of two such line operators: the tensor product of two infinite-dimensional irreducible representations generally results in a direct integral of irreps with infinite multiplicities, so fusion coefficients also seem to be ill-defined. To be more precise, if $\pi_1$ and $\pi_2$ are two unitary irreps of $\Lie{G}$ with respective orbits $\orb{1}$ and $\orb{2}$, their tensor product has the direct integral decomposition
\begin{equation}
    \pi_1 \otimes \pi_2 \simeq \int^\oplus d\sigma \, n_\sigma \sigma,
\end{equation}
where $\sigma$ ranges over all unitary irreps of $\Lie{G}$, and
\begin{equation}
    n_\sigma = \text{\# of $\Ad^*\Lie{G}_{\mathrm{diag}}$ orbits in } \qty{(\varphi_1, \varphi_2) \in \mathcal{O}_1 \times \mathcal{O}_2 \mid \varphi_1 + \varphi_2 \in \orb{\sigma}}
\end{equation}
is the multiplicity of the $\sigma$ representation \cite{Corwin88}. We stress that these multiplicities can be infinite depending on the structure of the orbits. To determine if these distributional line operators can then form a well-defined mathematical object akin to a fusion category (after proper regularization) is left as an open question.

\subsection{Example: CS theory with Lie algebra \texorpdfstring{$T^*\lie{h}_3$}{T*h\_3}} \label{subsec:T*h3CSexample}

To illustrate the general discussions presented above, let us study in detail the CS theory with Lie algebra $T^*\lie{h}_3$, following the notation introduced in Section \ref{subsec:T*h3}. Let us also choose the particular value $\lambda = k/2\pi$, for reasons that will become clear soon.

If we expand the CS connection $\mathcal{A}$ as $\sum_i A^iE_i + B^iF_i$, with $A^i$ and $B^j$ 1-form gauge fields, the CS action \ref{eq:CSaction} reads
\begin{equation}
    S_{\CS}[A,B] = \frac{k}{2\pi} \int_M \sum_{i=1}^3 B^i \wedge dA^i + \frac{k^2}{(2\pi)^2} \int_M A^1\wedge A^2 \wedge A^3.
\end{equation}
The finite gauge transformations of $\mathcal{A}$ are translated, thanks to the nilpotency of $T^*\lie{h}_3$, into finite gauge transformations for the $A$ and $B$ fields:
\begin{equation} \label{eq:T*h3gaugetransf}
    \begin{aligned}
        A^i & \longto A^i - d\lambda^i, \\
        B^i & \longto B^i - d\eta^i + \frac{k}{2\pi} \epsilon_{ijk} \qty(\lambda^j A^k - \frac{1}{2}\lambda^j d\lambda^k).
    \end{aligned}
\end{equation}
Note that, for $k = 2$, this is exactly the same action appearing in \cite{Wang:2016qhf, Putrov:2016qdo, He:2016xpi}, where it is claimed to be an equivalent formulation of the $D_4$ discrete gauge theory. More generally, for $k = p$ prime the theory naively corresponds to the DW theory with gauge group $\Z_p^3$ twisted by a particular 3-cocycle in $H^3(\Z_p^3,U(1))$, which makes it equivalent to the $\Heis(\F_p)$ DW theory. However, these claims must be reevaluated due to certain subtleties as we will now discuss.

The first comment concerns the nature of the $A^i$ and $B^j$ gauge fields. In the simply connected case, each basis element spans a non-compact direction of the Lie algebra. This implies that the gauge fields are all $\R$-valued, clashing with the statements in the literature where the theory is treated as a stack of BF theories (whose fields are $U(1)$ gauge fields) with a cubic interaction. In the non-simply connected case, where only the central directions can be compactified, the $B$ fields can be turned into $U(1)$ gauge fields, but the $A$ fields always remain real.

A consequence of the above is that no quantization condition arises for the charges associated to the $A$ fields, since there is no flux quantization. Thus, there is no reason to expect that only a finite set of charges is allowed, as assumed in \cite{Wang:2016qhf, Putrov:2016qdo, He:2016xpi}. This agrees with the fact that the global structure of the gauge group is non-compact, so that one generally expects an infinite amount of simple line operators. Discrete gauge theories, on the other hand, always have a finite number of simple lines.

As mentioned in the general discussion at the start of this section, there is no quantization condition on the CS level $k$. We can also understand this fact from the perspective of quadratic Lie algebra isomorphisms. The map
\begin{equation}
    E_i \longmapsto \lambda^{1/3} E_i, \qquad F_j \longmapsto \lambda^{2/3} F_j
\end{equation}
is an isomorphism between $T^*\lie{h}_3$ with its canonical metric and $T^*\lie{h}_3$ with the rescaled metric $\langle E_i, F_j \rangle = \frac{1}{\lambda} \delta_{ij}$. In particular, the map with $\lambda = k/2\pi$ can be used to absorb the overall coefficient in front of the CS action (\ref{eq:CSaction}). The level $k$ then becomes a parameter in the definition of the metric, which is unconstrained.

\textbf{Line operators.} Let us also compute explicitly the line operators in this theory. First of all, we need to expand the path-ordered exponential $\mathcal{P}e^{\oint_\gamma \mathcal{A}}$ through the Magnus expansion. A simple calculation yields
\begin{equation}
    \mathcal{P}e^{\oint_\gamma \mathcal{A}} = \exp\qty{\sum_i \oint_\gamma A^iE_i + \sum_i \qty(\oint_\gamma B^i - \sum_{jk} \frac{k}{4\pi} \epsilon_{ijk} A^j \phi^k) F_i}.
\end{equation}
Here we have introduced the function
\begin{equation}
    \phi^i(x) \equiv \int_{[\gamma(0), x]\subset \gamma} A^i
\end{equation}
for brevity. We now plug this into the characters of $T^*\lie{h}_3$ to find all possible line operators. The first kind of lines are those corresponding to the abelian (and therefore finite) representations $\pi_{(\alpha,0)}$. They read
\begin{equation}
    W_{(\alpha,0)}[\gamma] = \exp\qty{i \alpha_i \oint_\gamma A^i}.
\end{equation}
The second kind of lines correspond to the infinite-dimensional representations $\pi_{(\alpha,\beta)}$ with $\beta \neq 0$. These are distributional line operators of the form
\begin{equation} \label{eq:T*h3Wilsonline}
    W_{(\alpha,\beta)}[\gamma] = \frac{(2\pi)^2}{k|\beta|} \delta\qty(\beta \times \oint_\gamma A) \exp\qty{i \alpha_i \oint_\gamma A^i + i \beta_i \qty(\oint_\gamma B^i - \frac{k}{4\pi} \epsilon_{ijk} A^j \phi^k)}.
\end{equation}
This is a gauge invariant operator, as can be checked by plugging in the gauge transformations in Eq.\ (\ref{eq:T*h3gaugetransf}). We remark that, up to the delta function in front, these are exactly the same line operators found previously in \cite{Wang:2016qhf, Putrov:2016qdo, He:2016xpi}. While in their computations the condition that holonomies of $A^i$ vanish must be imposed by hand, from our construction the more general condition $\beta \times \oint_\gamma A = 0$ appears naturally from the representation theory of the underlying nilpotent gauge group. Indeed, this condition is enough to ensure that the line operator is gauge invariant.

\subsection{Brunnian link expectation values from Feynman diagrams} \label{sec:CSperturbative}

Feynman diagrams allow us to compute expectation values of link operators as a perturbative series in the Chern-Simons level $k$. These series have a particularly simple form when the gauge group is nilpotent. In this section, we will first exemplify this by computing the expectation values of 3-component Brunnian links in the CS theory with Lie algebra $T^*\lie{h}_3$, following the same logic of Rozansky \cite{rozansky1994reshetikhin} in the $SU(2)$ case. We will then expand on the general case of CS theories with Lie algebras $T^*\lie{g}$, for $\lie{g}$ nilpotent.

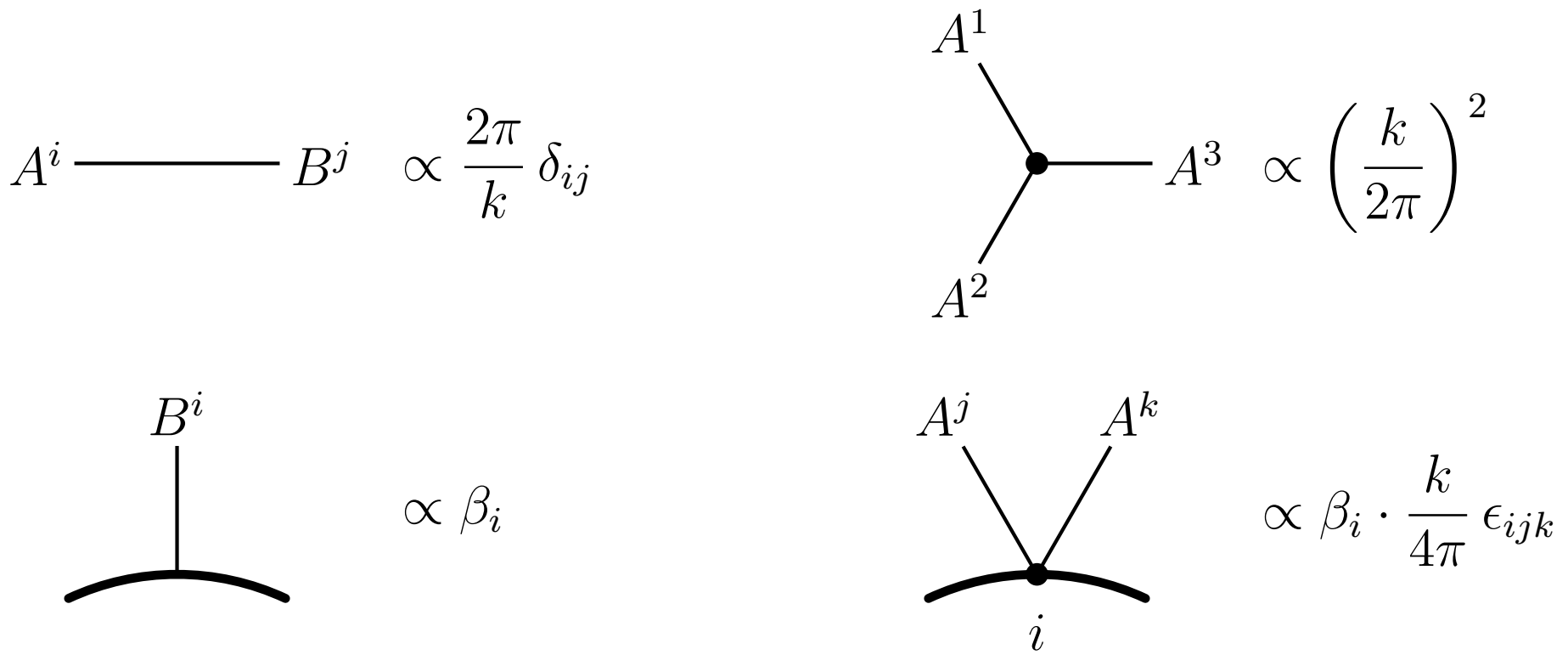
\begin{figure}[h!]
    \centering
    \begin{tikzpicture}[thick,
    boundary/.style={line width=2pt, line cap=round},
    leglabel/.style={inner sep=2pt},
    rule/.style={anchor=west}]
  \draw (0,0) -- (1.6,0);
  \node[leglabel, anchor=0]   at (0,0)   {$A^i$};
  \node[leglabel, anchor=180] at (1.6,0) {$B^j$};
  \node[rule] at (2.4,0) {$\displaystyle \propto \frac{2\pi}{k}\,\delta_{ij}$};

  \draw (7.5,0) -- ++(0:0.9)   node[leglabel, anchor=180] {$A^3$};
  \draw (7.5,0) -- ++(120:0.9) node[leglabel, anchor=-60] {$A^1$};
  \draw (7.5,0) -- ++(240:0.9) node[leglabel, anchor=60]  {$A^2$};
  \fill (7.5,0) circle (2.5pt);
  \node[rule] at (9.1,0) {$\displaystyle \propto \left(\frac{k}{2\pi}\right)^{2}$};

  \draw[boundary] ([shift={(0.8,-5.2)}]65:2) arc[start angle=65, end angle=115, radius=2];
  \draw (0.8,-3.2) -- (0.8,-2.2) node[leglabel, anchor=-90] {$B^i$};
  \node[rule] at (2.4,-2.7) {$\propto \beta_i$};

  \draw[boundary] ([shift={(7.5,-5.2)}]65:2) arc[start angle=65, end angle=115, radius=2];
  \draw (7.5,-3.2) -- ++(120:1.15) node[leglabel, anchor=-60]  {$A^j$};
  \draw (7.5,-3.2) -- ++(60:1.15)  node[leglabel, anchor=-120] {$A^k$};
  \fill (7.5,-3.2) circle (2.5pt);
  \node at (7.5,-3.65) {$i$};
  \node[rule] at (9.1,-2.7) {$\displaystyle \propto \beta_i \cdot \frac{k}{4\pi}\,\epsilon_{ijk}$};
\end{tikzpicture}
    \caption{Feynman rules of the CS theory with Lie algebra $T^*\lie{h}_3$. Thick lines represent Wilson line insertions.}
    \label{fig:T*h3feynmanrules}
\end{figure}

Let us start by laying down all the ingredients in the calculation for the $T^*\lie{h}_3$ CS theory. We consider the insertion of three Wilson lines, supported along the components of a 3-component Brunnian link, with pure $\beta$ charges as follows:
\begin{equation}
    W_i[\gamma_i] = W_{(0,\vec{\beta}_i)}[\gamma_i], \qquad (\vec{\beta}_i)_j = \beta_i\delta_{ij}.
\end{equation}
We make this choice in order to compare with the results in \cite{Putrov:2016qdo}. The expectation value of the link then reads
\begin{equation}
    \int \mathcal{D}A\mathcal{D}B \exp\qty{i \frac{k}{2\pi} \int B^i\wedge dA^i + i\frac{k^2}{(2\pi)^2} \int A^1 \wedge A^2 \wedge A^3 + i \beta_i \qty(\oint_{\gamma_i}B^i - \frac{k}{4\pi}\epsilon_{ijk}A^j\phi^k)},
\end{equation}
omitting overall normalization factors (but keeping in mind the constraints on the holonomies of $A$). In terms of Feynman diagrams, the first two terms give the propagators and a trivalent vertex, while the last term gives rise to two kinds of vertices supported on the curves $\gamma_i$. We depict these Feynman rules in Figure \ref{fig:T*h3feynmanrules}.

\begin{figure}[h!]
    \centering
    \begin{tikzpicture}[thick, scale=0.5,
    boundary/.style={line width=2pt, line cap=round}]
  \newcommand{\threearcs}[1]{%
    \begin{scope}[shift={(#1,0)}]
      \draw[boundary] ([shift={(90:3.4)}]220:1.2)  arc[start angle=220, end angle=320, radius=1.2];
      \draw[boundary] ([shift={(210:3.4)}]-20:1.2) arc[start angle=-20, end angle=80,  radius=1.2];
      \draw[boundary] ([shift={(330:3.4)}]100:1.2) arc[start angle=100, end angle=200, radius=1.2];
    \end{scope}}

  \threearcs{0}
  \draw (0,0) -- (0,2.2);
  \draw (0,0) -- (-1.905,-1.1);
  \draw (0,0) -- (1.905,-1.1);
  \fill (0,0) circle (4pt);

  \threearcs{8}
  \draw (6.095,-1.1) -- (8,2.2) -- (9.905,-1.1);
\end{tikzpicture}
    \caption{Leading-order tree-level diagrams contributing to expectation values of 3-component links in the $T^*\lie{h}_3$ CS theory.}
    \label{fig:T*h3treelevel3}
\end{figure}
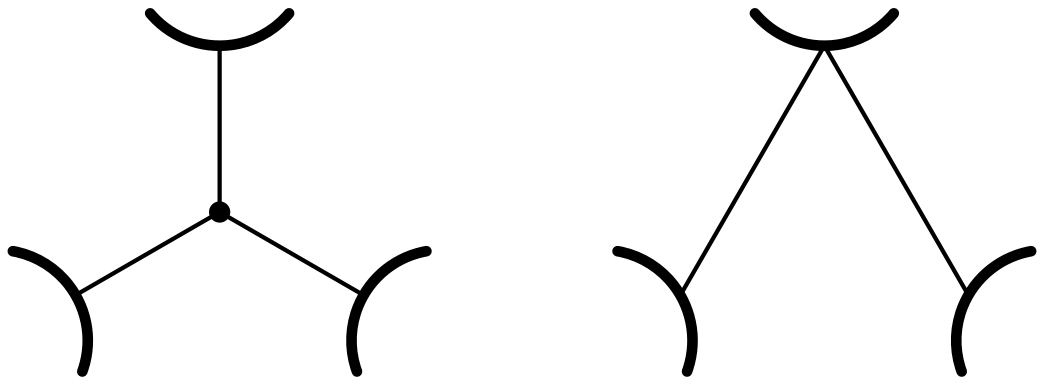

Notice that propagators connect $A$ fields to $B$ fields, while the trivalent vertex has all 3 legs of type $A$ --- this makes it impossible to form loop diagrams. Therefore, the path integral contains only tree-level contributions. Moreover, connected diagrams at tree-level are also constrained by the structure of vertices. At leading order, there are essentially two types of diagrams, as shown in Figure \ref{fig:T*h3treelevel3}. Note that diagrams of the form \wilsonchain\ also exist, but they are proportional to the pairwise linking numbers, which vanish due to the Brunnian property.

Let us compute the contributions to the path integral coming from the two diagrams in Figure \ref{fig:T*h3treelevel3}. First, following \cite{rozansky1994reshetikhin} we define the closed 1-forms $\omega_i$ on the link complement $S^3\backslash \mathcal{L}$ by the condition
\begin{equation}
    \oint_{\myemm_j} \omega_i = \delta_{ij},
\end{equation}
where $\myemm_j$ denotes the meridian around the $j$-th component $\gamma_j$. They have an explicit form in terms of the Green's function appearing in the propagator:\footnote{Another way to describe the 1-form $\omega_i$ is by considering the Poincaré dual of a surface $\Sigma_i$ with $\partial \Sigma_i = \gamma_i$.}
\begin{equation}
    \omega_i( - ) = \oint_{\gamma_i} dx\ \Omega(x, -).
\end{equation}
 Using these, we can write the contribution from the first diagram as
\begin{equation}
    -i\frac{2\pi}{k} \beta_1\beta_2\beta_3\int_{S^3}d^3x \ \omega_1(x)\omega_2(x)\omega_3(x),
\end{equation}
and the one from diagrams of the second type as
\begin{equation}
    -i\frac{2\pi}{k} \beta_1\beta_2\beta_3 \sum_{ijk} \frac{1}{2} \epsilon_{ijk} \oint_{\gamma_i} dy \int_{[x_i(0), y] \subset \gamma_i} dz \ \omega_j(y) \omega_k(z).
\end{equation}
As shown in \cite{rozansky1994reshetikhin}, the two contributions sum up to give the integral form of the triple Milnor invariant $\overline{\mu}(123)$. Moreover, while tree-level diagrams of higher-order exist, it is again impossible to find connected diagrams given our propagators and vertex. Hence, the exact expectation value reads, up to an overall normalization,
\begin{equation} \label{eq:pertexpval3brunn}
    \expval{\prod_{i=1}^3 W_i[\gamma_i]} \propto \exp\qty(-i \frac{2\pi}{k} \beta_1\beta_2\beta_3 \overline{\mu}(123)),
\end{equation}
just as expected from the path integral computations in \cite{Putrov:2016qdo}. We would like to stress that the Brunnian property is necessary for the integral form of $\overline{\mu}(123)$ to hold.

What about $(n > 3)$-component Brunnian links? Once again, the structure of the propagators and vertex do not allow for connected diagrams, even at tree-level. Expectation values are then trivial for these types of links, matching the behavior we found in discrete gauge theories (cf.\ Theorem \ref{th:nilpDWBrunnians}). This already hints at a correspondence between expectation values of Brunnian links in nilpotent CS theories versus those in nilpotent DW theories. We will come back to this in Section \ref{sec:6corresp}.

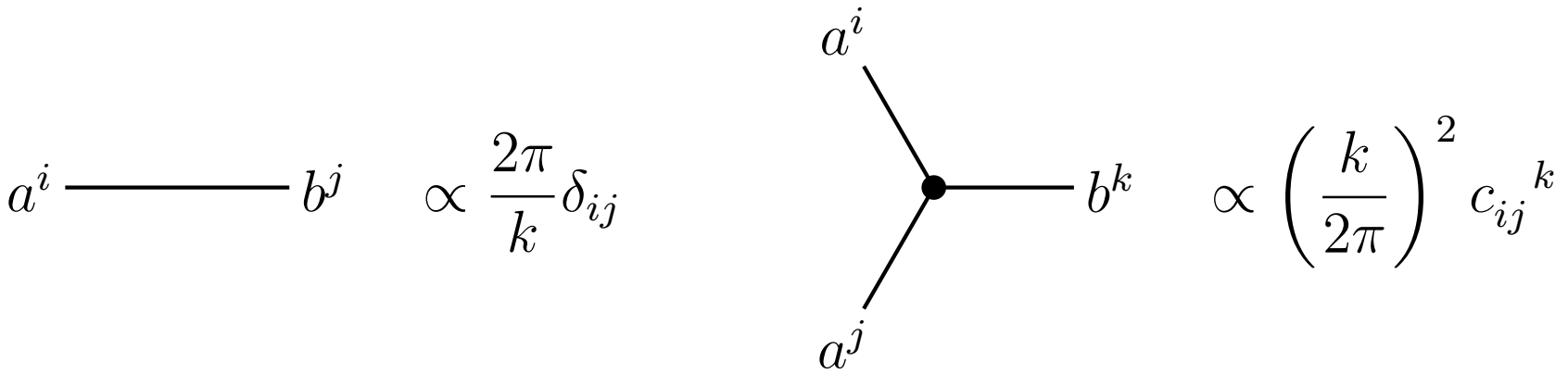
\begin{figure}[h!]
    \centering
    \begin{tikzpicture}[thick,
    baseline={([yshift=-.5ex]current bounding box.center)},
    leglabel/.style={inner sep=2pt}]
  \draw (0,0) -- (1.6,0);
  \node[leglabel, anchor=0] at (0,0) {$a^i$};
  \node[leglabel, anchor=180] (bj) at (1.6,0) {$b^j$};
  \node[anchor=base west] at ([xshift=0.3cm]bj.base east) {$\displaystyle \propto \frac{2\pi}{k} \delta_{ij}$};

  \draw (6.2,0) -- ++(0:1)   node[leglabel, anchor=180] (bk) {$b^k$};
  \draw (6.2,0) -- ++(120:1) node[leglabel, anchor=-60]      {$a^i$};
  \draw (6.2,0) -- ++(240:1) node[leglabel, anchor=60]       {$a^j$};
  \fill (6.2,0) circle (2.5pt);
  \node[anchor=base west] at ([xshift=0.3cm]bk.base east) {$\displaystyle \propto \left( \frac{k}{2\pi}\right)^2 c_{ij}^{\ \ k}$};
\end{tikzpicture}
    \caption{The propagators and vertices of the Chern-Simons theory with Lie algebra $T^*\lie{g}$.}
    \label{fig:T*gpropvert}
\end{figure}

The above results hold similarly for CS theories with Lie algebra $T^*\lie{g}$ for $\lie{g}$ nilpotent (cf.\ Section \ref{sec:DWtoT*gCS}). If we denote the basis of $\lie{g}$ by $\qty{X_i}$, its corresponding dual basis by $\qty{Y_i}$, and use the canonical metric $\langle X_i, Y_j \rangle = \delta_{ij}$, the propagators and vertices can be extracted by expanding the CS action in components, using $\mathcal{A} = a^iX_i + b^i Y_i$.\footnote{Note that in the example of $T^*\lie{h}_3$, $a$ fields correspond to $A_1, A_2$ and $B_3$, while $b$ fields to $B_1, B_2$ and $A_3$. This is just an unfortunate consequence of using the basis (\ref{eq:T^*h3EFbasis}).} We depict them graphically in Figure \ref{fig:T*gpropvert}, where $c_{ij}^{\ \ k}$ denote the structure constants of $\lie{g}$.

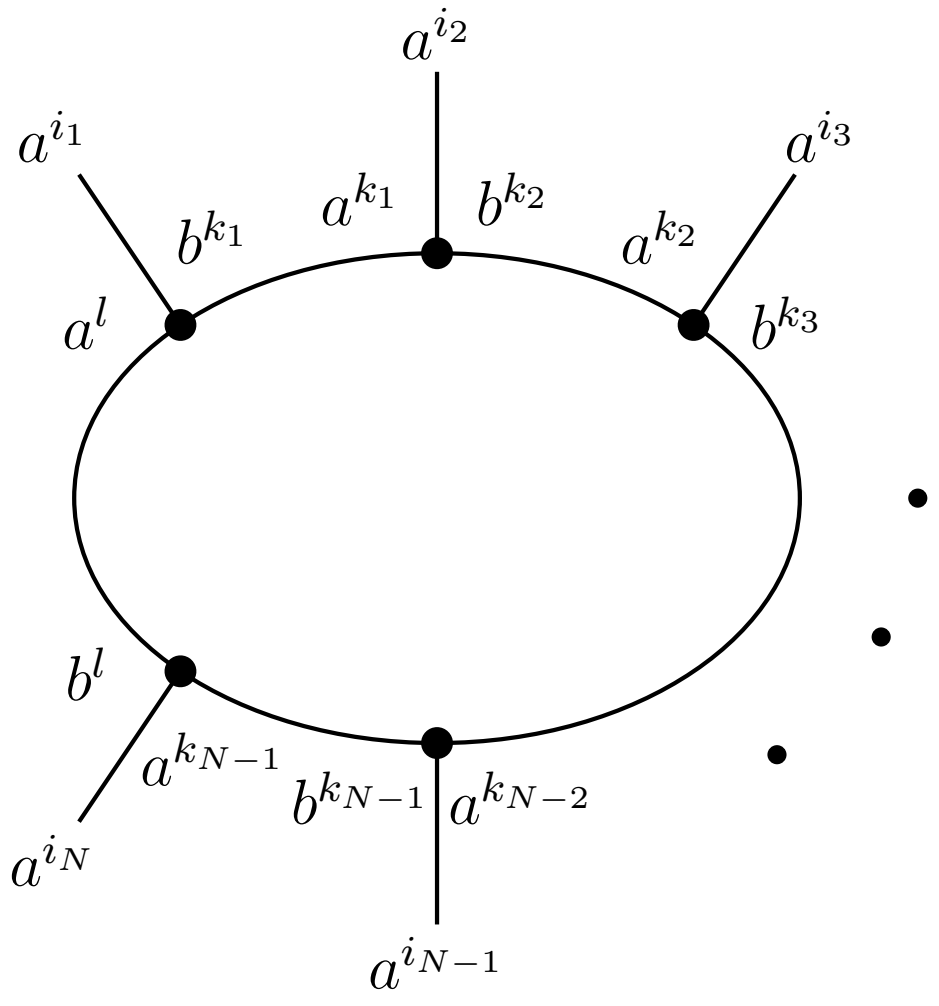
\begin{figure}[h!]
    \centering
    \begin{tikzpicture}[thick, scale=1.2]
  \newcommand{\loopvertex}[6]{%
    \begin{scope}[shift={(#1:2 and 1.35)}, rotate=#2]
      \draw (0,0) -- (1,0) node[anchor=#3, inner sep=2pt] {#4};
      \fill (0,0) circle (2.5pt);
      \node at (0.3,0.42) {#5};
      \node at (0.3,-0.42) {#6};
    \end{scope}}

  \draw (0,0) ellipse [x radius=2, y radius=1.35];
  \loopvertex{135}{124}{-56}{$a^{i_1}$}{$a^{l}$}{$b^{k_1}$}
  \loopvertex{90}{90}{-90}{$a^{i_2}$}{$a^{k_1}$}{$b^{k_2}$}
  \loopvertex{45}{56}{236}{$a^{i_3}$}{$a^{k_2}$}{$b^{k_3}$}
  \loopvertex{270}{270}{90}{$a^{i_{N-1}}$}{$\hspace{3pt} a^{k_{N-2}}$}{$b^{k_{N-1}}$}
  \loopvertex{225}{236}{56}{$a^{i_N}$}{$a^{k_{N-1}}$}{$b^{l}$}
  \foreach \t in {0, -22.5, -45}
    \fill (\t:2.65 and 2) circle (1.5pt);
\end{tikzpicture}
    \caption{An $N$-leg loop diagram in the Chern-Simons theory with Lie algebra $T^*\lie{g}$.}
    \label{fig:T*gloop}
\end{figure}

Let us first analyze the possible loop structures that could appear in expectation values of $N$-component links. As before, we only want to consider connected diagrams. The first easy observation is that tadpoles formed by closing the $a$ and $b$ legs of a trivalent vertex are not possible. Indeed, summing over its internal indices gives $c_{ik}^{\ \ k}$, which equals the trace of $X_i$ in the adjoint representation. By Engel's theorem (see e.g.\ \cite{Fulton2004}), there exists a basis in which these matrices are strictly upper triangular, so their traces vanish. Consider instead the loop diagrams with $N$ legs of type $a$, as shown in Figure \ref{fig:T*gloop}.
We can connect the external legs to Wilson line vertices, which is how these loops could in principle contribute to the expectation value. This loop comes with a weight of the form
\begin{equation} \label{eq:T*gloopweight}
    c_{i_1 m}^{\ \ \ \ k_1} c_{k_1 i_2}^{\ \ \ \ k_2} \cdots c_{k_{N-2} i_{N-1}}^{\qquad \quad \ k_{N-1}} c_{k_{N-1} i_N}^{\qquad \ \ m},
\end{equation}
which is up to a sign the trace of $X_{i_N}X_{i_{N-1}}\cdots X_{i_1}$ in the adjoint representation. Again, Engel's theorem implies that all of these traces vanish. Note also that
these are the only $N$-leg loop diagrams that can be built with our vertices, in the sense that it
is not possible to form these with one or more external $b$ legs.

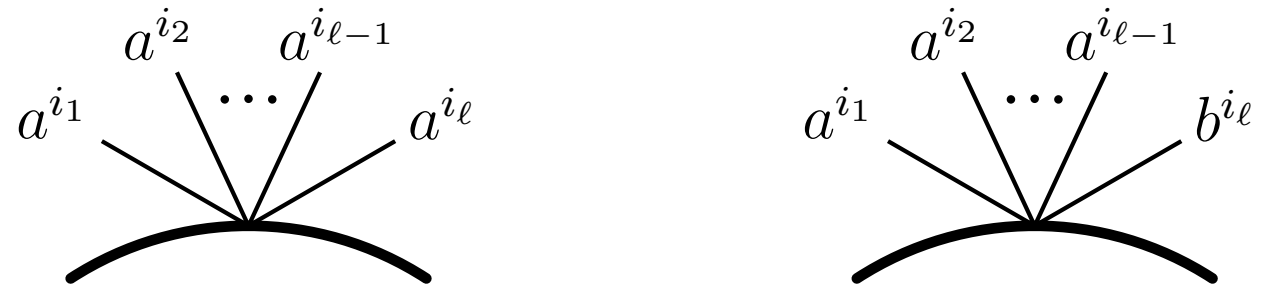
\begin{figure}[h!]
    \centering
    \begin{tikzpicture}[thick, scale=0.8,
    boundary/.style={line width=2pt, line cap=round},
    leglabel/.style={inner sep=2pt}]
  \newcommand{\wilsonpanel}[2]{%
    \begin{scope}[shift={(#1,0)}]
      \draw[boundary] ([shift={(0,-2.7)}]57:2.7) arc[start angle=57, end angle=123, radius=2.7];
      \draw (0,0) -- (150:1.4) node[leglabel, anchor=-30]  {$a^{i_1}$};
      \draw (0,0) -- (115:1.4) node[leglabel, anchor=-65]  {$a^{i_2}$};
      \draw (0,0) -- (65:1.4)  node[leglabel, anchor=-115] {$a^{i_{\ell-1}}$};
      \draw (0,0) -- (30:1.4)  node[leglabel, anchor=-150] {#2};

      \foreach \t in {87, 90, 93}
        \fill ([shift={(0,-2.7)}]\t:3.75) circle (1pt);
    \end{scope}}

  \wilsonpanel{0}{$a^{i_\ell}$}
  \wilsonpanel{6.5}{$b^{i_\ell}$}
\end{tikzpicture}
    \caption{Possible Wilson line vertices in the CS theory with Lie algebra $T^*\lie{g}$. If $\lie{g}$ is $n$-step nilpotent, $\ell$ ranges from $1$ to $n$.}
    \label{fig:T*gwilsonvertices}
\end{figure}

We can also consider forming loops by using the Wilson line vertices with two or more external legs, i.e.\ the ones arising from the Magnus expansion of the path-ordered exponential. In general, it is easy to see that if $\lie{g}$ is $n$-step nilpotent, the theory has vertices of this kind with number of legs at most $n$, and either all legs are of $a$ type, or all legs except one are of $a$ type, see Figure \ref{fig:T*gwilsonvertices}.\footnote{The reason why we get vertices of these form follows from the general remarks on irreducible characters made in Section \ref{subsec:distriblines}.}
Finally, the weight of such a vertex with $\ell$ external legs is a product of $\ell-1$ (suitably contracted) structure constants. We will say that these vertices are of order $\mathcal{O}(c^{\ell-1})$. They also come with a power of $k^{\ell-1}$ in the perturbative expansion. Once again, if we build loops using these vertices we will find contractions of structure constants as in Eq.\ (\ref{eq:T*gloopweight}), which vanish by Engel's theorem. We therefore conclude that there are no loop corrections in nilpotent $T^*\lie{g}$ CS theories.

\begin{figure}[h!]
    \centering
    \begin{tikzpicture}[thick,
    boundary/.style={line width=2pt, line cap=round}]

  \newcommand{\treearcs}[1]{%
    \begin{scope}[shift={(#1,0)}]
      \foreach \x in {0,1,2,3}
        \draw[boundary] ([shift={(\x,0.45)}]220:0.45) arc[start angle=220, end angle=320, radius=0.45];
      \draw[boundary] ([shift={(1.5,-2.95)}]40:0.45) arc[start angle=40, end angle=140, radius=0.45];
      \draw (1.5,-2) -- (1.5,-2.5);
      \fill (1.5,-2) circle (2.5pt);
    \end{scope}}

  \treearcs{0}
  \draw (0,0) -- (1.5,-2) -- (3,0);
  \draw (1,0) -- (0.5,-0.667);
  \draw (2,0) -- (1,-1.333);
  \fill (0.5,-0.667) circle (2.5pt);
  \fill (1,-1.333) circle (2.5pt);

  \begin{scope}[shift={(5.2,0)}]
    \treearcs{0}
    \draw (0,0) -- (1.5,-2) -- (3,0);
    \draw (1,0) -- (2,-1.333);
    \draw (2,0) -- (1.5,-0.667);
    \fill (1.5,-0.667) circle (2.5pt);
    \fill (2,-1.333) circle (2.5pt);
  \end{scope}

  \begin{scope}[shift={(10.4,0)}]
    \treearcs{0}
    \draw (0,0) -- (1.5,-2) -- (3,0);
    \draw (1,0) -- (0.5,-0.667);
    \draw (2,0) -- (2.5,-0.667);
    \fill (0.5,-0.667) circle (2.5pt);
    \fill (2.5,-0.667) circle (2.5pt);
  \end{scope}
\end{tikzpicture}
    \caption{Examples of ``tree" diagrams contributing to expectation values of $5$-component links, at leading order in $1/k$, in the CS theory with Lie algebra $T^*\lie{g}$.}
    \label{fig:T*gtrees}
\end{figure}
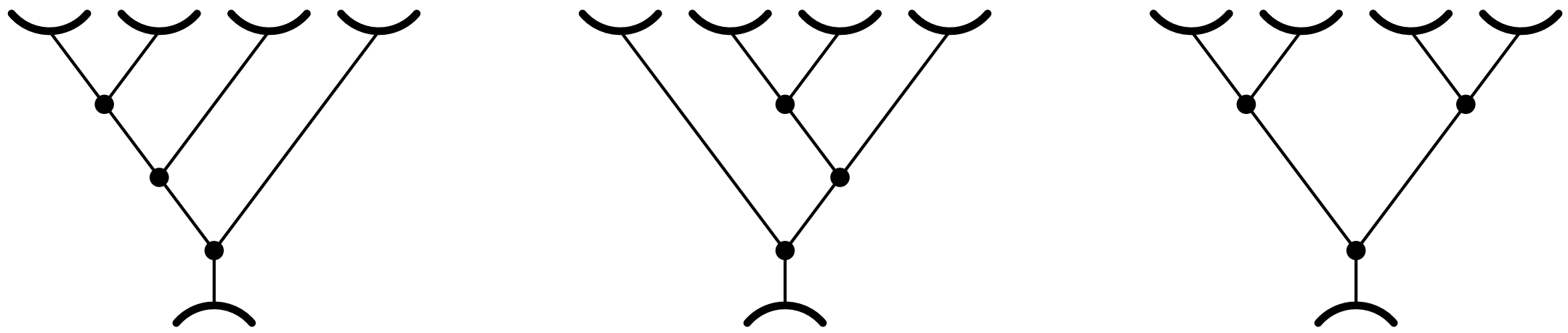

We now make a last comment regarding tree-level diagrams. As we saw in the example of $T^*\lie{h}_3$, the leading-order tree-level diagrams are actually the only ones that contribute to the correlation functions of 3-component Brunnian links. It is again not hard to believe that this will be generally true in the $T^*\lie{g}$ CS theory: expectation values of $(n+1)$-Brunnian links are completely determined by the leading-order tree-level diagrams if $\lie{g}$ is $n$-step nilpotent. Indeed, such diagrams are formed by using a combination of the trivalent vertices in Figure \ref{fig:T*gpropvert} and the Wilson line vertices in Figure \ref{fig:T*gwilsonvertices}. The simplest diagrams are ``trees" constructed purely out of trivalent vertices and single-leg Wilson line vertices, see Figure \ref{fig:T*gtrees} for examples when $n = 4$. Notably, these diagrams are all of order $\mathcal{O}(c^{n-1})$ and appear at order $1/k$ in the perturbative expansion.

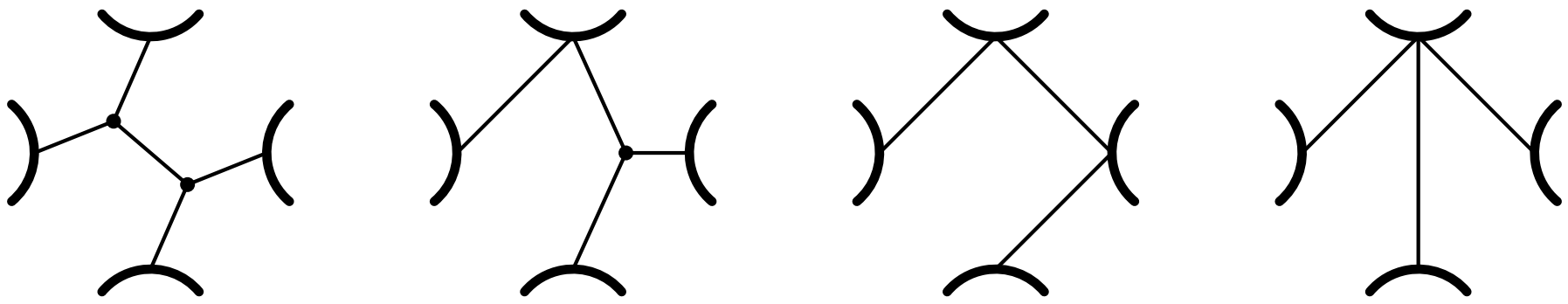
\begin{figure}[h!]
    \centering
    \begin{tikzpicture}[thick, scale=0.4,
    boundary/.style={line width=2pt, line cap=round}]
  \newcommand{\fourarcs}[1]{%
    \begin{scope}[shift={(#1,0)}]
      \draw[boundary] ([shift={(90:3.4)}]220:1.2)  arc[start angle=220, end angle=320, radius=1.2];
      \draw[boundary] ([shift={(180:3.4)}]-50:1.2) arc[start angle=-50, end angle=50,  radius=1.2];
      \draw[boundary] ([shift={(0:3.4)}]130:1.2)   arc[start angle=130, end angle=230, radius=1.2];
      \draw[boundary] ([shift={(270:3.4)}]40:1.2)  arc[start angle=40,  end angle=140, radius=1.2];
    \end{scope}}

  \fourarcs{0}
  \draw (0,2.2) -- (-0.7,0.6) -- (-2.2,0);
  \draw (-0.7,0.6) -- (0.7,-0.6);
  \draw (2.2,0) -- (0.7,-0.6) -- (0,-2.2);
  \fill (-0.7,0.6) circle (4pt);
  \fill (0.7,-0.6) circle (4pt);

  \fourarcs{8}
  \draw (5.8,0) -- (8,2.2) -- (9,0);
  \draw (10.2,0) -- (9,0) -- (8,-2.2);
  \fill (9,0) circle (4pt);

  \fourarcs{16}
  \draw (13.8,0) -- (16,2.2) -- (18.2,0) -- (16,-2.2);

  \fourarcs{24}
  \draw (21.8,0) -- (24,2.2) -- (26.2,0);
  \draw (24,2.2) -- (24,-2.2);
\end{tikzpicture}
    \caption{All leading-order tree-level diagrams contributing to expectation values of 4-component links in the CS theory with Lie algebra $T^*\lie{g}$.}
    \label{fig:T*g4comptreelevel}
\end{figure}

If we now use the Wilson line vertices with $\ell > 1$ legs, the decrease in number of propagators and trivalent vertices is compensated (in terms of the $k$-dependence of the diagram) by the fact that these vertices carry powers of $k$ themselves. For example, Figure \ref{fig:T*g4comptreelevel} depicts all leading-order, tree-level, connected diagrams for the $n = 3$ case. Remarkably all these diagrams appear at order $1/k$ --- moreover they are of order $\mathcal{O}(c^2)$ in the structure constants.
Without proof, we expect this to generalize to arbitrary $n$: all leading-order tree-level diagrams contributing to correlation functions of $(n+1)$-component links have weights of order $\mathcal{O}(c^{n-1})$, and those at subleading orders $1/k^2$ and beyond would have larger, and therefore vanishing orders in the structure constants.

Let us summarize our findings. In the 2-step nilpotent example of $T^*\lie{h}_3$, expectation values of 3-component Brunnian links depend only on the length-3 Milnor number $\mil(123)$, which arises solely from leading-order tree-level diagrams. In general, we expect that in a $T^*\lie{g}$ CS theory with $n$-step nilpotent $\lie{g}$, expectation values of $(n+1)$-component Brunnians depend only on their length-$(n+1)$ Milnor numbers, which as argued above should arise from the leading-order tree-level piece of the perturbative series. These statements are supported by the results in \cite{rozansky1994reshetikhin}: in Section VI therein, Rozansky proves that also the length-4 Milnor invariants arise from the leading-order tree-level diagrams contributing to the expectation values of 4-component Brunnian links. He then proceeds to conjecture more generally that for an $N$-component Brunnian link, the leading-order tree-level piece of the expectation value contains the length-$N$ Milnor invariants. In our nilpotent theories, where leading-order tree-level contributions are all there is, the Milnor numbers completely determine expectation values of Brunnian links, just as in nilpotent DW theories (see Theorem \ref{th:DWBrunnianMilnor}).

\section{Correspondence between nilpotent DW and CS theories} \label{sec:6corresp}

In the present section we will discuss the main result of this work: a conjectural correspondence between correlation functions in Dijkgraaf-Witten and Chern-Simons theories with nilpotent gauge groups.

\subsection{From \texorpdfstring{$\DW_G$}{DW\_G} theories to \texorpdfstring{$T^*\lie{g}$}{T*g} CS theories}
\label{sec:DWtoT*gCS}

Consider $\lie{g}_\Z$ to be a nilpotent Lie ring of nilpotency class $c$, whose additive group is free abelian of finite rank. We can use this Lie ring to construct both a finite group, which will be used in the DW side, and a real Lie algebra, which will be used in the CS side. In one case, let $p > c$ and reduce the Lie ring mod $p$ to find a Lie algebra over $\F_p$, which we denote by $\lie{g}_{\F_p}$, which is also nilpotent, of class at most $c$. In what follows we assume that the Lie ring $\lie{g}_\Z$ is chosen so that the nilpotency class is exactly $c$ after the reduction mod $p$. By the Lazard correspondence, this maps to a $p$-group $G = \exp(\lie{g}_{\F_p})$ of nilpotency class $c < p$ and exponent $p$, following the arguments in Section \ref{subsec:lazardfiniteorbits}. In the other case, we simply lift the Lie ring to a Lie algebra over $\R$, which we denote by $\lie{g}_\R$. Note that this is always possible because of the freeness condition on the additive group of $\lie{g}_\Z$.

Although by construction $\lie{g}_\R$ is a nilpotent Lie algebra (of nilpotency class $c < p$), it is not guaranteed that an ad-invariant metric on it exists, which is necessary to define a CS theory. There is however a construction which, starting from an arbitrary Lie algebra, yields a Lie algebra with a natural metric. Remember that the coadjoint representation $\ad^*: \lie{g}_\R \to \End(\lie{g}_\R^*)$ is given by
\begin{equation}
    (A_1 \cdot \varphi)(A_2) = - \varphi \cdot \ad_{A_1} (A_2),
\end{equation}
for all $A_1, A_2 \in \lie{g}_\R$ and $\varphi \in \lie{g}_\R^*$. The \textit{cotangent Lie algebra} of $\lie{g}_\R$, denoted $T^*\lie{g}_\R$, is the vector space $\lie{g}_\R \oplus \lie{g}_\R^*$ equipped with Lie bracket
\begin{equation}
    [(A_1, \varphi_1), (A_2, \varphi_2)]_{T^*\lie{g}_\R} = ([A_1, A_2]_{\lie{g}}, A_1 \cdot \varphi_2 - A_2 \cdot \varphi_1).
\end{equation}
This algebra has a natural ad-invariant metric, defined by
\begin{equation}
    \langle (A_1, \varphi_1), (A_2, \varphi_2) \rangle = \varphi_1(A_2) + \varphi_2(A_1).
\end{equation}
One can moreover check that $T^*\lie{g}_\R$ has the same nilpotency class as $\lie{g}_\R$. We will often denote the basis of $\lie{g}_\R$ by $\qty{X_i}$ and its associated dual basis by $\qty{Y_i}$, so that together they form a basis for $T^*\lie{g}_\R$ with commutation relations
\begin{equation}
    [X_i,X_j] = c_{ij}^{\ \ k}X_k, \qquad [X_i, Y_j] = -c_{ik}^{\ \ j}Y_k, \qquad [Y_i,Y_j] = 0.
\end{equation}

The cotangent Lie algebra construction described above applies also for Lie algebras over $\F_p$. In that case, $T^*\lie{g}_{\F_p}$ is itself an $\F_p$-Lie algebra with nilpotency class less than $p$, and the Lazard correspondence gives us a finite group
\begin{equation}
    \Lie{G}^{(p)} = \exp(T^*\lie{g}_{\F_p}),
\end{equation}
which as we know is a $p$-group of exponent $p$ and nilpotency class $c < p$. It turns out that there is an isomorphism of algebras
\begin{equation}
    \Phi: D(G) \longto \C[\Lie{G}^{(p)}],
\end{equation}
where $\C[\Lie{G}^{(p)}]$ denotes the group algebra of $\Lie{G}^{(p)}$, whose pullback $\Phi^*$ induces an isomorphism between simple $D(G)$-modules and simple $\C[\Lie{G}^{(p)}]$-modules, that is a 1-to-1 map from the simple objects in $\Rep(D(G))$ to the irreducible representations of $\Lie{G}^{(p)}$. If we compute the latter via the finite orbit method, we can classify them by coadjoint orbits of functionals $\varphi \in (T^*\lie{g}_{\F_p})^*$, which can be written in coordinates as
\begin{equation}
    \varphi = e \cdot X^* + q \cdot Y^*.
\end{equation}
Here, the ``charges" $e_i, q_i$ are $\F_p$-valued. In terms of these charges, the isomorphism $\Phi$ assigns to the simple object $(a,\chi)$ charges $(e,q)$ such that
\begin{equation} \label{eq:Phi*maprecipe}
    \log(a) = q \cdot X, \qquad e \cdot X^*\eval_{\log(C_G(a))} = \psi,
\end{equation}
where $\psi \in \log(C_G(a))^*$ is a representative of the coadjoint orbit associated to $\chi$ (i.e.\ $\chi \simeq \pi_\psi$ in the notation of Theorem \ref{thm:KirillovFinite}). Of course, there exist many choices for $e$ such that $e \cdot X^*$ restricts to $\psi$, but they are all related by the coadjoint action of $\exp({\lie{g}_{\F_p}}^*) \subset \Lie{G}^{(p)}$ on $(T^*\lie{g}_{\F_p})^*$. We leave all the mathematical details about the isomorphism $\Phi$ in Appendix \ref{app:Cdictionary}. For the purposes of this section, it suffices to know the explicit description given above.

We will now use these tools to formulate a conjecture that relates expectation values of line operators in the $\DW_G$ theory and the CS theory with Lie algebra $T^*\lie{g}_\R$.
\begin{noframetbox}
    \begin{conjecture}
        Let $\mathcal{L} = \gamma_1 \cup \cdots \cup \gamma_N$ be a framed $N$-component link in $S^3$, and $\lie{g}_\Z$ a $c$-step nilpotent Lie ring whose additive group is free abelian of finite rank. Then there exists a polynomial $P_\mathcal{L} \in \Z\qty[\frac{1}{c!}][e,q,s]$ such that, for any prime $p > c$, the following two statements hold:
        \begin{itemize}
            \item[(1)] In the $T^*\lie{g}_\R$ CS theory at level $k = p$, with $\lie{g}_\R$ a real lift of $\lie{g}_\Z$,
            \begin{equation}
                \expval{W_{(e_1,q_1)}[\gamma_1] \cdots W_{(e_N, q_N)}[\gamma_N]}_{\CS} = \qty(\prod_{i=1}^N d^{\CS}_{(e_i,q_i)}) \frac{1}{V^n} \int_{\R^n} \frac{d^ns}{p^n} \ e^{-\frac{2\pi i}{p} P_\mathcal{L}(e,q,s)},
            \end{equation}
            where $d^{\CS}_{(e_i,q_i)}$ is the expectation value of $W_{(e_i,q_i)}$ on the unknot with zero framing, and $V \equiv \delta(0) \equiv \int_{\mathbb{R}}\frac{ds}{p}\,e^{\frac{2\pi isx}{p}}|_{x=0}$ is the volume of $\R$.
            \item[(2)] In the $\DW_G$ theory, with $G = \exp(\lie{g}_{\F_p})$ where $\lie{g}_{\F_p}$ is the reduction mod $p$ of $\lie{g}_\Z$,
            \begin{equation}
                \expval{L_{(a_1,\chi_1)}[\gamma_1] \cdots L_{(a_N, \chi_N)}[\gamma_N]}_{\DW} = \qty(\prod_{i=1}^N d_{(a_i,\chi_i)}) \frac{1}{p^n} \sum_{s \in \F_p^n} e^{\frac{2\pi i}{p}P^{(p)}_\mathcal{L}(e,q,s)},
            \end{equation}
            where $d_{(a_i,\chi_i)}$ is the quantum dimension of $L_{(a_i,\chi_i)}$, $P^{(p)}_\mathcal{L}$ is the $\F_p$-reduction of $P_\mathcal{L}$, and the pairs $(e_i, q_i)$ are the images of $(a_i, \chi_i)$ under the isomorphism $(\Phi^{-1})^*$.
        \end{itemize}
    \end{conjecture}
\end{noframetbox}
In other words, in both theories the expectation value of the link, divided by the expectation value of the unlink with the same labels, is the average of the exponential over $s$. In order to support this conjecture, we will proceed to study two examples of correspondent DW and CS theories: the $\Heis(\F_p)$ DW theory against the $T^*\lie{h}_3$ CS theory, and the $\Fil_4(\F_p)$ DW theory against the $T^*\lie{f}_4$ CS theory. In the first example, we will show explicitly the values of ${\Phi^{*}}^{-1}$ on all simple lines in the DW side, and use these to compare two kinds of correlation functions: the Hopf link and the framed unknot expectation values, which up to a normalization correspond to the $S$ and $T$ matrices of the theories. In the second example, we will instead present selected entries for both the map $\Phi^{*}$ and the modular matrices, which provide further evidence for our conjecture.

\subsection{Example 1: \texorpdfstring{$\Heis(\F_p)$}{Heis(F\_p)} DW vs.\ \texorpdfstring{$T^*\lie{h}_3$}{T*h\_3} CS theories}

\subsubsection*{$\Heis(\F_p)$ Dijkgraaf-Witten theory}

Let us first write down the different line operators in the $\Heis(\F_p)$ DW theory. They come in 3 types:
\begin{equation} \label{eq:Heisallsimples}
    (g_c, \psi_{x,y}), \qquad (g_c, \chi_z), \qquad (g_{a,b}, \varphi_{m,n}^{(a,b)}),
\end{equation}
where in the notation of Section \ref{subsec:HeisFp}, $g_c \equiv [0,0,c]$, $g_{a,b} \equiv [a,b,ab/2]$, and $\varphi_{m,n}$ are the characters of $C_{\Heis(\F_p)}(g_{a,b}) \cong \Z_p^2$ which act like
\begin{equation}
    \varphi_{m,n}(g_{a,b}^s \cdot g_1^t) = \exp(\frac{2\pi i}{p}(ms+nt)),
\end{equation}
with $s, t \in \F_p$. Note that the theory has a total of $p^4+p^3-p$ line operators, with the identity line being $1 \equiv (g_0, \psi_{0,0})$. The lines of first type are all the abelian lines, as one can verify from their quantum dimensions.

The modular matrices of the theory can be computed by a readily (but lengthy) application of Eqs.\ (\ref{eq:DWSmatrix}) and (\ref{eq:DWTmatrix}). We start by spelling out the elements of the $T$-matrix. They read
\begin{align}
    T_{(g_c, \psi_{x,y})} & = 1, \\
    T_{(g_c, \chi_z)} & = \exp(\frac{2\pi i}{p}cz), \\
    T_{(g_{a,b}, \varphi_{m,n})} & = \exp(\frac{2\pi i}{p} m).
\end{align}
Here we have denoted $T_{\alpha, \beta} = \delta_{\alpha, \beta} T_\alpha$. The $S$-matrix on the other hand is given as follows:
\begingroup
\allowdisplaybreaks
\begin{align}
        S_{(g_c, \psi_{x,y}), (g_{c'}, \psi_{x',y'})} &= \frac{1}{p^3}, \\
        S_{(g_c, \psi_{x,y}), (g_{c'}, \chi_{z'})} &= \frac{1}{p^2}\exp(-\frac{2\pi i}{p} cz'), \\
        S_{(g_c, \psi_{x,y}), \qty(g_{a',b'}, \varphi_{m',n'}^{(a',b')})} &= \frac{1}{p^2} \exp(-\frac{2\pi i}{p} (xa'+yb'+cn')),
        \\
        S_{(g_c, \chi_z), (g_{c'}, \chi_{z'})} &= \frac{1}{p}\exp(-\frac{2\pi i}{p}(zc' + z'c)), \\
        S_{(g_c, \chi_z), \qty(g_{a',b'}, \varphi_{m',n'}^{(a',b')})} &= 0, \\
        S_{\qty(g_{a,b}, \varphi_{m,n}^{(a,b)}), \qty(g_{a',b'}, \varphi_{m',n'}^{(a',b')})} &= \delta_{(a,b,n) \times (a',b',n'), 0} \frac{1}{p} \exp(-\frac{2\pi i}{p}(m\lambda + m'\lambda^{-1})),
\end{align}
\endgroup
where $\lambda$ in the last entry is such that $(a',b',n') = \lambda (a,b,n)$, as they are constrained to be parallel. We remark that the Kronecker deltas are understood to impose vanishing conditions mod $p$.

Following the statement of our conjecture, the next step is to rewrite the $S$- and $T$-matrix elements in terms of charges $(e,q)$ as given by the isomorphism $\Phi^*$. In Table \ref{tab:HeisPhi*map} we write down the explicit values of $(\Phi^{-1})^*$ on the simple lines (\ref{eq:Heisallsimples}), computed by following the recipe given in Eq.\ (\ref{eq:Phi*maprecipe}).
\begin{table}[h!]
    \centering
    \begin{tabular}{Sc|Sc|Sc}
         $(g, \chi)$ & $e$ & $q$  \\
          \hline
         $(g_c, \psi_{x,y})$ & $e = (x,y,0)$ & $q = (0,0,c)$ \\
         $(g_c, \chi_z)$ & $e = (0,0,z)$ & $q = (0,0,c)$ \\
         $(g_{a,b}, \varphi_{m,n})$ & $e = (e_1, e_2, n)$ & $q = (a,b,0)$
    \end{tabular}
    \caption{Charges $e$ and $q$ corresponding to the values of $(\Phi^{-1})^*$ on the simple lines of the $\Heis(\F_p)$ DW theory. In the last entry, $e_1$ and $e_2$ must satisfy $e_1a+e_2b = m$ mod $p$.}
    \label{tab:HeisPhi*map}
\end{table}
Using this dictionary, and using the definition (\ref{eq:definitionST}) for the $S$ and $T$ matrices, one has for the framed unknot and Hopf link expectation values that
\begin{equation} \label{eq:conjectureT-DWHeis}
    \expval{L_{(a,\chi)}[\bigcirc^{+1}]} = d_{(a,\chi)} \exp(\frac{2\pi i}{p} e \cdot q),
\end{equation}
and
\begin{multline} \label{eq:conjectureS-DWHeis}
    \expval{L_{(a,\chi)}[\gamma]L_{(a',\chi')}[\gamma']} = d_{(a,\chi)} d_{(a',\chi')} \exp(-\frac{2\pi i}{p}(e \cdot q' + e' \cdot q)) \\
    \times \frac{1}{p^3} \sum_{s \in \F_p^3} \exp(\frac{2\pi i}{p}\qty(s_1(q_2e'_3-q'_2e_3) + s_2(e_3q_1'-e_3'q_1) + s_3(q_1q_2'-q_1'q_2))).
\end{multline}

\subsubsection*{$T^*\lie{h}_3$ Chern-Simons theory at level $p$}

As discussed in Section \ref{subsec:T*h3CSexample}, line operators in the theory are labeled by pairs $(\alpha,\beta)$ (with $\alpha_i, \beta_i \in \R$) and have explicit expressions $W_{(\alpha,\beta)}[\gamma]$ as given in Eq.\ (\ref{eq:T*h3Wilsonline}). Note that we have reserved the labels $e$ and $q$ for the charges in the $X$-$Y$ basis of $T^*\lie{h}_3$, which relate to $\alpha$ and $\beta$ as used in Section \ref{subsec:T*h3} by
\begin{equation}
    \alpha_{1,2} = e_{1,2}, \quad \beta_{1,2} = q_{1,2}, \quad \alpha_3 = q_3, \quad \beta_3 = e_3,
\end{equation}
which are instead the coefficients in the $E$-$F$ basis. We will hereon denote the line operators as $W_{(e,q)}[\gamma]$, taking into account this relabeling.

The framed unknot and Hopf link expectation values in this theory can be found exactly using path integral methods. We leave the details of this computation to Appendix \ref{app:BT*h3modular}. The final result for the framed unknot expectation value reads
\begin{equation} \label{eq:conjectureT-CST*h3}
    \expval{W_{(e,q)}\qty[\bigcirc^{+1}]} = d^{\CS}_{(e,q)} \exp(-\frac{2\pi i}{p} e \cdot q), \qquad d^{\CS}_{(e,q)} = \begin{cases} p \abs{\beta} V^2, & \beta \neq 0, \\ 1, & \beta = 0, \end{cases}
\end{equation}
where $\beta = (q_1,q_2,e_3)$ and $V \equiv \delta(0)$. The value of $d^{\CS}_{(e,q)}$ follows from the normalization of the line operators in Eq.\ (\ref{eq:T*h3Wilsonline}), with the delta function written through its integral representation.
For the Hopf link, we rewrite for convenience the result in Eq.\ (\ref{eq:CShopflinkresult}) as
\begin{multline} \label{eq:conjectureS-CST*h3}
    \expval{W_{(e,q)}[\gamma_1] W_{(e',q')}[\gamma_2]} = d^{\CS}_{(e,q)} d^{\CS}_{(e',q')} \exp(\frac{2\pi i}{p} (e\cdot q' + e' \cdot q)) \\
    \times \frac{1}{(pV)^3} \int_{\R^3} d^3s \exp(\frac{2\pi i}{p}\qty(s_1(q_2e'_3-q'_2e_3) + s_2(e_3q_1'-e_3'q_1) + s_3(q_1q_2'-q_1'q_2))).
\end{multline}
Here the square of the delta function in Eq.\ (\ref{eq:CShopflinkresult}) is understood with the formal rule $\delta(x)^2 = \delta(0)\delta(x)$, and the component of $s$ along $\beta$, which does not enter the exponent, is integrated to $pV$.

\subsubsection*{Validity of the conjecture}

Comparing Eqs.\ (\ref{eq:conjectureT-DWHeis}), (\ref{eq:conjectureS-DWHeis}) with (\ref{eq:conjectureT-CST*h3}), (\ref{eq:conjectureS-CST*h3}), we find that the polynomials satisfying the statements of the conjecture for the framed unknot and the Hopf link are
\begin{equation}
    P_{T}(e,q,s) = e \cdot q,
\end{equation}
and
\begin{equation}
    P_{S}(e,q,e',q',s) = - e \cdot q' - e' \cdot q + s_1(q_2e'_3-q'_2e_3) + s_2(e_3q_1'-e_3'q_1) + s_3(q_1q_2'-q_1'q_2),
\end{equation}
respectively. The normalization factors in both theories are as stated in the conjecture, with $n = 0$ for the framed unknot and $n = 3$ for the Hopf link. We thus verify our conjecture for these two types of links.

\subsubsection{Comments on \texorpdfstring{$p = 2$}{p = 2}: the \texorpdfstring{$D_4$}{D\_4} DW theory}

Let us make some remarks on a possible validity of the conjecture for $p = 2$, in which case the group $\Heis(\F_p)$ is isomorphic to the dihedral group $D_4$ of order 8. Note that this group does not satisfy the assumptions of the Kirillov method, given that the nilpotency class is equal to $p$. Therefore, we do not have a map analogous to $\Phi^*$ that we can use to assign charges to simple lines in a systematic way. In any case, we can still try to come up with a dictionary by inspection of e.g.\ the modular matrices.

The main problem arises when analyzing the structure of centralizers of conjugacy classes in $D_4$. If we work with the well-known group presentation
\begin{equation}
    D_4 = \langle s, r \mid s^2 = r^4 = 1, srsr=1 \rangle,
\end{equation}
the conjugacy class $[r]$ has as centralizer the subgroup $\langle r \rangle \cong \Z_4$. Thus, there are simple lines in the $D_4$ DW theory labeled by $([r], \rho_m)$, where $\rho_m$ ($m \in \Z_4$) is a character of $\Z_4$ acting as $\rho_m(r^j) = \exp(2\pi imj/4)$. By Eq.\ (\ref{eq:DWTmatrix}), the $T$-matrix element of this line reads
\begin{equation}
    T_{([r], \rho_m)} = \exp(\frac{2\pi i}{4}m).
\end{equation}
If we want the expression in part (2) of our conjecture to hold for some $P_\mathcal{L}^{(p)}$, we would need to replace $p$ by 4 and find charges $e, q$ valued in $\Z_4$ that can reproduce all values of the $T$-matrix.

At the same time, on the CS side, the calculations of Appendix $\ref{app:BT*h3modular}$ are independent of the choice of level. Therefore, at level $p = 2$ the phases of $T$-matrix elements are all of the form
\begin{equation}
    \exp(-\frac{2\pi i}{2} e \cdot q).
\end{equation}
In the CS theory with simply connected group $\exp(T^*\lie{h}_3)$ charges are allowed to be arbitrary real numbers, and even in the non-simply connected case we still have $\R$-valued electric charges.
It is still possible to match the expectation values of CS theory with those of the $D_4$ DW theory for a suitable choice of integral magnetic charges $q$ and half-integer electric charges $e$. Therefore, we conclude that there is an embedding of simple lines in the $D_4$ DW theory into the $T^*\lie{h}_3$ CS theory at level 2. However, we stress that these results are in contrast (and in fact, contradict) the claims in \cite{He:2016xpi, Putrov:2016qdo} stating that the $D_4$ simple lines can be understood as $W_{(e,q)}[\gamma]$ with $\Z_2$-valued charges since, as we just discussed, these lines do not recover all possible values of the $T$-matrix in the $D_4$ gauge theory.\footnote{The matching of the expectation values with integer electric charges $e$ in principle can still be achieved if instead of the regularization scheme that uses the normal push-off of the magnetic surfaces, as in Appendix \ref{app:BT*h3modular}, one uses the framed push-off for all the lines. The latter regularization scheme, however, breaks the gauge invariance in general.}

\subsection{Example 2: \texorpdfstring{$\Fil_4(\F_p)$}{Fil\_4(F\_p)} DW vs.\ \texorpdfstring{$T^*\lie{f}_4$}{T*f\_4} CS theories}

We now review the validity of our conjecture in a $3$-step nilpotent example. Since computations become quite intricate already at this nilpotency level, we will only make a few comparisons for particular entries of the $S$ and $T$ matrices.

In the $\Fil_4(\F_p)$ DW theory, consider the following two types of line:
\begin{equation}
    \begin{aligned}
        L_1 &= ([0,0,0,d], \chi_{y,w}^{(3)}), & w \neq 0,\\
        L_2 &= ([0,b,0,d], \sigma_{m,n,r}), & b \neq 0.
    \end{aligned}
\end{equation}
Here $\sigma_{m,n,r}$ is a character of the centralizer of $[0,b,0,d]$, acting as
\begin{equation}
    \sigma_{m,n,r}(e^{sX_2}e^{tX_3}e^{uX_4}) = \exp(\frac{2\pi i}{p}(ms+nt+ru)),
\end{equation}
see Equation (\ref{eq:FilZp3centralizer}). Under the map ${\Phi^{*}}^{-1}$, the corresponding charges of these lines are
\begin{equation}
    \begin{aligned}
        {\Phi^{*}}^{-1}(L_1): & \qquad e = (0,y,0,w), & \quad q = (0,0,0,d), \\
        {\Phi^{*}}^{-1}(L_2): & \qquad e = (0,m,n,r), & \quad q = (0,b,0,d).
    \end{aligned}
\end{equation}

Let us now discuss briefly the CS side, where the gauge algebra is $T^*\lie{f}_4$. Expanding the connection as $\mathcal{A} = a^iX_i + b^iY_i$, the CS action at level $k$ reads
\begin{equation}
    S_{\CS}[a,b] = \frac{k}{2\pi} \int_{S^3} b^i \wedge da^i + \frac{k^2}{(2\pi)^2} \int a^1 \wedge a^2 \wedge b^3 + a^1 \wedge a^3 \wedge b^4.
\end{equation}
In the path integral, insertions of line operators are now harder to deal with since they can contain terms up to third order in the Magnus expansion. Coadjoint orbits in $T^*\lie{f}_4$ are of dimension up to 4, so orbit integrals in the characters introduce further complicated dependencies on the gauge fields as well. This makes it hard to give a general recipe for path integral computations --- one should instead study them on a case-by-case basis. Generally speaking, since the fields $b^1, b^2$ and $a^4$ do not appear in the cubic interactions, one aims to integrate these fields out first and solve for their dual gauge fields (see Appendix \ref{app:BT*h3modular} for the analogous procedure in the $T^*\lie{h}_3$ theory). The reader can verify that following these guidelines leads to the $T$ and $S$ matrix elements we present below.

Having said this, let us compare first the $T$-matrix entry of $L_1$ versus the corresponding framed unknot expectation value of $W_{{\Phi^{*}}^{-1}(L_1)}$ in the continuous theory. The former reads
\begin{equation}
    T_{([0,0,0,d], \chi^{(3)}_{y,w})} = \exp(\frac{2\pi i}{p} dw).
\end{equation}
On the CS side, the charges of ${\Phi^{*}}^{-1}(L_1)$ are of type 5, so one must consider a Wilson line constructed from the character in Eq.\ (\ref{eq:T*f4type5char}). Without showing the details, one will find that
\begin{equation}
    \expval{W_{((0,e_2,0,e_4),(0,0,0,q_4))}[\bigcirc^{+1}]} = p \abs{e_4} V^2 \exp(- \frac{2\pi i}{p} e_4q_4),
\end{equation}
that is $d^{\CS}_{(e,q)} = p \abs{e_4} V^2$ for the lines of type 5.
We expect then, more generally, that the polynomial $P_T$ associated to the framed unknot is
\begin{equation}
    P_T(e,q,s) = e \cdot q,
\end{equation}
just as in the previous example. For a proof that this is the general expectation in DW theories whose gauge group is a $p$-group of nilpotency class $c<p$ and has exponent $p$, see Appendix \ref{subsec:appC-Phi*}.\footnote{This result in $T^*\mathfrak{g}$ CS theory is also not surprising from the point of view of Vassiliev knot invariants \cite{bar1995vassiliev}. Indeed, universal Lie algebra weight systems assign to the one chord diagram $\Theta$ the quadratic Casimir of the Lie algebra, which in our case can be shown to be exactly $e \cdot q$. The $T$-matrix element of the line is then, up to some factors, the exponential of this diagram \cite{LeMurakami96}.}

More interesting is to compare the $S$-matrix element between $L_1$ and $L'_2$ with the Hopf link expectation value of their images under the dictionary. On the DW side, we have
\begin{equation}
    S_{L_1, L'_2} = \frac{1}{p^3} \exp(-\frac{2\pi i}{p}(yb'+wd'+dr')) \sum_{s \in \F_p} \exp(-\frac{2\pi i}{p}\cdot \frac{1}{2}wb's^2),
\end{equation}
where $\frac{1}{2}$ in the last term refers to $2^{-1}$ mod $p$. We see that this expression splits into a product of a phase factor times a quadratic Gauss sum over $s \in \F_p$. Unfortunately, since we don't have the general form of the $S$-matrix, we can't quite write down the polynomial $P_S$ associated to the Hopf link. A natural guess is that it must include the term $e \cdot q' + e' \cdot q$, just as in the example of $T^*\lie{h}_3$, and we suspect that this is in fact a general statement for nilpotent $T^*\lie{g}$ CS theories. On the other hand, we also know that it must contain a term of the form $-\frac{1}{2}e_4q_2's^2$. In other words, we find that after evaluation of $P_S$ on $e_1 = e_3 = q_1 = q_2 = q_3 = 0$ and $e_1' = q_1' = q_3' = 0$, and possibly other parameters $s_1, s_2, \dots = 0$, it looks like
\begin{equation}
    P_S = - e \cdot q' - e' \cdot q - \frac{1}{2} e_4q_2' s^2.
\end{equation}
Now we compare this to the corresponding expectation value in the CS side. Remarkably, the theory recovers both the universal phase factor and a ``continuous" version of the quadratic sum here. Indeed, if we consider two lines with charges $e = (0,e_2,0,e_4)$, $q = (0,0,0,q_4)$ and $e' = (0,e'_2,e'_3,e'_4)$, $q' = (0,q'_2,0,q'_4)$, with $e_4' \neq 0$ so that the latter line also falls into the type-5 case, one may find that
\begin{multline}
    \expval{W_{(e,q)}[\gamma_1] W_{(e',q')}[\gamma_2]} = d^{\CS}_{(e,q)} d^{\CS}_{(e',q')} \exp(\frac{2\pi i}{p}(e\cdot q' + e' \cdot q)) \\
    \times \frac{1}{pV} \int_\R ds \exp(\frac{2\pi i}{p} \frac{1}{2} e_4q_2' s^2).
\end{multline}
As promised, the polynomial appearing in the exponential is exactly $P_S$, again restricted to the same subset of charges mentioned above. The normalization factors also agree with the conjecture, with $n = 1$. Namely, on the DW side $d_{L_1} = d_{L'_2} = p$, so that the expectation value of the Hopf link is $p^4 S_{L_1,L'_2} = d_{L_1} d_{L'_2} p^{-1} \sum_{s \in \F_p} (\dots)$.

\section*{Acknowledgments}

The authors acknowledge Michele Del Zotto, Nicola Dondi, Stefano C.\ Lanza, Lucrezia B.\ Lorenzi, Daniele Migliorati, and Antonio Santaniello for fruitful discussions. The research of T.N.\ is partially supported by the INFN Iniziativa Specifica GAST and Indam GNFM. The research of J.Q.R.\ is partially supported by the INFN Iniziativa Specifica ST\&FI.

\paragraph{Use of AI tools.} In the preparation of this paper the authors used the AI system Claude (Anthropic) for numerical and symbolic checks of the formulas, for generating all figures, for assistance with some derivations in Section \ref{sec:6corresp} and Appendix \ref{app:Cdictionary}, and for editing certain portions of the text. All results were verified by the authors, who take full responsibility for the content of the paper.

\appendix
\section{Brunnian links in abelian TQFTs} \label{app:Aabelian}

In this appendix we give a proof of the fact that abelian TQFTs are exactly the ones that do not detect any 3-component Brunnian links. As in the introduction, a TQFT is \textit{abelian} if all its simple line operators have quantum dimension one, and it \textit{detects} a link if for some choice of the colors the expectation value of the link differs from the expectation value of the unlink with the same number of components and the same colors.
\begin{noframetbox}
    \begin{theorem} \label{th:abelianBrunnian}
        Let $\mathcal{T}$ be a unitary TQFT with a finite number of simple line operators. Then $\mathcal{T}$ is abelian if and only if it does not detect any 3-component Brunnian link.
    \end{theorem}
\end{noframetbox}
Let us first assume that $\mathcal{T}$ does not detect any 3-component Brunnian link, and show that it is abelian. The proof will make use of the properties of the TQFT partition function under \textit{surgery}. Namely, consider removing a tubular neighborhood of a knot $K$ inside a 3-manifold $M$, and then gluing back a solid torus via a diffeomorphism $U \in \SL(2, \Z)$ of its boundary torus. In this way we obtain a new manifold $M_U$, whose partition function can be written as a sum over insertions of lines in $M$, supported on $K$:
\begin{equation} \label{eq:surgeryprop}
    Z(M_U) = \sum_{\ell } U_{\ell, 1} Z(M; K_\ell).
\end{equation}
Here $\ell$ ranges over all the simple lines of the TQFT. This same logic applies to links by performing surgery individually to each of their components.

We now apply this machinery to $M = S^3$ and $\mathcal{L}$ the Borromean rings. Indeed, it is well-known that if we perform surgery via $S \in \SL(2,\Z)$ to each of its components, the resulting 3-manifold is the 3-torus. Equation (\ref{eq:surgeryprop}) then implies that
\begin{equation}
    Z(\T^3) = \sum_{\ell_1, \ell_2, \ell_3} S_{\ell_1, 1} S_{\ell_2, 1} S_{\ell_3, 1} Z(S^3; \mathcal{L}_{\ell_1, \ell_2, \ell_3}).
\end{equation}
Since by assumption the theory doesn't detect the Borromean rings, the following equality holds:
\begin{equation}
    Z(S^3; \mathcal{L}_{\ell_1, \ell_2, \ell_3}) = Z(S^3) \expval{\bigcirc_{\ell_1} \sqcup \bigcirc_{\ell_2} \sqcup \bigcirc_{\ell_3}} = \frac{d_{\ell_1}d_{\ell_2}d_{\ell_3}}{\mathcal{D}}.
\end{equation}
Using that $S_{\ell,1} = d_\ell/\mathcal{D}$ and $\mathcal{D} = \sqrt{\sum_\ell d_\ell^2}$, we find that $Z(\T^3) = \sum_\ell d_\ell^2$. At the same time, we know that $Z(\T^3)$ must give the dimension of the 2-torus Hilbert space, which equals the number of simple lines. Unitarity of the TQFT requires all quantum dimensions to be equal or greater than one --- we conclude that all lines have unit quantum dimension, implying that the TQFT is abelian.

For the converse statement, namely that an abelian TQFT does not detect any 3-component Brunnian link, we can give a geometrical proof as follows. Since the fusion (vector) spaces of lines are one-dimensional (in the sense that the only non-zero fusion coefficients are $N_{a,b}^{a+b} = 1$), the \textit{braiding matrix} of two lines $a$ and $b$, defined by
\begin{equation}
    R_{ab} = 
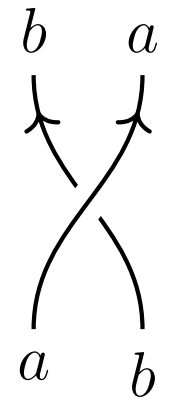
\begin{tikzpicture}[scale=0.6, thick,
    baseline={([yshift=-.5ex]current bounding box.center)},
    overstrand/.style={preaction={draw, white, line width=7pt,
        shorten <=1pt, shorten >=1pt}},
    uparrow/.style={decoration={markings,
        mark=at position 0.85 with {\arrow[xshift=1pt]{>}}},
      postaction=decorate}]
  \draw[uparrow] (1.2,0) to[out=90, in=-90] (0,2.8);
  \draw[overstrand, uparrow] (0,0) to[out=90, in=-90] (1.2,2.8);
  \node[below] at (0,0)     {$a$};
  \node[below] at (1.2,0)   {$b$};
  \node[above] at (0,2.8)   {$b$};
  \node[above] at (1.2,2.8) {$a$};
\end{tikzpicture},
\end{equation}
is $U(1)$-valued. This allows us to exchange positive crossings by negative crossings (and vice versa) inside expectation values at the cost of a phase factor:\footnote{Note that even if $R_{ab}$ is not a ``gauge-invariant'' quantity, the combination $R_{ab}R_{ba}$ is. In fact, up to the total quantum dimension of the TQFT, it is the $S$-matrix element $S_{a\bar{b}}$.}
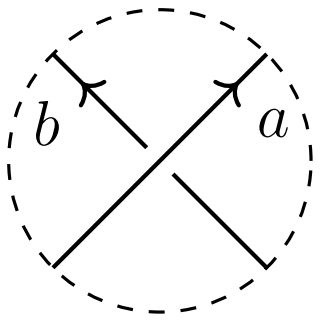
\begin{equation}
    {\begin{tikzpicture}[thick,
    baseline={([yshift=-.5ex]current bounding box.center)},
    overstrand/.style={preaction={draw, white, line width=7pt,
        shorten <=1pt, shorten >=1pt}},
    uparrow/.style={decoration={markings,
        mark=at position 0.85 with {\arrow[xshift=1pt]{>}}},
      postaction=decorate}]
  \draw[dashed, semithick] (0,0) circle (1);
  \draw[uparrow] (-45:1) -- (135:1); %
  \draw[overstrand, uparrow] (225:1) -- (45:1); %
  \node at (-0.75,0.25) {$b$};
  \node at (0.75,0.25)  {$a$};
\end{tikzpicture}} \ = R_{ab} R_{ba} ~ 
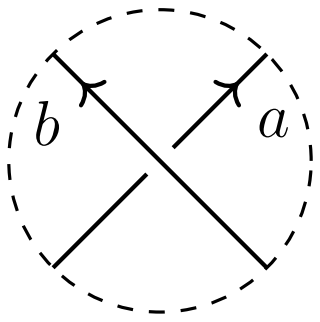
{\begin{tikzpicture}[thick,
    baseline={([yshift=-.5ex]current bounding box.center)},
    overstrand/.style={preaction={draw, white, line width=7pt,
        shorten <=1pt, shorten >=1pt}},
    uparrow/.style={decoration={markings,
        mark=at position 0.85 with {\arrow[xshift=1pt]{>}}},
      postaction=decorate}]
  \draw[dashed, semithick] (0,0) circle (1);
  \draw[uparrow] (225:1) -- (45:1); %
  \draw[overstrand, uparrow] (-45:1) -- (135:1); %
  \node at (-0.75,0.25) {$b$};
  \node at (0.75,0.25)  {$a$};
\end{tikzpicture}} \ .
\end{equation}

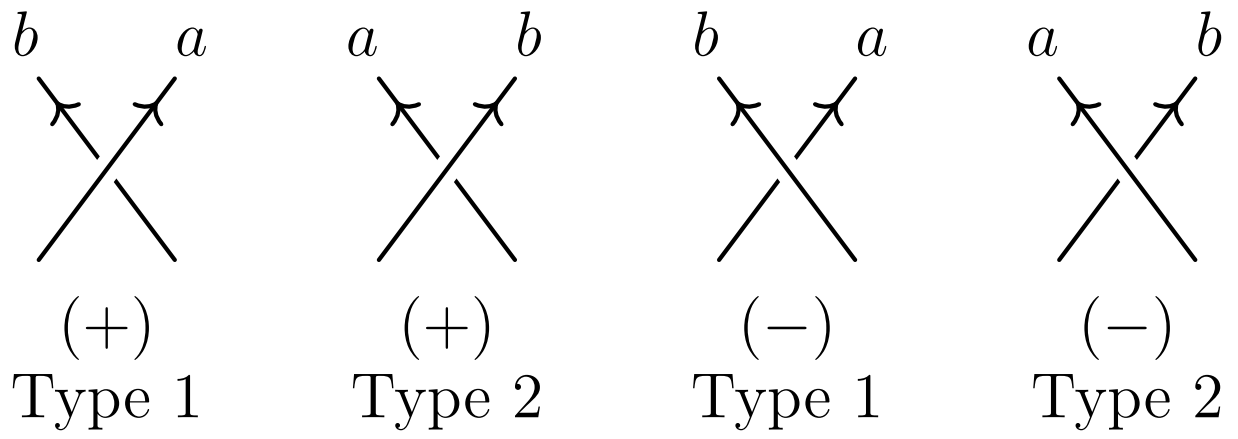
\begin{figure}[h!]
    \centering
    \begin{tikzpicture}[scale=0.75, thick, line cap=round,
    overstrand/.style={preaction={draw, white, line width=5pt,
        shorten <=1pt, shorten >=1pt}},
    uparrow/.style={decoration={markings,
        mark=at position 0.85 with {\arrow[xshift=1pt]{>}}},
      postaction=decorate}]
  \newcommand{\crossing}[6]{%
    \begin{scope}[shift={(#1,0)}]
      \ifnum#2=1
        \draw[uparrow] (1.2,0) -- (0,1.6);
        \draw[overstrand, uparrow] (0,0) -- (1.2,1.6);
      \else
        \draw[uparrow] (0,0) -- (1.2,1.6);
        \draw[overstrand, uparrow] (1.2,0) -- (0,1.6);
      \fi
      \node[above left]  at (0.2,1.6)   {$#3$};
      \node[above right] at (1,1.6) {$#4$};
      \node at (0.6,-0.6) {$(#5)$};
      \node at (0.6,-1.25) {Type #6};
    \end{scope}}

  \def\sep{3}
  \crossing{0}{1}{b}{a}{+}{1}
  \crossing{\sep}{1}{a}{b}{+}{2}
  \crossing{2*\sep}{2}{b}{a}{-}{1}
  \crossing{3*\sep}{2}{a}{b}{-}{2}
\end{tikzpicture}
    \caption{Positive $(+)$ and negative $(-)$ crossings of type 1 and type 2 between two lines $a$ and $b$.}
    \label{fig:posnegcrossings}
\end{figure}

Remember that one of the properties of a Brunnian link (with more than 2 components) is that pairwise linking numbers vanish, meaning there is an equal number of positive and negative crossings between any two components. Note also that positive and negative crossings come each in two flavors, which we refer to as type 1 and type 2, as shown in Figure \ref{fig:posnegcrossings}. We denote the total numbers of positive crossings of type 1 and 2 by $p_1$ and $p_2$, and those of negative crossings of type 1 and 2 by $n_1$ and $n_2$, respectively. If the linking number between the two components is zero, it is moreover true that $p_1 = n_2$ and $p_2 = n_1$.

Consider now a Brunnian link $\mathcal{L}$. We can turn it into the unlink by choosing any one component, whose label we denote by $a$, and replacing every undercrossing with the other components by an overcrossing. In other words, we replace all positive crossings of type 2 (negative crossings of type 1) by negative crossings of type 2 (positive crossings of type 1). After the change the component can be pulled away from the rest. The resulting link must then be an unlink by the Brunnian property. At the same time, the net effect on the expectation value of the link is trivial, since for any component $b \neq a$ of the link, we obtain a factor of
\begin{equation}
    (R_{ab}R_{ba})^{p_2}(R_{ab}R_{ba})^{-n_1} = 1.
\end{equation}
We conclude that the expectation value of $\mathcal{L}$ is the same as that of the unlink. \qed

\section{Computations of expectation values in the \texorpdfstring{$T^*\lie{h}_3$}{T*h\_3} CS theory} \label{app:BT*h3modular}

\subsection{Framed unknot}

We first compute the framed unknot expectation value in the $T^*\lie{h}_3$ CS theory at level $k$. For a Wilson line $W_{(\alpha,\beta)}[\gamma]$, with $\gamma$ the unknot, the path integral reads (see Section \ref{sec:5CSnilp})
\begin{multline}
    \int \mathcal{D}A\mathcal{D}B \exp \biggl\{ i \frac{k}{2\pi} \int_{S^3} B^i\wedge dA^i + i\frac{k^2}{(2\pi)^2} \int_{S^3} A^1 \wedge A^2 \wedge A^3 \biggr. \\
    \biggl. + i \alpha_i \oint_\gamma A^i + i \beta_i \qty(\oint_{\gamma}B^i - \frac{k}{4\pi}\epsilon_{ijk}A^j\phi^k) \biggr\}
\end{multline}
normalized by the partition function, times the factor $\frac{(2\pi)^2}{k\abs{\beta}} \delta\qty(\beta \times \oint_\gamma A)$ of Eq.\ (\ref{eq:T*h3Wilsonline}). It is convenient to write this factor through the integral representation of the delta function:
\begin{equation} \label{eq:appB-deltaintegral}
    \frac{(2\pi)^2}{k\abs{\beta}} \, \delta\qty(\beta \times \oint_\gamma A) = \frac{\abs{\beta}}{k} \int_{\beta^\perp} d^2s \, \exp\qty{i (s \times \beta) \cdot \oint_\gamma A},
\end{equation}
where the integral is over the plane orthogonal to $\beta$. The first step is to integrate out the $B$ fields, which imposes the equations of motion
\begin{equation}
    dA^i = -\frac{2\pi}{k}\beta_i \PD[\gamma].
\end{equation}
On $S^3$, these equations can be solved (up to exact terms) by setting
\begin{equation}
    A^i = -\frac{2\pi}{k}\beta_i \PD[\Sigma],
\end{equation}
where $\Sigma$ is a surface with boundary $\gamma$. Plugging this back into the path integral yields
\begin{multline}
    \exp \biggl\{ -\frac{2\pi i}{k}(\alpha \cdot \beta) \oint_\gamma \PD[\Sigma] -\frac{2\pi i}{k} \beta_1\beta_2\beta_3 \int_{S^3} \PD[\Sigma]^{\wedge 3} \biggr. \\
    \biggl. - \frac{i}{2} \frac{2\pi}{k} \epsilon_{ijk} \beta_i \beta_j \beta_k \oint_\gamma \PD[\Sigma] \int_{[\gamma(0), x]} \PD[\Sigma] \biggr\}.
\end{multline}
Note that $A^i$ immediately satisfies the Dirac delta constraints, so the factor (\ref{eq:appB-deltaintegral}) becomes
\begin{equation} \label{eq:appB-qdim}
    \frac{\abs{\beta}}{k} \int_{\beta^\perp} d^2s = k \abs{\beta} V^2,
\end{equation}
where $V \equiv \delta(0) \equiv \int_{\R}\frac{ds}{k}\,e^{\frac{2\pi isx}{k}}|_{x=0}$ is the volume of $\R$.

The various integrals in the above equation are ill-defined unless a regularization procedure is prescribed, which we do by means of the framing. The first term can be easily regularized by pushing off $\gamma$ along the +1 framing vector field --- this new curve $\gamma^{+1}$ then intersects $\Sigma$ once (with positive sign), thus
\begin{equation}
    \oint_{\gamma^{+1}} \PD[\Sigma] = +1.
\end{equation}

On the other hand, regularizing the quadratic and cubic terms requires some care. As seen in Section \ref{sec:CSperturbative} (from the point of view of perturbation theory), the sum of these two contributions yields the Milnor invariant $\mil(123)$ if mutual linking numbers of the (regularized) components vanish. Otherwise, the more detailed discussion in \cite{Mellor_2003} shows that this quantity is parametrization dependent --- more specifically, it depends on the position of the chosen basepoint for $\gamma$. Then, in order to have a gauge invariant regularization procedure, we push off $\gamma$ along its zero framing to compute these two terms (rather than the +1 framing), so that all pairwise linking numbers vanish.\footnote{In that sense, the framing of the line $W_{(\alpha,\beta)}[\gamma]$ affects only its ``electric'' component.} A more natural formulation of this regularization scheme is that for quantities directly depending on the surface $\Sigma$ (and not just on its boundary loop $\gamma$) we consider push-offs of $\Sigma$ in the normal direction, while for the quantities that involve $\gamma$ only we use its framing to define the push-off. It is easy to verify that, under this prescription, these terms vanish.

All in all, we find as a result that the framed unknot expectation value is
\begin{equation}
    \expval{W_{(\alpha,\beta)}[\gamma]} = k \abs{\beta} V^2 \exp(-\frac{2\pi i}{k}\alpha \cdot \beta).
\end{equation}

\subsection{Hopf link}

Now we proceed to find the expectation value of a Hopf link formed by two lines $W_{(\alpha^I,\beta^I)}[\gamma^I]$, $I = 1, 2$. The path integral reads
\begin{multline}
    \int \mathcal{D}A\mathcal{D}B \exp \biggl\{ i \frac{k}{2\pi} \int_{S^3} B^i\wedge dA^i + i\frac{k^2}{(2\pi)^2} \int_{S^3} A^1 \wedge A^2 \wedge A^3 \biggr. \\
    \biggl. + \sum_{I=1}^2 i \alpha^I_i \oint_{\gamma^I} A^i + i \beta^I_i \qty(\oint_{\gamma^I}B^i - \frac{k}{4\pi}\epsilon_{ijk}A^j\phi^k) \biggr\},
\end{multline}
normalized by the partition function, times the factors $\prod_{I=1}^2 \frac{(2\pi)^2}{k\abs{\beta^I}} \delta(\beta^I \times \oint_{\gamma^I} A)$, which we again write as in Eq.\ (\ref{eq:appB-deltaintegral}). Integrating out the $B$ fields enforces
\begin{equation}
    A^i = -\frac{2\pi}{k} \sum_{I=1}^2 \beta^I_i \PD[\Sigma^I]
\end{equation}
up to exact terms, where $\partial \Sigma^I = \gamma^I$. Plugging back, we get
\begin{multline}
    \exp \biggl\{ -\frac{2\pi i}{k} (\alpha^I \cdot \beta^J) \oint_{\gamma^I} \PD[\Sigma^J] -\frac{2\pi i}{k} \beta^I_1\beta^J_2\beta^K_3 \int_{S^3} \PD[\Sigma^I] \wedge \PD[\Sigma^J] \wedge \PD[\Sigma^K] \biggr. \\
    \biggl. - \frac{i}{2} \frac{2\pi}{k} \epsilon_{ijk} \beta^I_i \beta^J_j \beta^K_k \oint_{\gamma^I} \PD[\Sigma^J] \int_{[\gamma^I(0), x]} \PD[\Sigma^K] \biggr\},
\end{multline}
times the factors (\ref{eq:appB-deltaintegral}) evaluated on this solution, which give the ``square'' of the Dirac delta function imposing $\beta \times \beta' = 0$, as we will see below. Here and in what follows we denote $(\alpha,\beta) \equiv (\alpha^1,\beta^1)$ and $(\alpha',\beta') \equiv (\alpha^2,\beta^2)$. As in our previous computations, each term here must be properly regularized by considering pushoffs of the curves $\gamma^I$ along their zero framings. Let us analyze this term-by-term.

The integrals in the first term are zero whenever $I = J$, because pushing off $\gamma^I$ along the zero framing leads to a curve that does not intersect $\Sigma^I$. Whenever $I \neq J$, the integral is well-defined and gives $-1$, since $\gamma^I$ always intersects $\Sigma^J$ negatively (using the same orientation of the Hopf link as in Eq.\ (\ref{eq:definitionST})). The cubic term always vanishes by the following simple observation: it measures the triple intersection between (appropriate pushoffs of) the surfaces $\Sigma^{I,J,K}$, but if we use the zero framing this never leads to such an intersection. Finally, the quadratic term also vanishes due to the following reason. First, one can check that choosing $I = J$ or $I = K$ makes the integral immediately vanish, as $\gamma^I$ never intersects $\Sigma^{J}$ or $\Sigma^{K}$ in that case. The only surviving terms are those with $I \neq J$ and $J = K$, but then the integral immediately vanishes by symmetry: $\epsilon_{ijk}\beta_j^J\beta_k^K = 0$.

With the same regularization, the holonomies entering the factors (\ref{eq:appB-deltaintegral}) are $\oint_{\gamma^1} A = \frac{2\pi}{k} \beta'$ and $\oint_{\gamma^2} A = \frac{2\pi}{k} \beta$, so that these factors become
\begin{equation}
    \frac{\abs{\beta}\abs{\beta'}}{k^2} \int_{\beta^\perp} d^2s \int_{\beta'^\perp} d^2s' \, \exp\qty{\frac{2\pi i}{k} (s - s') \cdot (\beta \times \beta')} = k^2 \abs{\beta} \abs{\beta'} \, \delta^{(2)}(\beta \times \beta')^2.
\end{equation}
Indeed, the vector $\beta \times \beta'$ is orthogonal to both $\beta$ and $\beta'$, and each of the two integrals gives $k^2 \delta^{(2)}(\beta \times \beta')$. For $\beta \times \beta' = 0$ the right-hand side is the product of two factors of the form (\ref{eq:appB-qdim}).

Summing up, we find the final result
\begin{equation} \label{eq:CShopflinkresult}
    \expval{\prod_{I=1}^2 W_{(\alpha^I,\beta^I)}[\gamma^I]} = k^2 \abs{\beta} \abs{\beta'} \, \delta^{(2)}(\beta \times \beta')^2 \exp(\frac{2\pi i}{k} (\alpha \cdot \beta' + \alpha' \cdot \beta)).
\end{equation}

\section{The algebra isomorphism \texorpdfstring{$D(G) \simeq \C[\Lie{G}^{(p)}]$}{D(G) \textasciitilde{} C[G\textasciicircum(p)]}} \label{app:Cdictionary}

In this appendix we write the map $\Phi$ of Section \ref{sec:6corresp} explicitly, show that it is an isomorphism of algebras, and check that $\Phi^*$ acts on simple modules as stated in Eq.\ (\ref{eq:Phi*maprecipe}).

Before anything, let us fix some notation. As before, $G$ is a $p$-group of exponent $p$ and nilpotency class $c < p$, $\lie{g}^{(p)} = \log(G)$ is its Lie algebra over $\F_p$ with a basis $\qty{X_i}_{i=1}^n$, so that $\abs{G} = p^n$, and $\Lie{G}^{(p)} = \exp(T^*\lie{g}^{(p)})$. We also reserve the letters $X$ and $Y$ for the basis of $\lie{g}^{(p)}$ and the dual basis of $\lie{g}^{(p)*}$, as in Section \ref{sec:6corresp}. Finally, for a vector space $V$ over $\F_p$, we identify the dual space $V^*$ with its Pontryagin dual by sending $f \in V^*$ to the character $\zeta^{f( \cdot )}$ (with $\zeta = e^{2\pi i/p}$), following the discussion after Theorem \ref{thm:KirillovFinite}.

Let us remind the reader what the algebras involved in the isomorphism look like. First, the Drinfeld double $D(G)$ \cite{Roche:1990hs} is the algebra with basis $\qty{\delta_g h}_{g,h \in G}$ and product defined as
\begin{equation} \label{eq:appC-DG-product}
    \delta_g h \cdot \delta_{g'} h' = \delta_{g, hg'h^{-1}} \delta_g hh'.
\end{equation}
One can think of $\delta_g$ as a projector onto states with flux $g$, and of $h$ as a gauge transformation which acts on the fluxes by conjugation\footnote{The physics-inclined reader may find an illuminating presentation of Drinfeld doubles in the paper by Kitaev \cite{Kitaev_2003}, therein referred to as ``quantum doubles".}. The simple $D(G)$-module $M(a,\chi)$ (with $a$ and $\chi$ as in Section \ref{sec:DWnilp}) has the following explicit form. Let $\rho$ be the irreducible representation of $C_G(a)$ with character $\chi$, and let $V_\chi$ be its representation space. For each $x$ in the conjugacy class $[a]$, fix $g_x \in G$ such that $x = g_x a g_x^{-1}$. Then the module $M(a,\chi)$ has basis $\ket{x,v}$, where $x \in [a]$ and $v$ runs over a basis of $V_\chi$, and the action is
\begin{equation} \label{eq:appC-DG-module}
    U_{(a,\chi)}(\delta_g h) \ket{x,v} = \delta_{g,\, hxh^{-1}} \ket{hxh^{-1},\, \rho\qty(g_{hxh^{-1}}^{-1}\, h\, g_x) v}.
\end{equation}

Let us compute the character of the module $M(a, \chi)$. This is defined as the map $D(G) \to \C$ given by
\begin{equation}
    \delta_g h \longmapsto \Tr\qty(U_{(a,\chi)}(\delta_g h)).
\end{equation}
From Eq.\ (\ref{eq:appC-DG-module}), the only basis elements that contribute to the trace are $\ket{x,v}$ with $x = g$ and $hxh^{-1} = x$. For such $x$ and $h$, the element $g_x^{-1} h g_x$ belongs to $C_G(a)$, and the sum over $v$ gives the value of $\chi$ on it. For any $x \in [a]$, denote also
\begin{equation} \label{eq:appC-chi-x-def}
    \chi_x(h) \equiv \chi\qty(g_x^{-1}\, h\, g_x), \qquad h \in C_G(x),
\end{equation}
which is a character of $C_G(x) = g_x C_G(a) g_x^{-1}$. The character of the module is then
\begin{equation} \label{eq:appC-DG-character}
    \Tr\qty(U_{(a,\chi)}(\delta_x h)) = \begin{cases} \chi_x(h), & x \in [a], \ h \in C_G(x), \\ 0, & \text{otherwise}. \end{cases}
\end{equation}

On the other side of the map, we have the group algebra $\C[\Lie{G}^{(p)}]$. This is an algebra with basis $\{\ket{g}\}_{g \in \Lie{G}^{(p)}}$ and product
\begin{equation}
    \ket{g_1} \cdot \ket{g_2} = \ket{g_1g_2}.
\end{equation}
We can parametrize these basis elements in a more convenient way as follows. Note that the BCH formula implies that elements of the form $e^f$ generate an abelian subgroup of $\Lie{G}^{(p)}$, and moreover $\Lie{G}^{(p)}$ includes a copy of $G$ given by $\exp(\lie{g}^{(p)})$. Applying Eq.\ (\ref{eq:lazardconjadjintertwine}) to $\Lie{G}^{(p)}$, we have
\begin{equation} \label{eq:appC-semidirect}
    e^f e^{f'} = e^{f+f'}, \qquad h\, e^f h^{-1} = e^{f \circ \Ad_{h^{-1}}},
\end{equation}
for $f,f' \in \lie{g}^{(p)*}$, and $h \in G$. The $p^{2n}$ elements of the form $e^f h$ are all distinct, so every element of $\Lie{G}^{(p)}$ can be written in this form. The basis of the group algebra can then be parametrized as $\ket{e^fh}$. Note, finally, that the simple modules of the group algebra are exactly the irreducible representations of $\Lie{G}^{(p)}$.

\subsection{The map \texorpdfstring{$\Phi: D(G) \to \C[\Lie{G}^{(p)}]$}{Phi: D(G) -> C[G\textasciicircum(p)]}}

We define the map $\Phi$ on basis elements as follows:
\begin{equation} \label{eq:appC-Phi}
    \Phi(\delta_g h) = \frac{1}{p^n} \sum_{f \in \lie{g}^{(p)*}} \zeta^{-f(\log g)} \ket{e^f h}.
\end{equation}
Note that this map can be understood as a Fourier transform with respect to the flux. We can then use the orthogonality property
\begin{equation}
    \sum_{f\in \lie{g}^{(p)*}} \zeta^{f(A)} = p^n \delta_{A,0},
\end{equation}
to write down the inverse map explicitly:
\begin{equation} \label{eq:appC-Phi-inverse}
    \Phi^{-1}(\ket{e^f h}) = \sum_{g \in G} \zeta^{f(\log g)}\, \delta_g h.
\end{equation}
We now prove that $\Phi$ is an isomorphism of algebras. Since the map is invertible, we only need to show that it respects products. Denote $P_A \equiv \Phi(\delta_{e^A} 1) = p^{-n} \sum_f \zeta^{-f(A)} \ket{e^f}$, so that in general $\Phi(\delta_g h) = P_{\log g} \ket{h}$. See that
\begin{equation}
    P_A P_{A'} = \frac{1}{p^{2n}} \sum_{f,f'} \zeta^{-f(A) - f'(A')} e^{f+f'} = \frac{1}{p^{2n}} \sum_{\tilde{f}} \zeta^{-\tilde{f}(A')} e^{\tilde{f}} \sum_{f} \zeta^{f(A'-A)} = \delta_{A,A'} P_A.
\end{equation}
After changing the summation variable to $f' = f \circ \Ad_{h^{-1}}$ (which is possible since the adjoint action on the Lie algebra is an automorphism), we obtain
\begin{equation}
    h P_A h^{-1} = \frac{1}{p^n} \sum_{f} \zeta^{-f(A)} e^{f \circ \Ad_{h^{-1}}} = \frac{1}{p^n} \sum_{f'} \zeta^{-f'(\Ad_h A)} e^{f'} = P_{\Ad_h A}.
\end{equation}
Finally, taking into account that $\Ad_h \log g' = \log(hg'h^{-1})$, it now follows that
\begin{multline}
    \Phi(\delta_g h)\, \Phi(\delta_{g'} h') =  P_{\log g} h\, P_{\log g'} h' \\
    = P_{\log g} \qty(h P_{\log g'} h^{-1}) hh' = \delta_{g,\, hg'h^{-1}}\, P_{\log g}\, hh' = \Phi\qty((\delta_g h)(\delta_{g'} h')),
\end{multline}
as promised.

The proof above uses only the fact that $\exp$ intertwines the adjoint action with conjugation: the multiplication in $G$ does not enter. The latter is instead important in the definition of the coproduct of $D(G)$,
\begin{equation}
    \Delta(\delta_g h) = \sum_{g'g'' = g} \delta_{g'} h \otimes \delta_{g''} h,
\end{equation}
which endows it with the structure of a \textit{Hopf algebra}. The coproduct of $\C[\Lie{G}^{(p)}]$ transported by $\Phi$ is given by the same formula, but with the sum over pairs satisfying $\log g' + \log g'' = \log g$. Therefore, $\Phi$ is not an isomorphism of Hopf algebras, unless $G$ is abelian. This means that although $\Phi$ identifies the simple objects of $\Rep(D(G))$ with the irreducible representations of $\Lie{G}^{(p)}$, it does not respect their tensor product structure.

\subsection{Coadjoint orbits of \texorpdfstring{$\Lie{G}^{(p)}$}{G\textasciicircum(p)}}

By Theorem \ref{thm:KirillovFinite} the irreducible representations of $\Lie{G}^{(p)}$ are labelled by the coadjoint orbits in $(T^*\lie{g}^{(p)})^*$. A functional $\varphi$ in the dual Lie algebra takes the general form $e\cdot X^* + q\cdot Y^*$. We will denote hereafter
\begin{equation} \label{eq:appC-EQ}
    \mathcal{E} = e \cdot X^* \in \lie{g}^{(p)*}, \qquad \mathcal{Q} = q \cdot X \in \lie{g}^{(p)},
\end{equation}
so that $\varphi((A, f)) = \mathcal{E}(A)+f(\mathcal{Q})$, and will also identify $\varphi$ with its associated pair $(\mathcal{E},\mathcal{Q})$.\footnote{Note we are using the identification $(\lie{g}^{(p)*})^* \cong \lie{g}^{(p)}$ to regard $q \cdot Y^*$ as the element $q \cdot X \in \lie{g}^{(p)}$.} We will refer to these as the \textit{electric} and \textit{magnetic} charges, respectively. It will also be useful to write down explicitly the coadjoint action of elements $h \in G$ and $e^f$ (with $f \in \lie{g}^{(p)*}$) (understood as elements of $\Lie{G}^{(p)}$) on these functionals:
\begin{equation} \label{eq:appC-coadjoint}
    \Ad^*_h (\mathcal{E},\mathcal{Q}) = \qty(\mathcal{E} \circ \Ad_{h^{-1}}, \Ad_h \mathcal{Q}), \qquad \Ad^*_{e^f} (\mathcal{E},\mathcal{Q}) = \qty(\mathcal{E} + f \circ \ad_{\mathcal{Q}}, \mathcal{Q}).
\end{equation}
As we mentioned previously any element of $\Lie{G}^{(p)}$ is uniquely expressed as a product of two such $h$ and $e^f$, so this determines the whole coadjoint action.

Let us now describe the coadjoint orbit $\orb{\varphi}$ of $\varphi = (\mathcal{E}, \mathcal{Q})$. Denote $a \equiv e^{\mathcal{Q}}$. From Eq.\ (\ref{eq:appC-coadjoint}) the elements $e^f$ do not change $\mathcal{Q}$, while $h \in G$ has the effect of sending $\mathcal{Q} \mapsto \Ad_h\mathcal{Q} = \log(hah^{-1})$. Therefore magnetic charges of functionals in $\orb{\varphi}$ are all of the form $\log(x)$, for $x \in [a]$.

Consider now    those functionals with a fixed magnetic charge $\mathcal{Q}$. They are obtained from $\varphi$ by the action of elements $e^f h$ with $h \in C_G(a)$:
\begin{equation} \label{eq:appC-fixed-flux}
    \Ad^*_{e^f h} (\mathcal{E},\mathcal{Q}) = \qty(\mathcal{E} \circ \Ad_{h^{-1}} + f \circ \ad_{\mathcal{Q}}, \mathcal{Q}).
\end{equation}
Denote by $\lie{c}(\mathcal{Q})$ the Lie subalgebra $\log(C_G(a))$, c.f.\ Equation (\ref{eq:lazardcentralizer}). Then the functional $f \circ \ad_{\mathcal{Q}}$ vanishes on $\lie{c}(\mathcal{Q})$, and conversely any functional in $\lie{g}^{(p)*}$ that vanishes on this subalgebra is of this form. That is,
\begin{equation} \label{eq:appC-perp}
    \lie{c}(\mathcal{Q})^\perp = \qty{f \circ \ad_{\mathcal{Q}} \mid f \in \lie{g}^{(p)*}} = \qty{\nu \in \lie{g}^{(p)*} \mid \nu|_{\lie{c}(\mathcal{Q})} = 0}.
\end{equation}
Therefore, when $f$ runs over $\lie{g}^{(p)*}$ at fixed $h$, the electric charge runs over all functionals with the same restriction to $\lie{c}(\mathcal{Q})$ as $\mathcal{E} \circ \Ad_{h^{-1}}$. For $h \in C_G(a)$ the map $\Ad_{h^{-1}}$ further preserves $\lie{c}(\mathcal{Q})$, so this restriction is $\psi \circ \Ad_{h^{-1}}$, with
\begin{equation}
    \psi = \mathcal{E}\eval_{\lie{c}(\mathcal{Q})} \in \lie{c}(\mathcal{Q})^* = \log(C_G(a))^*.
\end{equation}
When $h$ runs over $C_G(a)$, the functionals $\psi \circ \Ad_{h^{-1}}$ form the coadjoint orbit of $\psi$ with respect to the group $C_G(a)$, which we denote by $\widetilde{\mathcal{O}}_\psi$. To summarize, $(\mathcal{E}', \mathcal{Q})$ belongs to the orbit $\orb{\varphi}$ if and only if $\mathcal{E}'\eval_{\lie{c}(\mathcal{Q})} \in \widetilde{\mathcal{O}}_\psi$.

The last thing to consider is elements of $\orb{\varphi}$ with magnetic charge $\log(x)$, with $x = gag^{-1}$ for some $g \in G$. They can be obtained by considering arbitrary elements $(\mathcal{E}', \mathcal{Q})$ with fixed magnetic charge (as in the previous paragraph) and acting by $\Ad^*_{g}$. This action maps $\lie{c}(\log(x))$ isomorphically to $\lie{c}(\mathcal{Q})$, so the condition on electric charges becomes that its restriction to $\lie{c}(\log(x))$ belongs to the set
\begin{equation} \label{eq:appC-orbit-x}
    \widetilde{\mathcal{O}}^x_\psi = \qty{\lambda \circ \Ad_{g^{-1}} \mid \lambda \in \widetilde{\mathcal{O}}_\psi} \subset \lie{c}(\log(x))^*.
\end{equation}
This set depends on $x$, but not on the particular element $g$ we use to obtain it from $a$ by conjugation. Combining everything together, we find that the coadjoint orbit of $\varphi$ can be understood as
\begin{equation} \label{eq:appC-orbit}
    \orb{\varphi} = \qty{(\mathcal{E}', \mathcal{Q}') \mid x = e^{\mathcal{Q}'} \in [a], \mathcal{E}'\eval_{\lie{c}(\mathcal{Q}')} \in \widetilde{\mathcal{O}}^x_\psi}.
\end{equation}
It follows that $\abs{\orb{\varphi}} = |[a]|^2 |\widetilde{\mathcal{O}}_\psi|$: there are $|[a]|$ possible magnetic charges, $|\widetilde{\mathcal{O}}_\psi|$ restrictions for each one of them, and $\abs{\lie{c}(\mathcal{Q})^\perp} = |G/C_G(a)| = |[a]|$ electric charges for each restriction.

The following will be useful later on. To apply the character formula (\ref{eq:finiteorbitchar}) to $e^f h$, we also need its logarithm. If $h = e^A$, then
\begin{equation} \label{eq:appC-log}
    e^f e^A = \exp(A + f \circ S_A), \qquad S_A \equiv \frac{\ad_A}{1 - e^{-\ad_A}} = 1 + \frac{1}{2} \ad_A + \frac{1}{12} \ad_A^2 - \frac{1}{720} \ad_A^4 + \dots.
\end{equation}
Namely, all terms of the BCH formula with two or more entries of $f$ vanish, because $\lie{g}^{(p)*}$ is an abelian ideal. The terms with a single entry of $f$ are known in a closed form and sum up to $\frac{\ad_A}{e^{\ad_A} - 1} f$, and $\ad_A^m f = (-1)^m f \circ \ad_A^m$. We will only need the fact that $S_A \mathcal{Q}' = \mathcal{Q}'$ when $[A,\mathcal{Q}'] = 0$.

\subsection{The map \texorpdfstring{$\Phi^*: \Rep(\C[\Lie{G}^{(p)}]) \to \Rep(D(G))$}{Phi*: Rep(C[G\textasciicircum(p)]) -> Rep(D(G))}} \label{subsec:appC-Phi*}

In this section, we will prove the following statement.

\begin{proposition} \label{prop:appC-dictionary}
    Let $(a,\chi)$ be a simple $D(G)$-module, $\psi \in \log(C_G(a))^*$ be a representative of the coadjoint orbit corresponding to $\chi$, and $\varphi = e \cdot X^* + q \cdot Y^*$ be as in Eq.\ (\ref{eq:Phi*maprecipe}), that is $\mathcal{Q} = \log a$ and $\mathcal{E}|_{\lie{c}(\mathcal{Q})} = \psi$. Then
    \begin{equation} \label{eq:appC-characters}
        \Tr U_{(a,\chi)}\qty(\Phi^{-1}(\ket{k})) = \pi_\varphi(k), \qquad \forall k \in \Lie{G}^{(p)},
    \end{equation}
    where $\pi_\varphi$ is the character (\ref{eq:finiteorbitchar}) of the irreducible representation of $\Lie{G}^{(p)}$ associated to the coadjoint orbit of $\varphi$. In other words, $\Phi^*$ maps this representation to $(a,\chi)$.
\end{proposition}

Let $k = e^f h$ with $h = e^A$. From Eqs.\ (\ref{eq:appC-DG-character}) and (\ref{eq:appC-Phi-inverse}) we have
\begin{equation} \label{eq:appC-F}
    \Tr U_{(a,\chi)}\qty(\Phi^{-1}(e^f h)) = \sum_{x \in G} \zeta^{f(\log x)}\, \Tr U_{(a,\chi)}(\delta_x h) = \sum_{\substack{x \in [a] \\ h \in C_G(x)}} \zeta^{f(\log x)}\, \chi_x(h).
\end{equation}
The centralizer $C_G(a)$ is a $p$-group of nilpotency class less than $p$ with Lie algebra $\lie{c}(\mathcal{Q})$. By Theorem \ref{thm:KirillovFinite} we have $\chi(e^B) = |\widetilde{\mathcal{O}}_\psi|^{-1/2} \sum_{\lambda \in \widetilde{\mathcal{O}}_\psi} \zeta^{\lambda(B)}$ for $B \in \lie{c}(\mathcal{Q})$. Applying this to $\eta = \Ad_{g_x^{-1}} \xi$ in the definition (\ref{eq:appC-chi-x-def}), and using Equation (\ref{eq:appC-orbit-x}), we obtain
\begin{equation} \label{eq:appC-chi-x}
    \chi_x(e^A) = \frac{1}{|\widetilde{\mathcal{O}}_\psi|^{1/2}} \sum_{\lambda \in \widetilde{\mathcal{O}}_\psi^x} \zeta^{\lambda(A)},
\end{equation}
for $e^A \in C_G(x)$. On the other hand, the character formula (\ref{eq:finiteorbitchar}) for $\Lie{G}^{(p)}$, together with Eqs.\ (\ref{eq:appC-log}), (\ref{eq:appC-EQ}) and (\ref{eq:appC-orbit}), gives
\begin{multline} \label{eq:appC-orbit-sum}
    \pi_\varphi(e^f e^A) = \frac{1}{\abs{\orb{\varphi}}^{1/2}} \sum_{(\mathcal{E}',\mathcal{Q}') \in \orb{\varphi}} \zeta^{\mathcal{E}'(A) + f(S_A \mathcal{Q}')} = \\
    = \frac{1}{\abs{[a]} |\widetilde{\mathcal{O}}_\psi|^{1/2}} \sum_{x \in [a]} \zeta^{f(S_A \log x)} \sum_{\lambda \in \widetilde{\mathcal{O}}_\psi^x} \; \sum_{\mathcal{E}'|_{\lie{c}(\log x)} = \lambda} \zeta^{\mathcal{E}'(A)}.
\end{multline}
The functionals $\mathcal{E}' \in \lie{g}^{(p)*}$ with a given restriction $\lambda$ form a coset of $\lie{c}(\log x)^\perp$, which is the dual space of $\lie{g}^{(p)}/\lie{c}(\log x)$. By orthogonality the last sum vanishes unless $A \in \lie{c}(\log x)$, in which case it is equal to $\abs{[a]} \zeta^{\lambda(A)}$, and moreover $S_A \log x = \log x$. By Eq.\ (\ref{eq:lazardcentralizer}) the condition $A \in \lie{c}(\log x)$ is equivalent to $h \in C_G(x)$. Therefore
\begin{equation}
    \pi_\varphi(e^f h) = \sum_{\substack{x \in [a] \\ h \in C_G(x)}} \zeta^{f(\log x)}\, \frac{1}{|\widetilde{\mathcal{O}}_\psi|^{1/2}} \sum_{\lambda \in \widetilde{\mathcal{O}}_\psi^x} \zeta^{\lambda(A)},
\end{equation}
which coincides with the right-hand side of Eq.\ (\ref{eq:appC-F}). \qed

A few remarks are in order. First, the charges $e$ are not fixed uniquely by Eq.\ (\ref{eq:Phi*maprecipe}). Different extensions $\mathcal{E}$ of the same $\psi$ differ by elements of $\lie{c}(\mathcal{Q})^\perp$, and by Eqs.\ (\ref{eq:appC-coadjoint}) and (\ref{eq:appC-perp}) they are related by the coadjoint action of elements $e^f \in \Lie{G}^{(p)}$. A different choice of $\psi$ in its orbit, or of the representative $a$ of the conjugacy class, corresponds to the coadjoint action of elements in $G \subset \Lie{G}^{(p)}$. Second, the dimensions agree, since $\abs{[a]}\, \chi(1) = \abs{[a]} |\widetilde{\mathcal{O}}_\psi|^{1/2} = |\orb{\varphi}|^{1/2}$. Finally, $\mathcal{Q}$ is central in $\lie{c}(\mathcal{Q})$, so all $\lambda \in \widetilde{\mathcal{O}}_\psi$ take the same value on it. The $T$-matrix element (\ref{eq:DWTmatrix}) of the line $(a,\chi)$ is then
\begin{equation} \label{eq:appC-twist}
    \frac{\chi(a)}{\chi(1)} = \zeta^{\psi(\mathcal{Q})} = \zeta^{\mathcal{E}(\mathcal{Q})} = \zeta^{e \cdot q}.
\end{equation}
Note that $e \cdot q$ is a function on $(T^*\lie{g}^{(p)})^*$ invariant under the coadjoint action.

\bibliographystyle{JHEP}
\bibliography{biblio.bib}

\end{document}